\documentclass[%
 reprint,
superscriptaddress,
nofootinbib,
nobibnotes,
 amsmath,amssymb,
 aps,
prd,
floatfix,
longbibliography
]{revtex4-1}

\usepackage{color}
\usepackage{array} 
\usepackage{amsmath}
\usepackage{float}
\usepackage{graphicx}
\usepackage{epstopdf}
\usepackage[labelfont=bf,labelsep=period,justification=raggedright,singlelinecheck=false,font=small]{caption}
\usepackage{array}

\usepackage{wrapfig}
\usepackage{verbatim}
\usepackage{amssymb}
\usepackage[normalem]{ulem}
\usepackage{hyperref}
\usepackage[T1]{fontenc}
\usepackage[utf8]{inputenc}
\usepackage{comment}
\usepackage{multirow}
\usepackage{placeins} % For \FloatBarrier
\usepackage{subfigure}
\usepackage{threeparttable}
\usepackage{booktabs}
\usepackage{orcidlink}
\usepackage{dcolumn}% Align table columns on the decimal point
\usepackage{bm}% bold math
\usepackage[mathlines]{lineno}% Enable numbering of text and display math
\usepackage{tabularx}
\usepackage[export]{adjustbox}
\usepackage{url}
\usepackage{academicons}
\usepackage{xcolor}

\newcommand{\msun}{{\rm M}_{\odot}}

\hypersetup{
    pdfnewwindow=true,    % links in new window
    colorlinks=true,      % false: boxed links; true: colored links
    linkcolor=blue,       % color of internal links
    citecolor=blue,       % color of links to bibliography
    filecolor=blue,       % color of file links
    urlcolor=blue         % color of external links
}

\begin{document}

\title{Cosmological Zoom-In Simulations of Milky Way Host Mass Dark Matter Halos with a Blue-Tilted Primordial Power Spectrum}

\author{Jianhao Wu \orcidlink{0009-0000-7431-7885}}
 \email{contact author: jianhao.wu@link.cuhk.edu.hk} 
\affiliation{
 Department of Physics, The Chinese University of Hong Kong, Shatin, Hong Kong, China
}
\author{Tsang Keung Chan \orcidlink{0000-0003-2544-054X}}
 \email{contact author: tsangkeungchan@cuhk.edu.hk}
\affiliation{
 Department of Physics, The Chinese University of Hong Kong, Shatin, Hong Kong, China
}
\author{Victor J. Forouhar Moreno \orcidlink{0000-0003-1308-9908}}
\affiliation{
Leiden Observatory, Leiden University, PO Box 9513, NL-2300 RA Leiden, the Netherlands
}

\date{\today}

% ===== Abstract ===== #
\begin{abstract}

Recent observations from the James Webb Space Telescope revealed a surprisingly large number of galaxies at high redshift, challenging the standard Lambda Cold Dark Matter cosmology with a power-law primordial power spectrum. Previous studies alleviated this tension with a blue-tilted primordial power spectrum ($P(k)\propto k^{m_s}$ with $m_s>1$ at small scales $>1~{\rm cMpc}^{-1}$).

In this study, we examine whether the blue-tilted model can boost dark matter substructures especially at low redshift, thereby addressing other potential challenges to the standard cosmology. First, substructures in the standard cosmological model may not be sufficient to explain the anomalous flux ratio problem observed in strong gravitational lensing. Second, the number of observed nearby satellite galaxies could be higher than the theoretical predictions of the standard cosmology, after completeness correction and tidal stripping by baryonic disks.

To study the impact of a blue-tilted primordial power spectrum on substructures, we perform high-resolution cosmological zoom-in dark matter-only simulations of Milky Way host mass halos, evolving to redshift $z=0$. At $z=0$, we find that the blue-tilted subhalo mass functions can be enhanced by more than a factor of two for subhalo masses $M_{\rm sub} \lesssim 10^{10}~\msun$, whereas the subhalo $V_{\rm max}$ functions can be enhanced by a factor of four for maximum circular velocities $V_{\rm max}\lesssim 30 ~{\rm km/s}$. The blue-tilted scaled cumulative substructure fraction can be an order of magnitude higher at $\sim$10\% of the virial radius. The blue-tilted subhalos also have higher central densities, since the blue-tilted subhalos reach the same $V_{\rm max}$ at a smaller distance $R_{\rm max}$ from the center. We have also verified these findings with higher-resolution simulations.

\end{abstract}

\maketitle

\section{Introduction\label{intro}}

The standard cosmology model includes the single-field slow-roll inflation model \cite{Guth81inflation,Lidd93inflation,Stei84inflation,Salopek1990inflation,Lidd1992inflation}, which predicts the approximately single power-law (PL) primordial power spectrum (PPS) \citep{Lyth99inflation}, and the lambda cold dark matter model (LCDM, \cite{Peeb93cosmology,Fren1985LCDMsim}), which governs the later evolution of our universe. It has achieved great successes on large scales, e.g. supported by the cosmic microwave background (CMB) \cite{Planck20CMB,Planck20cmbcosmology}, galaxy surveys \cite{Troxel17DesCosmicShear, Blanton17SDSSsurvey} and Lyman-alpha forest \cite{Chabanier19LyalphaPS}. However, there are relatively sparse constraints for the scales smaller than $\sim 1~{\rm cMpc}^{-1}$. In particular, there have been active debates over the so-called small-scale crisis of LCDM \cite{Bullock17SmallScaleCrisis}, including the Missing Satellite Problem (MSP, \citep{Klypin99MSP,Moor99MSP}), the Core-Cuspy problem \citep{Salucci2000CoreCuspyObserve, Flores94CoreCuspySimulation},  and the Too-Big-To-Fail problem \citep{Baylan11TBTF}, etc. This leaves room for other models beyond the standard cosmology, including different models of dark matter, effects of baryons, and modifications to the PPS. 

One possible modification is to enhance the PPS on small scales (${\rm > 1 ~{\rm cMpc}^{-1}}$) \citep{Covi99BTPS, Martin01BTPS, Gong11BTPS}, deviating from the PL PPS. In these small-scale enhanced models, PPS will follow the traditional PL model on large scales, preserving the success of the PL PPS, while having an enhancement on small scales. For example, Ref. \cite{Hira15bluetilt} considered blue-tilted (BT) models, which are broken power law PPSs with a higher power law index in scales $> 1 ~{\rm cMpc}^{-1}$. They found a significant boost in the number of dark matter halos with BT PPS, especially at high redshift. Hence, using the BT model,  Ref. \cite{Hira15bluetilt} explained the large number of high redshift massive galaxies observed by the James Webb Space Telescope (JWST) \cite{Gardner06JWST}, which is unexpected from the PL model. More complicated forms of enhancements also exist, such as several bumps in the PPS \cite{Tkac24bumpyps}.

Instead of focusing on high redshift, in this study, we examine the impact of the BT model on the low redshift dark matter substructures. There are two major motivations for this: (1) the anomalous flux ratio problem in strong gravitational lensing and (2) potentially the too many satellite problem.

The strong gravitational lensing observations can probe invisible substructures, so it can constrain dark matter substructures and, thus, the cosmological model. For example, the asymptotic flux ratio relation, which serves as a general principle in smooth lensing potentials \citep{Mao92SmoothPotentialLensing}, is often observed to be violated. This is taken as evidence for the presence of the substructures around lensing galaxies \citep{Mao98LensingSubstructure}. However, in the PL cosmology, there could be not enough low mass substructures in lensing galaxies to explain the violations \citep{Macc06CDMlensing,Xu09AquariusLensing,Xu15lensingCDM}. This could suggest extra structures not present in dark matter-only simulations under the standard PL cosmology.  On the other hand, Ref. \cite{Gilm22stronglensingPPS} constrained sub-galactic PPS to be consistent with PL with the 11 strongly lensed and quadruply image quasars, albeit with possibilities with blue/red tilted PPS on small scales.

Another motivation for the BT model is that the number of satellite galaxies could be higher than the expected from the PL model. Thanks to the recent improvements in observations,  many more ($\gtrsim 50$) faint satellite galaxies have been discovered in Milky Way (MW) and nearby MW mass hosts \cite{Drli15DES,Bech15DES,Kopo15UFS,Homm23MWmanydwarf,Mull24M83}. After correcting for the detection efficiency (completeness correction), Ref. \cite{Kim18NoMSP} 
argued that there could be too many satellite galaxies inferred from observations than expected in simulations/models. This happens if some mechanisms can reduce the number of satellite galaxies in simulations / models. For example, a baryonic disk can tidally strip subhalos \cite{Garrison-Kimmel17notsolumpy}, and/or the minimum subhalo mass to host an observable galaxy is large \cite{Grau19,Jeth18minhalo}. This ``too many satellites problem'' motivates us to explore mechanisms to boost the number of subhalos, which includes the BT model.

A related constraint on the primordial power spectrum, though not the focus of this paper, is the density profiles of nearby low-mass dwarf galaxies. For example, the observed central densities of dwarf galaxies can provide insights into the nature of dark matter \cite{Roch13SIDM,Yang24SIDM}  or constraining baryonic physics \cite{DiCi14cuspcore,Chan15,Toll16NIHAOcore,Laza20coreddarkmatter}. Refs. \cite{Este23lumpy,Dekker24dwarfgalaxyBT} employed semi-analytic models to constrain the BT PPS parameter space, using the observed relation between half-light radius and V-band luminosity function in nearby dwarf galaxies.

Motivated by strong lensing, and nearby galaxy observations, we explore the impact of the BT PPS on the substructures in dark matter halos at low redshift, which, to our knowledge, has not been previously investigated with numerical simulations. We perform cosmological zoom-in dark matter-only simulations of MW host mass halos with BT PPS to study its substructures at $z=0$. Note that we only modify the PPS but still consider only cold dark matter (i.e. not warm or self-interacting dark matter).

In \autoref{sec:theory}, we will give a theoretical overview, laying down the framework for the BT PPS and our simulation. Then in \autoref{sec:numerical}, we introduce our numerical codes for initial conditions, cosmological simulations, and halo identification. We present our main results in \autoref{sec:result} and conclude in \autoref{sec:conclusion}. Additionally, as the first work to use HBT-HERONS (see \autoref{sub:halofinder}) for zoom-in simulations, we show the comparison between HBT-HERONS and another halo finder VELOCIraptor in \autoref{sec:VR}. We also run higher-resolution simulations and compare the fiducial results in \autoref{sec:res}.

\section{Theoretical Overview}
\label{sec:theory}

\begin{figure*}
    \centering
    \includegraphics[scale=0.25]{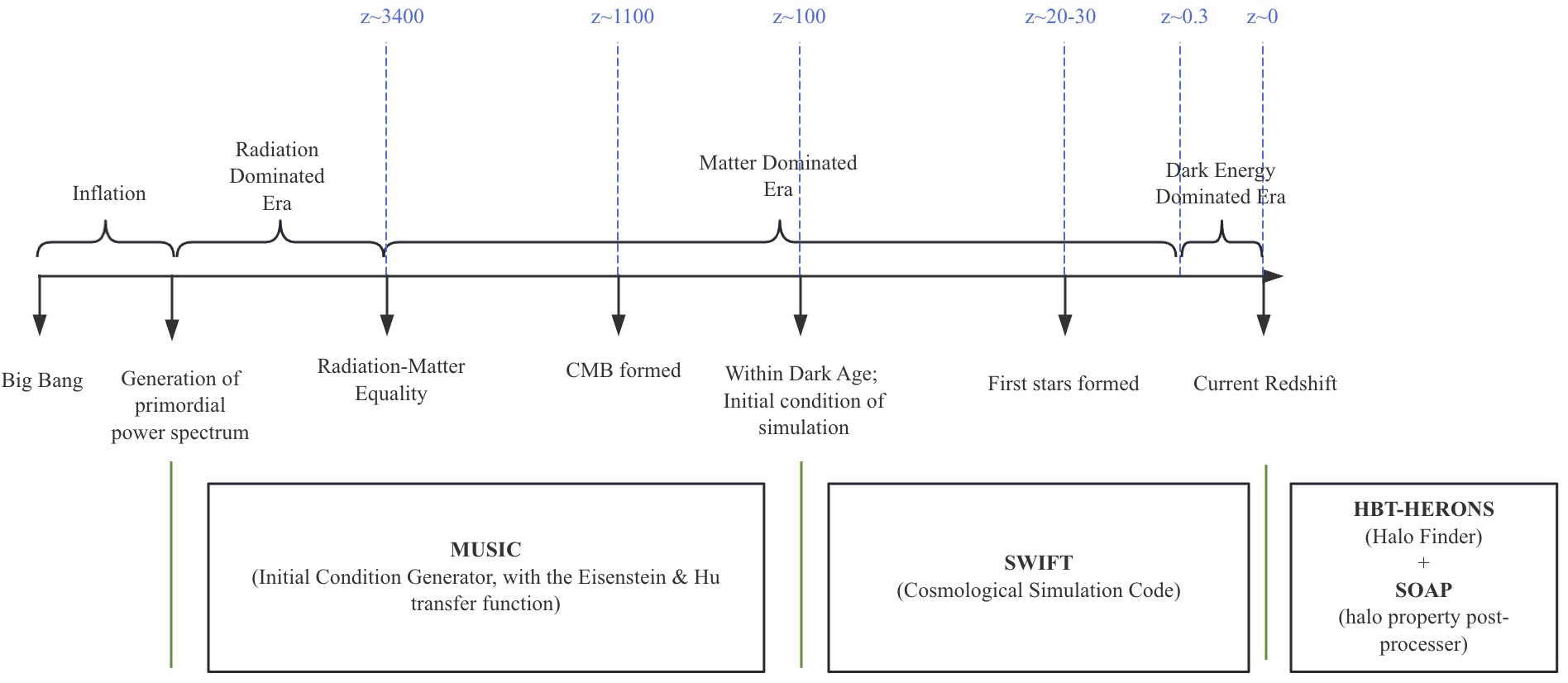}
    \caption{The conceptual steps of the cosmological stages and our simulation steps. The \textit{upper panel} shows the universe's chronology, along with redshifts of some critical eras; the \textit{lower panel} shows the numerical tools we used.}
    \label{fig:conceptual_flow}
\end{figure*}

The upper panel of \autoref{fig:conceptual_flow} shows the cosmic chronology. Shortly after the Big Bang, the primordial power spectrum (PPS) and small inhomogeneities were established during the inflation era.  At the cosmic recombination $z\sim 1100$, electrons are combined with protons to form neutral hydrogens, allowing radiation to travel freely and resulting in the cosmic microwave background (CMB). Under the influence of gravity, the small density fluctuations are amplified, which can still be described by Lagrangian perturbation theories. By $z\sim 100$, non-linearities started to grow and more dark matter collapsed into halos, requiring N-body simulations to model accurately. Eventually, stars and galaxies are born at the centers of dark matter halos, leading to the structure we observe today. Our study connects the very early universe to the present-day nearby galaxy structures and statistics.

The matter power spectrum in the post-recombination era is connected to the primordial power spectrum $P_i(k)$ through \cite{Mo10galaxybook}:
\begin{align}
    P(k,t) = P_i(k) T^2(k) D^2(t),
    \label{eq:P_MD}
\end{align}
where $T(k)$ is the linear transfer function and $D(t)$ is the growth factor. In the traditional single-field slow-roll inflation, the PPS follows the PL model:
\begin{align}
    P_i(k) \propto k^{n_s},
    \label{eq:Pk_PL}
\end{align}
with the spectral index $n_s \sim 0.96$ (see \autoref{subsub:cosmo_params}).

Ref. \cite{Hira15bluetilt} gave the following formalism for the BT models:
\begin{align}
     P_i(k) \propto \left\{\begin{matrix}
 k^{n_s}, &  ({\rm for}\; k\leq k_p), \\
 k^{n_s} \cdot {(\frac{k}{k_p})}^{m_s-n_s}, & ({\rm for}\; k > k_p), \\
\end{matrix}\right.  
\end{align}
which is a broken power law modification of \autoref{eq:Pk_PL}. We introduce $k_p$ as the pivot scale and $m_s$ as the spectral index on small scales. We require $m_s$ to be larger than $n_s$ to give an enhancement (i.e. blue-tilted, BT). When $m_s$ is smaller than $n_s$, the corresponding PPS model is called the red tilted (RT) model \cite{Dekker24dwarfgalaxyBT}.

In this paper, along with the PL model, we pick two BT models from Ref. \cite{Hira24bluetilt}\footnote{For $k_p$, the value in our table is a float number to several digits because original models use $h~\rm{cMpc}^{-1}$ as a unit of $k_{p}$, but we convert it into ${\rm cMpc}^{-1}$.}: BT\_deep and BT\_soft, listed in \autoref{tab:Params_all_models}. BT\_soft with a reasonable star formation efficiency $\sim 0.2$ can explain the large number of galaxies formed at high redshift as observed by JWST \cite{Hira24bluetilt}. However, to match the JWST observations, BT\_deep requires a high star formation efficiency, comparable to PL \cite{Hira24bluetilt}. Although BT\_deep does not help to explain the JWST observations, it is still interesting to examine its impact on the substructures of dark matter halo.

Both models are within the allowable parameter space in Ref. \cite{Dekker24dwarfgalaxyBT} (constrained with nearby dwarf central densities), although BT\_soft is the marginal case. We also confirm that our models are within the parameter space constrained with strong lensing \cite{Gilm22stronglensingPPS}.

\begin{figure}
    \centering
    \includegraphics[width={0.45\textwidth}]{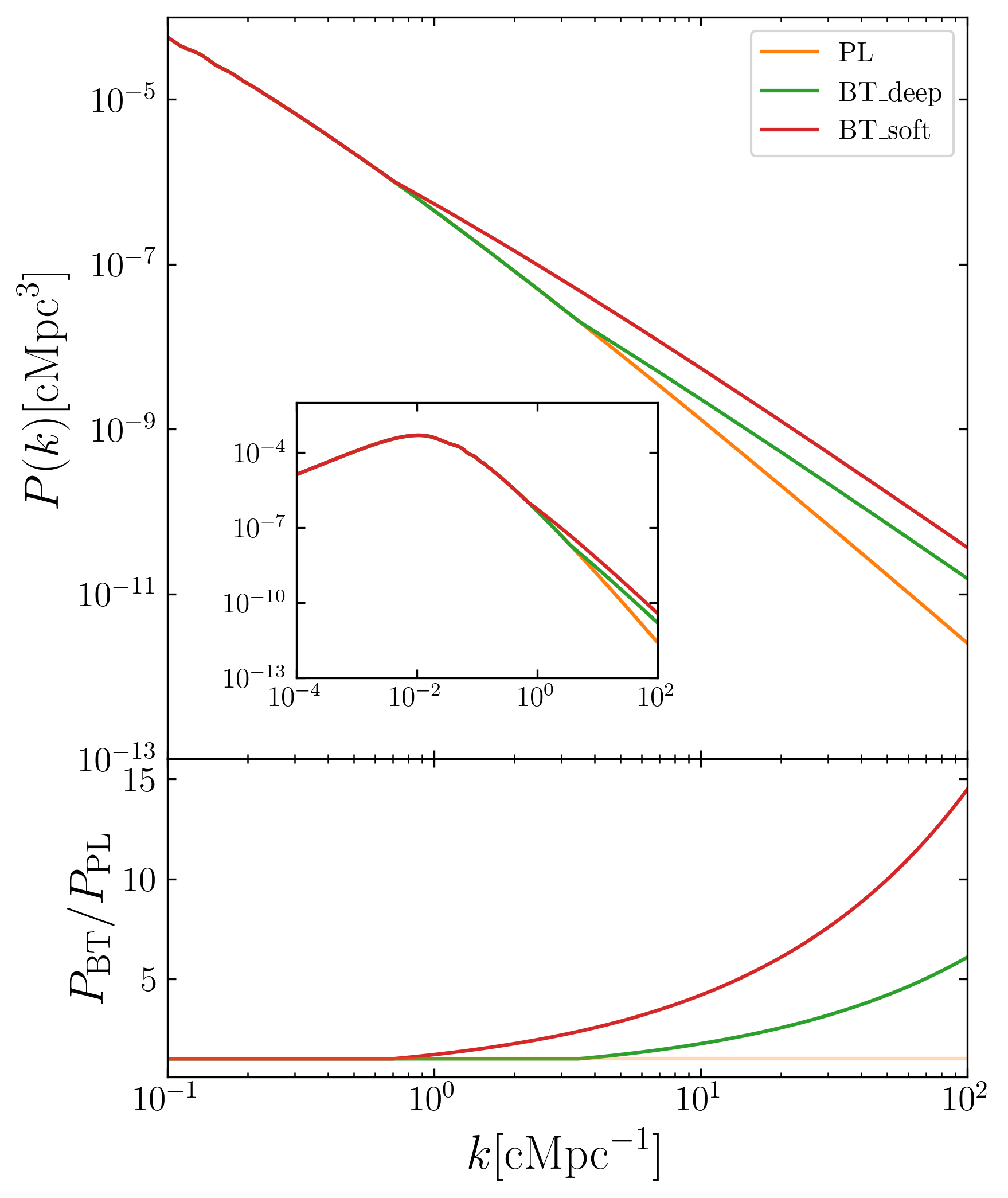}
    \caption{The \textit{upper panel} shows the power spectra for the matter density perturbation at z=1089 with PL (orange), BT\_deep (green), and BT\_soft (red). The \textit{inset} is for a wider wave vector $k$ range. The \textit{bottom panel} shows the ratios of the BT to PL power spectrum. The model parameters are listed in \protect\autoref{tab:Params_all_models}.}
    \label{fig:high_z_ps}
\end{figure}

For all of our PPS models, the matter power spectra at z=1089 and the parameter settings are shown in \autoref{fig:high_z_ps} and \autoref{tab:Params_all_models} respectively. BT\_soft deviates from PL at $0.7~{\rm cMpc}^{-1}$, whereas BT\_deep deviates at $3.51~{\rm cMpc}^{-1}$, but they have the same power law index afterward. 

\begin{table}[H]
\scalebox{1.0}{
\begin{tabular}{@{}lll@{}}
\toprule
           \multicolumn{1}{l}{Models} 
           &    
           \multicolumn{2}{c}{Related parameters} \\ \midrule
PL &  \multicolumn{2}{c}{Power Law Primordial Power Spectrum}    \\
& $n_s=0.961$ \\
  
BT\_deep &    $k_p = 3.51 \ \rm{cMpc}^{-1}$ & $ m_s = 1.5$   \\  
BT\_soft &    $k_p = 0.702 \ \rm{cMpc}^{-1}$ & $ m_s = 1.5$   \\  

\bottomrule

\end{tabular}
}
\caption{The parameters of all the chosen models. $k_p$ is the wave vector at which the BT PPS would deviate from the PL PPS. $m_s$ is the enhanced spectral index for $k>k_p$, at the small scales. For other cosmological parameters, see \protect\autoref{subsub:cosmo_params}.
}
\label{tab:Params_all_models}
\end{table}

\begin{figure}
    \centering
    \includegraphics[width={0.45\textwidth}]{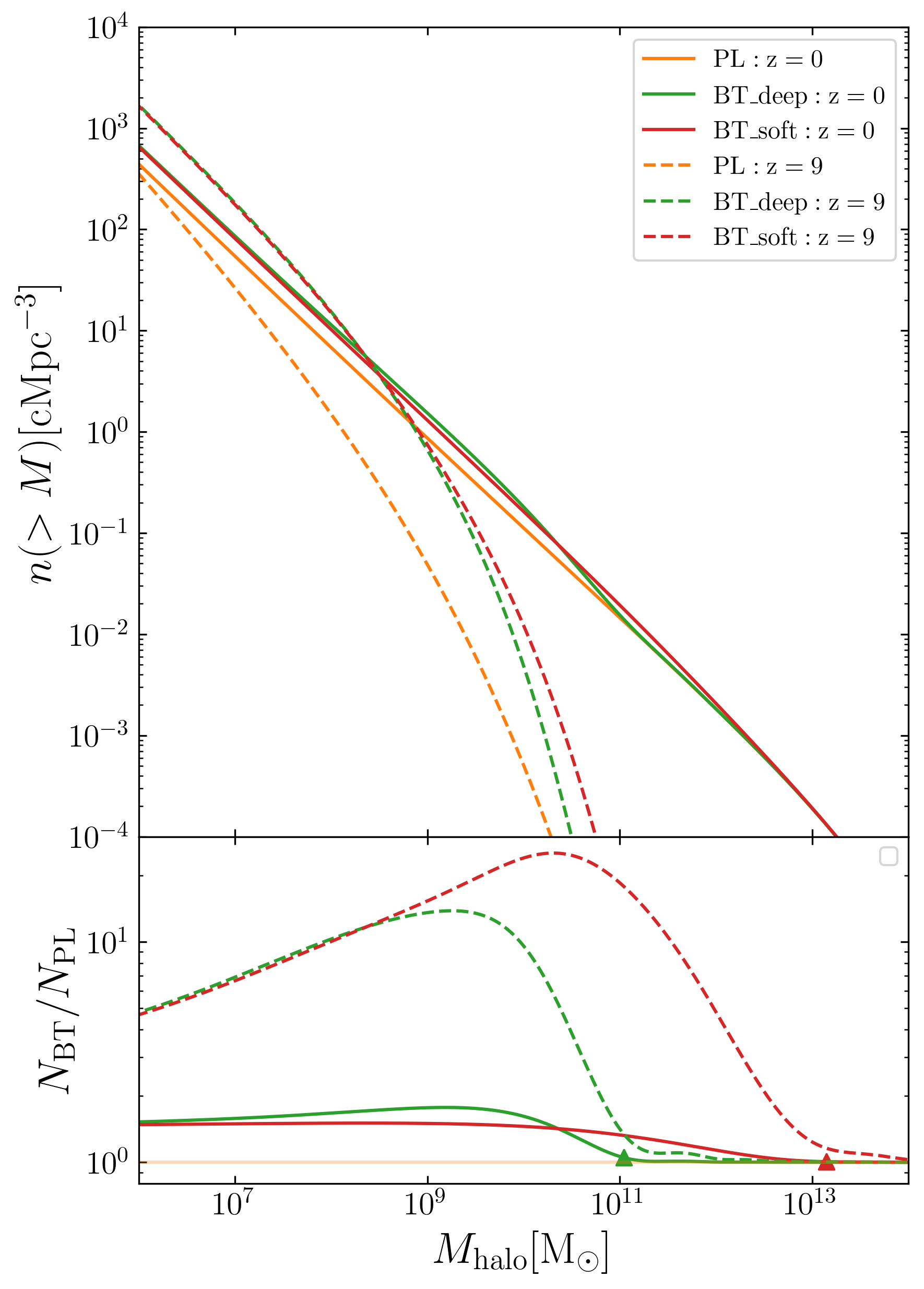}
    \caption{The cumulative halo mass function generated by genmf, at z=9 (dashed line) and z=0 (solid line), with PL (orange), BT\_deep (green), and BT\_soft (red). The \textit{upper panel} shows the cumulative halo mass function per unit comoving volume. The \textit{bottom panel} shows the ratios of BT to PL. The triangle symbols on the solid lines show the pivot masses $M_p$ for BT\_deep (green) and BT\_soft (red), respectively.}
    \label{fig:HMF_genmf}
\end{figure}

We can associate a comoving wavenumber $k$ with a mass scale $M_l(z)$ at a given redshift $z$ by considering a sphere with a comoving Lagrangian radius $r_l = \pi/k$ \cite{Bullock17SmallScaleCrisis}:
\begin{align}
    M_l(z) = & \frac{4\pi}{3} {r_l}^3 \rho_m(z)\\
    = & \frac{4\pi}{3} {r_l}^3 \rho_{crit}(z) \Omega_m(z)\notag \\
    = & \frac{4\pi}{3} {r_l}^3 \notag \\ & \rho_{crit,0} \left[\Omega_{m,0} ({1+z})^3 + \Omega_{r,0} ({1+z})^4 + \Omega_{L,0} \right] \notag \\ & \frac{\Omega_{m,0}}{\left[\Omega_{m,0} ({1+z})^3 + \Omega_{r,0} ({1+z})^4 + \Omega_{L,0} \right]}\notag \\
    = & \frac{4\pi}{3} ~ {r_l}^3 ~\rho_{crit,0} ~ \Omega_{m,0},
\end{align}
where $\rho_{crit}(z)$,$\rho_{m}(z)$ are the comoving critical and matter densities respectively. $\Omega_m(z)$ is the cosmological density parameter for matter at redshift z. $\Omega_{m,0}, \Omega_{r,0}, \Omega_{L,0}$ are the cosmological density parameters at $z=0$ for matter, radiation, and dark energy respectively. We assume a flat universe so the curvature density parameter at $z=0$ is zero, i.e. $\Omega_{k,0} = 0$. As a result, the mass scale is independent of redshift $z$ because both the comoving Lagrangian radius $r_l$ and the comoving average matter density $\rho_{m}(z)$ do not depend on redshift. 

We could further name a \textit{pivot mass} $M_p(k_p)$ for BT model, structures less massive than which could be enhanced while structures more massive should not be affected:
\begin{align}
    M_p (k_p) = & 5.29 \times 10^{12} \left(\frac{\Omega_{m,0}}{0.3}\right) \left(\frac{H_0}{70}\right)^2 \left(\frac{ k_p}{1~{\rm cMpc}^{-1}}\right)^{-3} \msun.
\label{eq:Mp}
\end{align}

For the BT\_deep and BT\_soft models, $M_p$ would be $\sim 1.1\times10^{11} ~\msun$ and $\sim 1.4\times10^{13} ~\msun$ respectively, using the cosmological parameters in our paper (see \autoref{subsub:cosmo_params}). The host halo of an MW-size galaxy is believed to be around $M_{\rm halo} \sim 10^{12} ~\msun$. Therefore, while both the BT\_deep and BT\_soft models can enhance the substructures of MW host mass halos, only in BT\_soft the main halo can be affected.

To illustrate the impact of the BT model on the dark matter halo statistics across different redshifts, we apply a public semi-numerical code {\small genmf} \cite{Reed07hmf} \footnote{Publicly available at \url{https://icc.dur.ac.uk/Research/PublicDownloads/genmf_readme.html}}. It can predict halo mass functions in cosmological simulations at a given redshift. Ref. \cite{Hira24bluetilt} verified that {\small genmf} is accurate within a factor of 2-3, by comparing it with numerical N-body simulations (with PL and BT).

We use {\small genmf} to show the cumulative halo mass function at $z=0$ and $z=9$ for the BT and PL models in \autoref{fig:HMF_genmf}. The figure shows, at $z=9$, the BT models can produce an order-of-magnitude more halos, especially for $M_{\rm halo}< 10^{10-11}\msun$. At $z=0$, the enhancement in the number of halos reduces to a factor of 2-3 for $M_{\rm halo}< 10^{10-11}\msun$. Ref. \cite{Reed07hmf} also found that the high redshift halo mass function is more sensitive to cosmology, due to the stiffness of the function. This high redshift enhancement motivates Ref. \cite{Hira24bluetilt} to explain the early formation of galaxies discovered by JWST with the BT models.

However, at both redshifts, there is no enhancement at the high-mass end, since the BT models can only enhance the structures lower than the pivot mass $M_p$ (\autoref{eq:Mp}), shown as the triangles in the figure.

While {\small genmf} can estimate the halo mass functions in the BT models within a factor of two, we still require cosmological N-body simulations to accurately capture the sub-halo distributions and statistics in the BT models. This will help us explore BT PPS's impact on nearby galaxies and strong lensing observations.

\section{Numerical Methods}
\label{sec:numerical}

Our cosmological simulation pipeline is outlined in the bottom panel of \autoref{fig:conceptual_flow}. We will introduce more details in the following three subsections.

\subsection{Zoom-in initial conditions (IC)}

We carry out a suite of cosmological zoom-in simulations of MW host mass halos. We increase the number of particles in a selected zoom-in region, which contains high-resolution (HRS) particles. The particles outside the zoom-in region are called background or low-resolution (LRS) particles. We chose the public code {\small MUSIC} - \textit{multi-scale cosmological initial conditions}  \citep{Hahn11}, which can handle multi-scale IC, to generate initial conditions for our cosmological zoom-in simulations. We adopted the Eisenstein and Hu \citep{Eise98transfer} transfer function and modified it according to \autoref{sec:theory} to mimic the blue-tilted primordial power spectra.

For the convenience of comparison, we adopt the {\small MUSIC} configuration files of the existing simulation project FIRE \citep{FIRE2,Wetz23FIREpublic}. We choose the zoom-in initial conditions at the mass of MW host mass halos (i.e., $M_h\sim 10^{12} ~\msun$ in mass), specifically, m12i \citep{Wetz16m12i} and m12f \citep{Garr17m12f}. We obtain those configuration files from the Flatiron Institute Data Exploration and Comparison Hub \footnote{Publicly available at \url{https://flathub.flatironinstitute.org/fire}}.

Several properties of the m12i and m12f host haloes and simulations are listed in \autoref{tab:result_simulations}.  m12f is 30\% more massive and more extended than m12i, resulting in also more substructures. m12i is more concentrated than m12f as expected from its lower halo mass \citep{Diem15massconcentration}. m12i has a series of mergers around redshift $z \gtrsim 1$, while m12f is slightly later forming than m12i.

\subsection{Simulation code}

We perform several zoom-in simulations using the state-of-the-art cosmological simulation code {\small SWIFT} \citep{ascl18SWIFT,Scha23SWIFT}\footnote{ publicly available at \url{https://swiftsim.com/}}. For N-body simulations, {\small SWIFT} uses the Fast Multipole Method (FMM) at small scales \citep{Dehn00FMM}, which can reduce the complexity of gravity calculation (including both the potential and force calculation) to $O(N)$. The Particle Mesh Method is also coupled at large scales to handle periodic volumes \citep{Hock88PM}.

\subsubsection{Cosmological parameters}
\label{subsub:cosmo_params}

For all the cosmological simulations in this paper, we adopt the WMAP-7 result \citep{Komatsu12wmap7}, which is also used in the original m12i \citep{Wetz16m12i} and m12f \citep{Garr17m12f} simulations.

\begin{enumerate}
    \item Universe density parameters $\Omega_m=0.272$, $\Omega_b=0.0455$, $\Omega_L= 0.728$.
    \item Hubble constant $H_0 = 70.2 \ \rm{km}\ \rm{s}^{-1}\ \rm{cMpc}^{-1}$
    \item Scalar spectral index $n_s = 0.961$
    \item Root-mean-square matter fluctuation averaged over a sphere of radius $8 h^{-1}~\rm{cMpc}$, $\sigma_8 = 0.807$.
\end{enumerate}

\subsection{Halo finder and halo property post-processor}
\label{sub:halofinder}
Current halo finders can be categorized into three types based on working mechanism: (A) the configuration space finders, (B) the phase space finders (which utilize both spatial and velocity space information), and (C) the tracking finders (which build the particle list of a certain subhalo based on its progenitors).

In the main text of the paper, we adopt HBT-HERONS \citep{HanJiaxin17HBTplus,Victor25HBTHERONS} \footnote{This is an updated version of HBT+ that improves the tracking of subhalos. It can be found at \url{https://github.com/SWIFTSIM/HBT-HERONS}} as our halo finder, which is a tracking finder. We also explore the result with a phase space finder, VELOCIraptor \citep{Elahi19VELOCIraptor}, and compare their results in \autoref{sec:VR}. In this paper, we adopt 20 as the minimum number of particles to be identified as a dark matter halo for both halo finders. We choose the position of the most bound particle as the halo center.

We use SOAP (\textit{Spherical Overdensity and Aperture Processor; McGibbon et al. submitted})\footnote{\url{https://github.com/SWIFTSIM/SOAP}} to calculate the spherical overdensity properties for the results from halo finders.

\subsubsection{Convergence radius}
\label{subsub:converg_radius}

The dark matter halo's radial density profile would be underestimated within a resolution limit radius in N-body simulations, due to the two-body relaxation effect \cite{Powe03}. This would artificially turn a cuspy profile into a core profile. This radius is called the convergence radius and defined as the smallest $r$ fulfilling \autoref{eq:convergence_radius}:
\begin{align}
    \frac{t_{\rm relax}(r)}{t_0} = \frac{\sqrt{200}}{8} \frac{N(r)}{\ln N(r)} \left[\frac{\rho(r)}{\rho_{\rm crit}}\right]^{-1/2} > 0.6,
\label{eq:convergence_radius}
\end{align}
where $\rho_{\rm crit}$ is the critical density of universe, $\rho(r)$ is the average density within radius $r$, $N(r)$ is the number of enclosed particles within radius $r$. $t_0$ is the integration time, which is close to the age of the Universe. We will indicate the regions within the convergence radius with shades in \autoref{fig:Radial_Density_Profile}.

\section{Result}
\label{sec:result}

We run nine simulations using the numerical pipeline mentioned in \autoref{sec:numerical}. Their numerical settings are listed in \autoref{tab:result_simulations}.

The six fiducial-resolution simulations are used to compare the BT and the PL models. Two initial conditions, m12i and m12f, are used to validate the conclusions. Thus, it would give six fiducial-resolution simulations in total. Additionally, we run three high-resolution simulations with m12i for the resolution studies in \autoref{sec:res}.

\subsection{Terminology\label{subsec:Terminology}}

We give a summary of the concepts and terminologies used in our paper:

\begin{enumerate}
    \item Dark Matter Halo and Subhalo: a dark matter halo is a high-density region in the universe composed of dark matter particles that are gravitationally bound to each other. Smaller overdense regions can also exist inside the halo, which are also self-bound. Those smaller halos are called the subhalos of the host halo.
    In zoom-in simulation, the most massive halo is called \textbf{main halo}, which hosts the largest number of subhalos in the zoom-in region.

    \item  $M_{\rm sub}$: the total mass of all the bound particles in a subhalo (as defined by the halo finders). In this paper, we use this mass definition for the subhalo mass function. 
        
    \item Spherical overdensity (SO) properties: these properties are calculated based on all the enclosed particles within a sphere around the center of the selected dark matter halo. This sphere's radius is usually where the average density for the enclosed region reaches an integer (like 50, 100, 200, ...) multiple of the critical density (labeled with 50c, 100c, 200c, ...) or the background mean density (labeled with 50m, 100m, 200m, ...). The properties include the spherical overdensity radius ($R_{200c}, R_{200m}, ...$), the spherical overdensity mass ($M_{200c}, M_{200m}, ...$), the spherical overdensity concentration parameter of the NFW fitting ($C^{NFW}_{200c}, C^{NFW}_{200m}, ...$). With the SO radius and SO mass properties, we can calculate the SO circular velocity properties($V_{200c}, V_{200m}, ...$) via \autoref{eq:Circular_Velocity}.

    \item  $R_{\rm vir}~\&~M_{\rm vir}$: Special kinds of the SO properties. $R_{\rm vir}$ is the virial radius of a dark matter halo. The most commonly used virial radius definition is given by Ref. \cite{Brya98}. $M_{\rm vir}$ is then the total mass within a sphere of radius $R_{\rm vir}$. We use this definition for the subhalo mass in \autoref{sub:Mvir-Vmax}.

    \item $V_{\rm max},~V_{\rm peak}~\&~R_{\rm max}$:
    The circular velocity is defined as:
    \begin{align}
        V_{\rm cir}(R) = \sqrt{\frac{GM(<R)}{R}},
        \label{eq:Circular_Velocity}
    \end{align}
    where $R$ is the distance to the halo center and $M(<R)$ the enclosed mass that is bound to the halo, within radius $R$. The $R - V_{\rm cir}(R)$ graph is called the galaxy rotation curve, and $V_{\rm max}$ is simply the maximum circular velocity of the rotation curve. In the simulation, the subhalo can lose its mass due to tidal forces from the host halo. However, $V_{\rm max}$ is a more robust measurement of subhalo mass, since it reflects the subhalo's central, gravitationally bound part, which is less sensitive to how the boundary of the subhalo is defined.

    During the evolution history of a dark matter halo, its $V_{\rm max}$ may change, and the largest $V_{\rm max}$ is called the peak circular velocity, $V_{\rm peak}$.

    The distance $R$ corresponding to the maximum circular velocity $V_{\rm max}$ is denoted as $R_{\rm max}$. For a central galaxy whose rotation curve is usually flat, $R_{\rm max}$ is quite sensitive to the simulation details even though $V_{\rm max}$ changes only slightly.
    
\end{enumerate}

\subsection{Comparison between the PL and the BT models\label{subsec:comparison_PL_BT}}

\begin{figure*}
    \centering
    \includegraphics[width=0.325\textwidth]{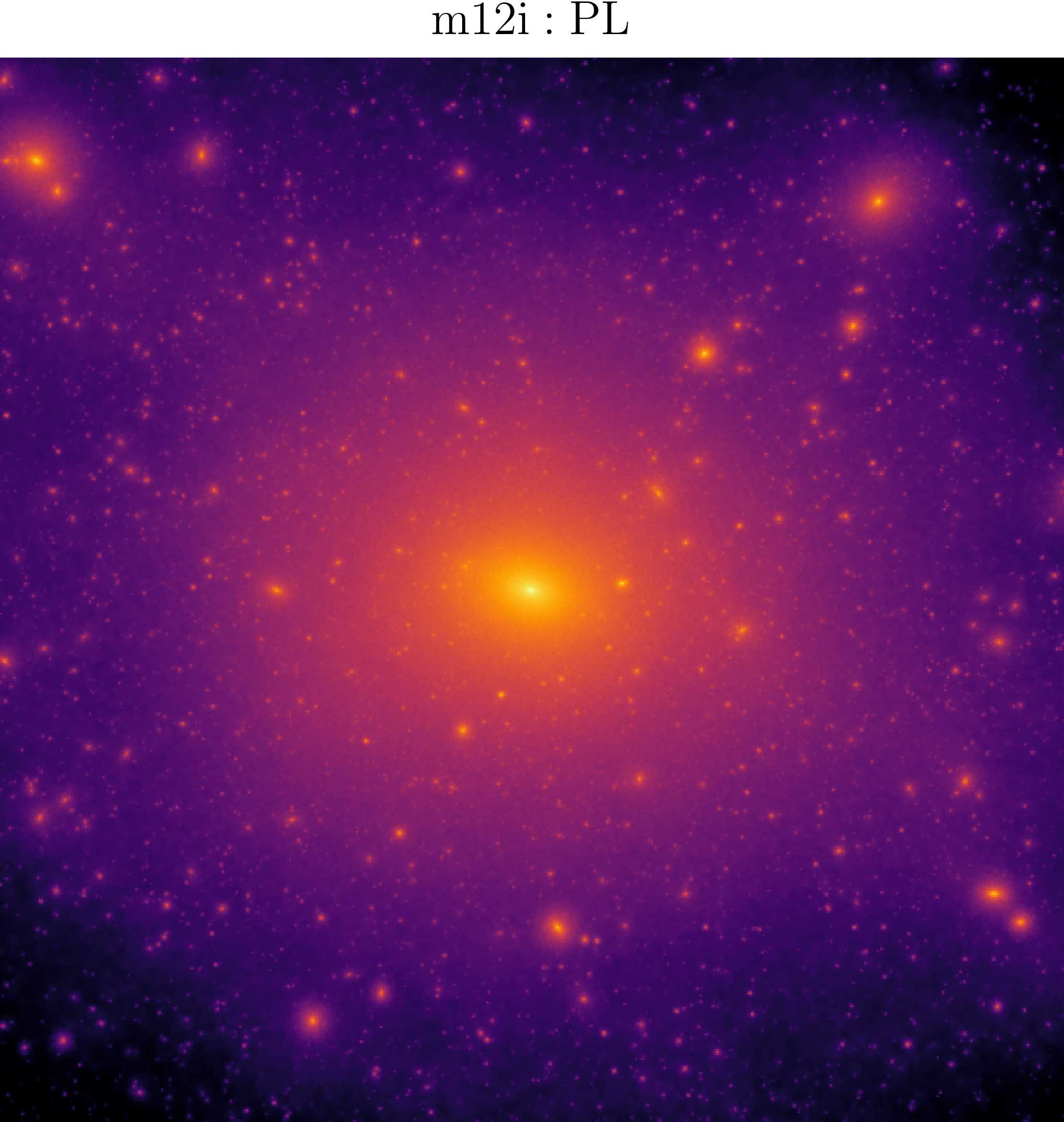}
    %\hspace{0.2in}
    \includegraphics[width=0.325\textwidth]{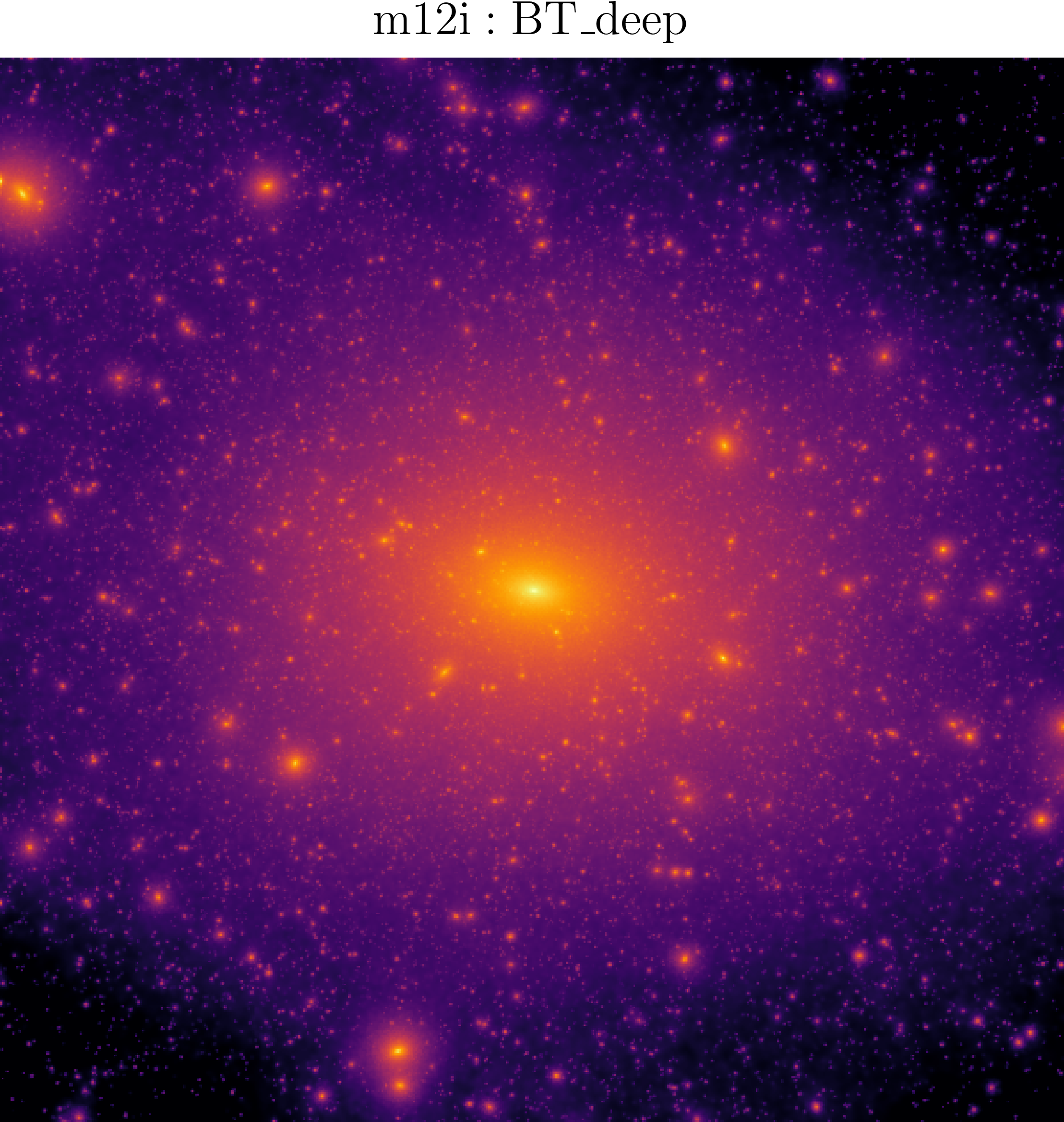}
    %\hspace{0.2in}
    \includegraphics[width=0.325\textwidth]{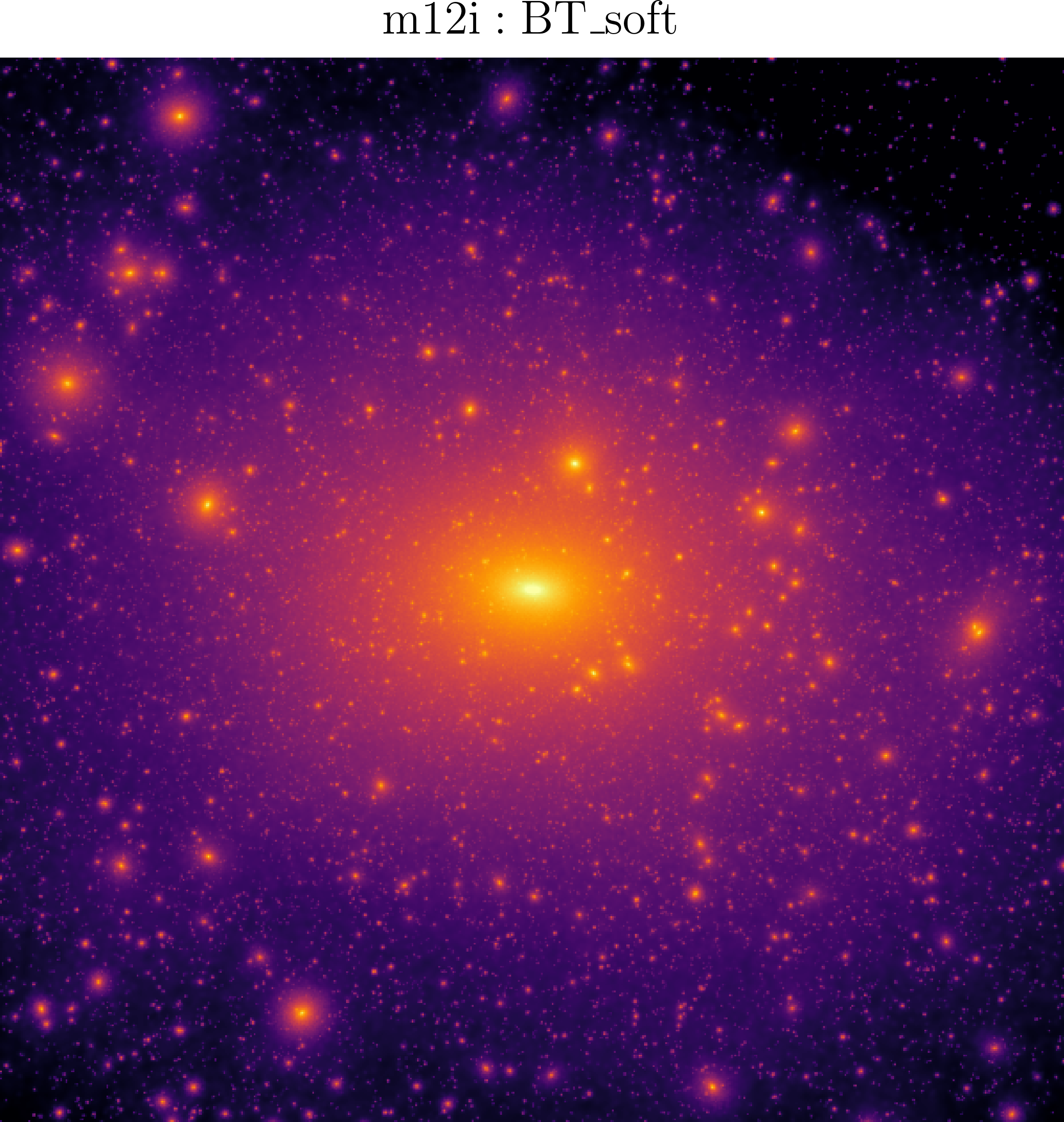}
    \caption{The comparison between the dark matter mass projection maps with the PL and BT (deep/soft) PPS for m12i at z=0. The color scales are the same across the three panels. All the images are 2D plane projections of cubic volumes with a side length of $400~\rm{kpc}$, centered at the main halo's most bound particle identified by HBT-HERONS.}
    \label{fig:comparison_projection_maps}
    \vspace{-2mm}
\end{figure*}

\autoref{tab:result_simulations} shows some selected properties for the main halos in all the simulations. We learn that the main halo of BT\_soft is larger than that of the PL model in mass, radius, concentration, and $V_{\rm max}$. However, the BT\_deep model behaves similarly to the PL model, because the mass of the halo is well above the pivot mass of $M_p$ based on \autoref{eq:Mp}.

\autoref{fig:comparison_projection_maps} shows the dark matter projection maps of the PL, BT\_deep and BT\_soft m12i simulations. Since their sizes are similar, the main halos in the projection map look alike in these three cases. However, there are significantly more subhalos in BT than in PL, including both the small subhalos and the massive ones. For the spatial distribution, the substructures are also more centrally concentrated in BT than in PL.

\begin{table*}[]
\centering
\resizebox{0.9\textwidth}{!}{%
\begin{tabular}{l|lll|lll|lll}
\midrule
\multirow{2}{*}{Property} & \multicolumn{3}{c}{m12i-fiducial resolution} & \multicolumn{3}{c}{m12i-high resolution} & \multicolumn{3}{c}{m12f-fiducial resolution} \\
                          & PL  & BT:deep  & BT:soft & PL  & BT:deep  & BT:soft  & PL  & BT:deep  & BT:soft\\
                          \midrule
$M_{200c}\hfill[10^{12}~\msun]$                      &  $0.950$   &    $0.938$      &$1.12$& $0.955$ &  $0.945$ & $1.13$ & $1.32$ & $1.35$ & $1.50$ \\

$M_{\rm vir}\hfill[10^{12}~\msun]$                      &  $1.14$   & $1.13$ & $1.28$        &  $1.14$   &    $1.11$      &    $1.28$  &    $1.61$  & $1.61$ & $1.73$\\

$M_{200m}\hfill[10^{12}~\msun]$       &  $1.26$   & $1.26 $  &   $1.41$      & $1.26$    &    $1.25$      &     $1.42$  &     $1.82$ & $1.81$ & $1.87$ \\

$R_{200c}\hfill{\rm [kpc]}$                      &   202  &    202      &  214       &  203   &     202     &    214  & 226  & 228 & 236 \\

$R_{\rm vir}\hfill{\rm [kpc]}$                     &   273  &    273      & 284  &   273  &     271     &    284  & 307  & 307 & 314\\

$R_{200m}\hfill{\rm [kpc]}$                      &  344   &    343      &  357  &  344   &   343     &    357  & 388  & 387 & 391\\

$V_{200c}\hfill{\rm [km/s]}$                      &   142  &    141      & 150 &  142   &    142      &   151  & 159  & 160 & 165\\

$C^{NFW}_{200m}$                      &  14.4   &    14.0      & 20.7 &  14.7   &    14.6      &    19.9  & 11.5 & 13.1 & 22.8\\

$V_{\rm max}\hfill{\rm [km/s]}$                      &   155  & 154 & 174 & 166& 167 & 188 & 176 & 171 & 194\\

%$R_{\rm max} [kpc]$                      &   56.7 & 43.4 & 39.2 & 24.9 & 24.6 & 19.6 & 16.8 & 15.1 & 42.8 \\

$m_{p}^{HRS}\hfill{\rm [10^{3} \msun]}$                      & $42.3$    &  $42.3$  &  $42.3$     & $5.28$  & $5.28$  & $5.28$  &  $42.3$   &   $42.3$       &  $42.3$   \\

$l_{\rm soft}\hfill{\rm [kpc]}$                      &  0.208   &   0.208       &     0.208    &  0.104   &  0.104        &  0.104   & 0.208 & 0.208 & 0.208   \\

\midrule
\end{tabular}
}
\vspace{-2mm}
\caption{Some selected properties for the main halos and the basic parameters of the simulations. The first and second rows give the simulation names, encoding the main halo (m12i or m12f), the resolution levels (fiducial or high resolution), and the PPS models (PL, BT\_deep or BT\_soft). The rest are the properties and basic parameters in the simulations. The properties (from 3rd to 11th rows) are all explained in \autoref{subsec:Terminology}. $m_{p}^{HRS}$ is the mass of high-resolution particle in the simulation. $l_{\rm soft}$ is the gravitational softening length (in terms of co-moving value), which is 0.02 times the mean inter-particle distance \protect\cite{Zhang19OptimalSoftLength}.}
\label{tab:result_simulations}
\end{table*}

In the rest of this subsection, we quantify these differences in terms of the main halo radial density profile, the subhalo mass function, the subhalo $V_{\rm max}$ function, the subhalo radial distribution (subhalo distance function), the cumulative substructure mass fraction and the subhalo $R_{\rm max}-V_{\rm max}$ relationship.

\subsubsection{Main Halo Radial Density Profile\label{subsub:RadialDensityProfile}}

\begin{figure*}
    \centering
    \includegraphics[width=0.45\linewidth]{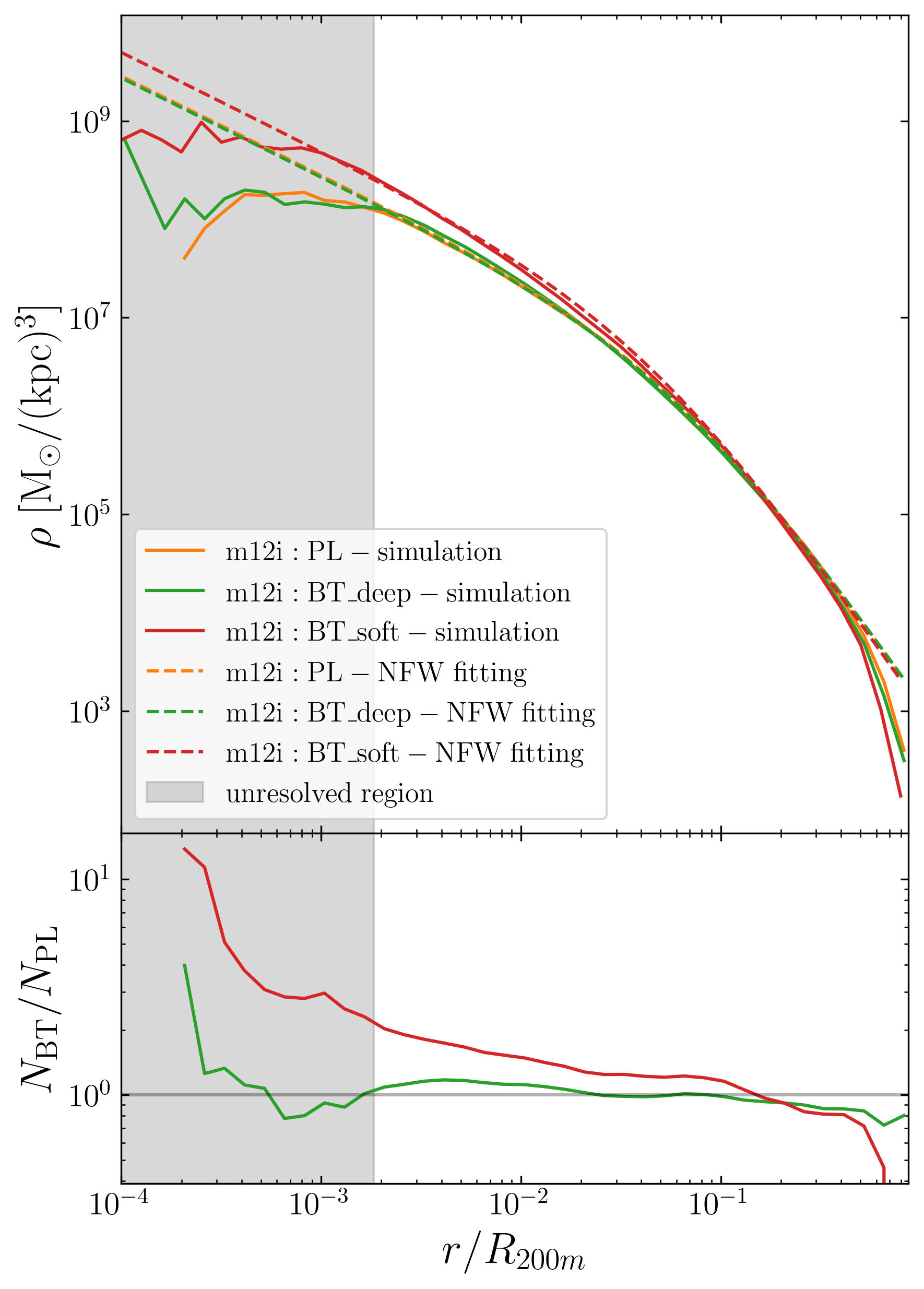}
    \includegraphics[width=0.45\linewidth]{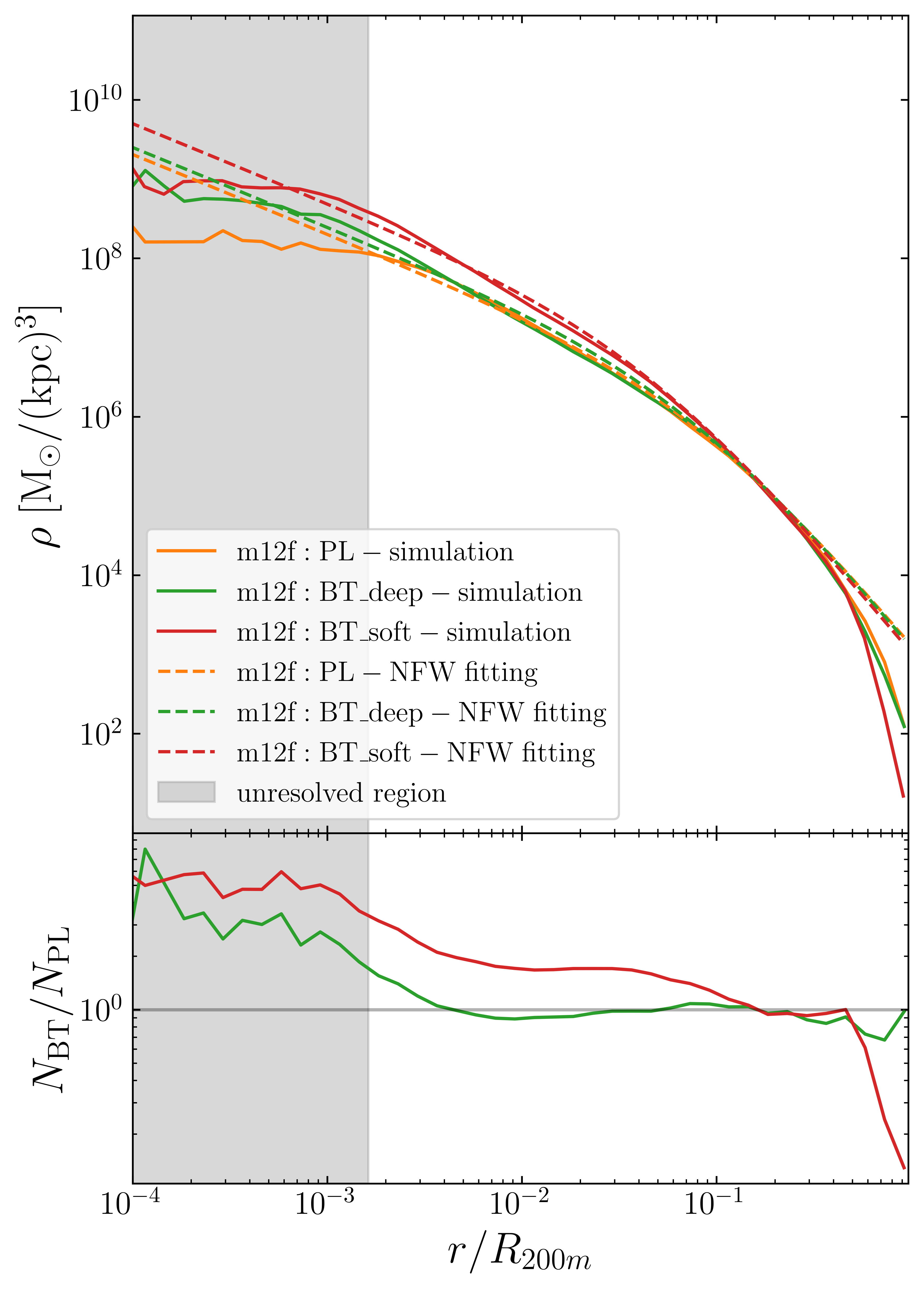}
    \caption{The density profiles of the main halos for m12i (left) and m12f (right) respectively. For both figures:\protect\\
    The \textit{upper panel} shows the density profiles of the main halo with the PL (orange), BT\_deep (green), and BT\_soft (red) models; the NFW fitting line is also shown for each model as a dashed line. 
    The \textit{bottom panel} shows the ratios of the BT density to the PL density.
    The \textit{shaded area} shows the radial range expected to be affected by the numerical two-body relaxation, given by \autoref{eq:convergence_radius}.}
    \label{fig:Radial_Density_Profile}
\end{figure*}

We present the radial density profiles of the main halos in \autoref{fig:Radial_Density_Profile}, for m12i and m12f, respectively. It is an essential property of the dark matter halo inner structure, which is well-characterized by the Navarro–Frenk–White (NFW) profile \cite{NFW}. We show the NFW fitting result in dashed lines and our simulation results in solid lines.

The upper panels of \autoref{fig:Radial_Density_Profile} \footnote{ For m12f, the main halo center identified by HBT-HERONS is shifted from the real center by $\leq 2 {\rm kpc}$, resulting from a bug existing in the code version we used (see Appendix A7 of \cite{Victor25HBTHERONS}), but which has since been fixed. Given this, we adopt the better centering offered by another halo finder, VELOCIraptor, \cite{Elahi19VELOCIraptor} in this figure. Other figures are not affected by this small shift. } show a good match between the simulation results and the NFW fitting lines. Thus we can use the concentration parameters given by the NFW profile to describe the radial density profile. The concentration parameters of BT\_deep and PL are pretty close (see \autoref{tab:result_simulations}), consistent with their similar radial density profiles for the main halos. Meanwhile, the BT\_soft model gets a larger concentration parameter than others, consistent with its higher radial density profiles. The unresolved regions are based on the criterion described by \autoref{subsub:converg_radius}.

In the bottom panels, the BT\_deep model behaves like the PL model in the resolved region, while the BT\_soft model is enhanced by a factor of two in the inner part of the main halo ($r< 0.01 R_{200m}$). It is also consistent with our calculation of the pivot masses for the deep and soft models. As the radius increases, BT and PL approach a similar density, indicating that the enhancement in the main halo density mainly occurs in the inner region.

\subsubsection{Subhalo Mass Function\label{subsub:HMF}}

\begin{figure}[H]
    \centering
    \includegraphics[width={0.45\textwidth}]{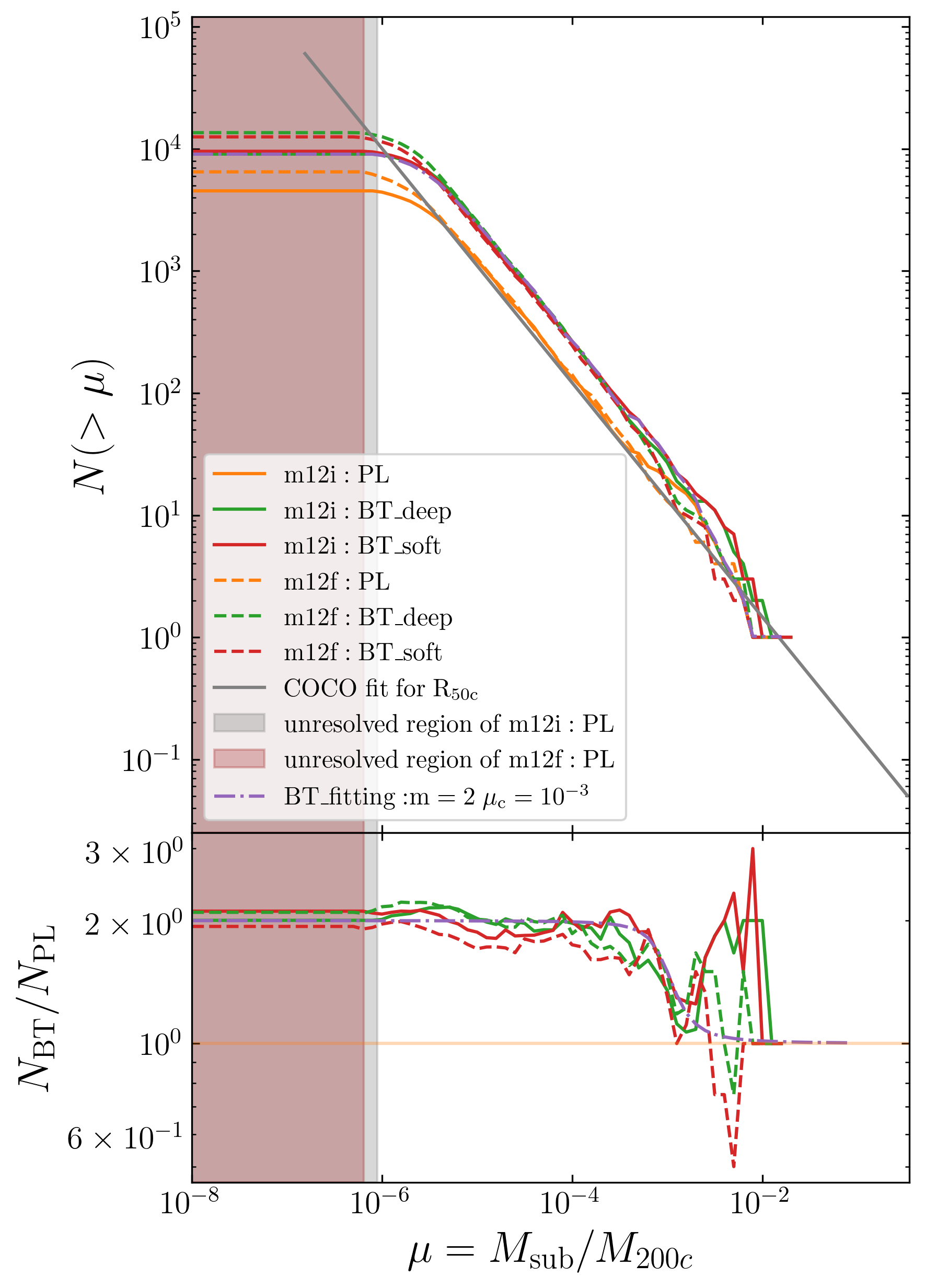}
    \caption{The cumulative scaled subhalo mass functions for m12i (colored solid) and m12f (colored dashed) respectively.\protect\\
    The \textit{upper panel} shows the subhalo mass function for all the subhalos within $R_{50c}$ from the main halo center in the PL (orange), BT\_deep (green), and BT\_soft (red) simulations, with the subhalo mass  $M_{\rm sub}$ scaled by $M_{200c}$ of the main halo. The \textit{bottom panel} shows the ratios of the BT to PL numbers. The \textit{shaded areas} show the subhalo mass cutoff due to the halo finder's setting for minimum halo size, where we set it to 20 particles. However, based on our resolution study in \autoref{sec:res}, the subhalo mass function is reliable on $\mu \geq 10^{-5.5}$, i.e. for subhalos with at least 50 particles. $\mu \sim 10^{-5.5}$ is also approximately the $\mu$ point where the subhalo mass function starts to deviate from a power-law function, i.e. a straight line in the log-log graph.
    }
    \label{fig:HMF_coco}
\end{figure}

We present the subhalo mass function in \autoref{fig:HMF_coco}. We show the subhalo count within $R_{50c}$ from the main halo center, $N(>\mu)$, as a function of $\mu = {M_{\rm sub}}/{M_{200c}}$, namely the ratio between the gravitationally bound subhalo masses and the $M_{200c}$ of the main halos. We adopt this scaled mass definition because it can make the subhalo mass function insensitive to the host halo mass \citep{Kravtsov04HMF_scaled, Hellwing16cocoproject}. The grey ``COCO fit'' line is the best fitting power function for the subhalo mass ($N(>\mu)\sim\mu^{-s}$ with $s=0.95$) in Ref. \citep{Hellwing16cocoproject}, for all the subhalos within $R_{50c}$\footnote{In COCO paper, they used the empirical relation to calculate $R_{50c}$, $\sim 1.66 \cdot R_{200c}$.} of the main halo. 

The upper panel of \autoref{fig:HMF_coco} shows a reasonable match between the COCO fit and our PL simulation, showing that we are consistent with COCO. The two BT models' lines diverge clearly from the PL line, with a higher number. The BT\_deep and BT\_soft subhalo mass functions almost overlap, indicating that the enhancement in subhalo mass is similar between these two models.

The flattening of the subhalo mass functions in \autoref{fig:HMF_coco} is due to the limited mass resolution: we cannot resolve subhalos with fewer than $\sim 20$ particles. The
shaded area shows the mass range cannot be covered due to the halo finder’s setting for the minimum halo size, which is 20 particles. Though both m12i and m12f are using the same parameters in the halo finder and thus the same mass of the minimum size halo, m12f has a larger $M_{200c}$ so its scaled subhalo mass limit $\mu = M_{\rm sub}/M_{200c}$ would be smaller. 

In the bottom panel, the BT models have an apparent enhancement in the number of dark matter subhalos at different masses. Especially at the low mass range ($\mu < 10^{-4}, M_{\rm sub}<10^8 ~\msun$), the BT effect can double the subhalo number. At the high mass range ($10^{-4}< \mu < 10^{-3}, 10^8 ~\msun<M_{\rm sub}<10^9 ~\msun$), the BT models still have tens of percent enhancement. At an even larger mass ($\mu > 10^{-3}, M_{\rm sub} > 10^9 ~\msun$), the ratio curve is dominated by the noise due to a lack of statistics. We can use an inverse S shape function to fit the ratio of BT to PL, by introducing two free parameters $m, \mu_c$:
\begin{align}
    f(x) = \frac{1}{2} \cdot (m-1) \cdot (\frac{-x}{\sqrt{1+x^2}}+1) + 1,
\label{eq:inverse_S_shape_HMF}
\end{align}
where
\begin{align}
    x = 4 \cdot \log_{10}\frac{\mu}{\mu_c}.
\label{eq:x_def_HMF}
\end{align}

\begin{figure}[H]
    \centering
    \includegraphics[width={0.45\textwidth}]{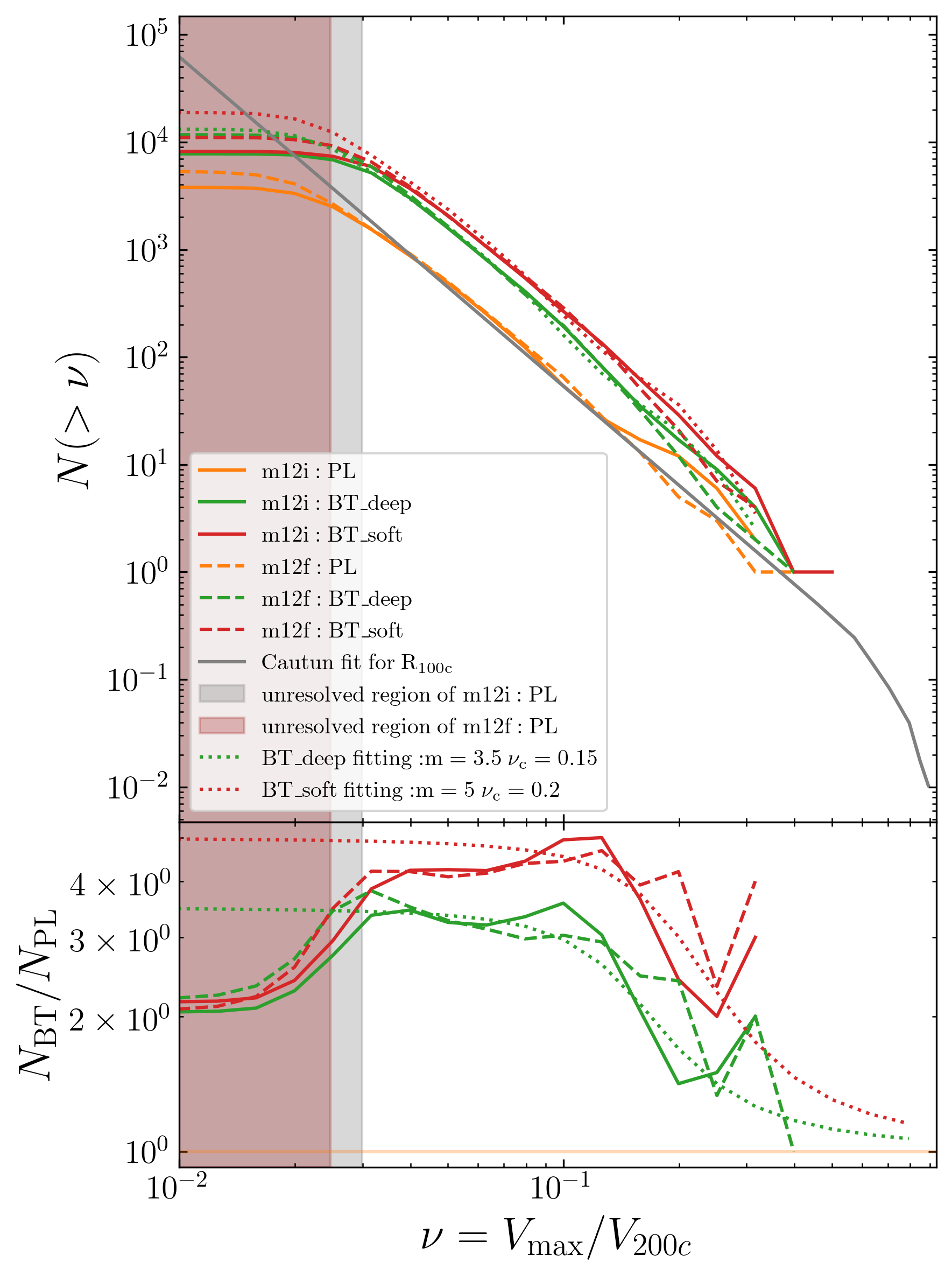}
    \caption{The cumulative scaled subhalo $V_{\rm max}$ functions for m12i (colored solid) and m12f (colored dashed) respectively.
    The \textit{upper panel} shows the subhalo $V_{\rm max}$ function for all the subhalos within $R_{100c}$ from the main halo center in the PL (orange), BT\_deep (green), and BT\_soft (red) simulations, with $V_{\rm max}$ of the subhalo scaled by $V_{200c}$ of the main halo. The \textit{bottom panel} shows the ratios of the BT to PL numbers. The \textit{shaded areas} show the unresolved $V_{\rm max}$ range, the limit of which is the median $V_{\rm max}$ value among all the subhalos with 100 particles in the PL simulations. 100 is an empirical lower limit of particle number for dark matter halo to be resolved \protect\citep{Puebla16BolshoiPMDPL}. Our resolution limit study in \autoref{sec:res} also shows that for subhalos with at least 100 particles, the relative deviation in the subhalo $V_{\rm max}$ function does not exceed 30\%, which is within acceptable limits.}
    \label{fig:HVF_cautun}
\end{figure}

Here, $m$ indicates the maximum relative ratio the BT model can reach compared to the PL model, and $\mu_c$ is the middle point between $\mu$ starting to drop and $\mu$ value finishing the dropping. According to the definition of \autoref{eq:inverse_S_shape_HMF}, $\mu_c$ is also where the inverse S shape function drops to half. Since the deep and soft models perform pretty closely in the subhalo mass function, we can use one set of parameters to fit: $m = 2, \mu_c = 10^{-3}$. After getting this fitting function from the bottom ratio panel, we draw it on the upper panel of \autoref{fig:HMF_coco}, showing a good fit with the simulation results.

\subsubsection{Subhalo \texorpdfstring{$V_{\rm max}$}{Vmax} Function\label{subsub:HVF}}

In \autoref{fig:HVF_cautun}, we present the subhalo $V_{\rm max}$ function. We show the subhalo count within $R_{100c}$ from the main halo center, $N(>\nu)$, as a function of $\nu = V_{\rm max}/V_{200c}$, which is the subhalo maximum circular velocity ($V_{\rm max}$) in the units of its main halo circular velocity at $R_{200c}$ ($V_{200c}$). We adopt this scaled $V_{\rm max}$ definition because it can make the function insensitive to the main halo masses \citep{Moore99HVF_scaled, Hellwing16cocoproject}. The grey ``Cautun fit'' line results from Ref. \cite{Cautun14r100fit} under the PL LCDM cosmology, which includes all the subhalos within $R_{100c}$ of the main halo. It is also adopted by Ref. \cite{Hellwing16cocoproject} as a fiducial reference line.

Note that when the number of particles is too few, the $V_{\rm max}$ of this subhalo is not resolved. The usual choice is not to trust subhalos with fewer than 100 particles \cite{Puebla16BolshoiPMDPL}, which corresponds to the shaded area in the figure.

The upper panel of \autoref{fig:HVF_cautun} shows a good match between the Cautun fit and our PL simulation, verifying the correctness of our pipeline. While the BT\_soft and BT\_deep lines are higher than the PL line, the soft model has an even stronger enhancement than the deep model. It is because the BT\_soft subhalos are more concentrated than the BT\_deep subhalos (see \autoref{subsub:Rmax-Vmax}). Thus, the subhalo $V_{\rm max}$ function can help us to differentiate the BT\_soft and BT\_deep models, although the subhalo mass function cannot.

In the bottom panel, the BT models increase the number of dark matter subhalos across a wide range of $V_{\rm max}$, especially between 0.03$V_{200c}$ and 0.1$V_{200c}$. Compared to the PL model, the BT\_deep model gets more than 300\% enhancement, while the BT\_soft model gets more than 400\%. The ratio begins to drop at a $V_{\rm max}$ slightly larger than 0.1$V_{200c}$ and has the trend to approach the unity at the high $V_{\rm max}$ end. Similarly, we can use the inverse S shape function in \autoref{eq:inverse_S_shape_HMF} to mimic the ratios of BT to PL. Only that we need to use $\nu$ instead of $\mu$ to define x:
\begin{align}
    x = 4 \cdot \log_{10}\frac{\nu}{\nu_c}.
\label{eq:x_def_HVF}
\end{align}
And we have different parameter sets for the two BT models: for BT\_soft, $m=5 ~ \nu_c = 0.2$; for BT\_deep, $m=3.5 ~ \nu_c = 0.15$. This indicates that the BT\_soft model has a stronger enhancement for a wider range of halo sizes than the BT\_deep model.

\subsubsection{Subhalo Radial Distribution\label{subsub:HRF}}

\begin{figure*}
    \centering
    \includegraphics[width=0.45\textwidth]{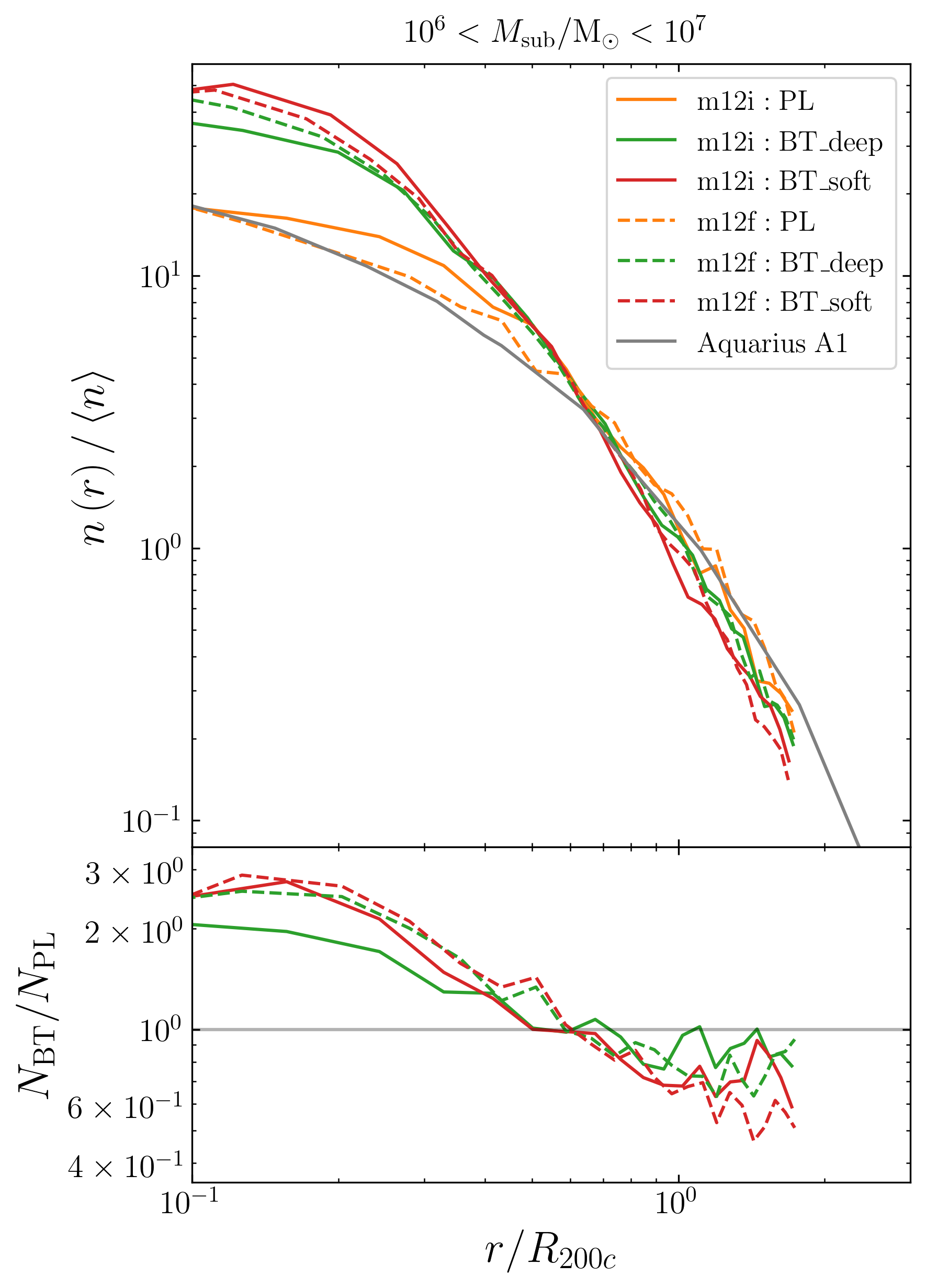}
    \includegraphics[width=0.45\textwidth]{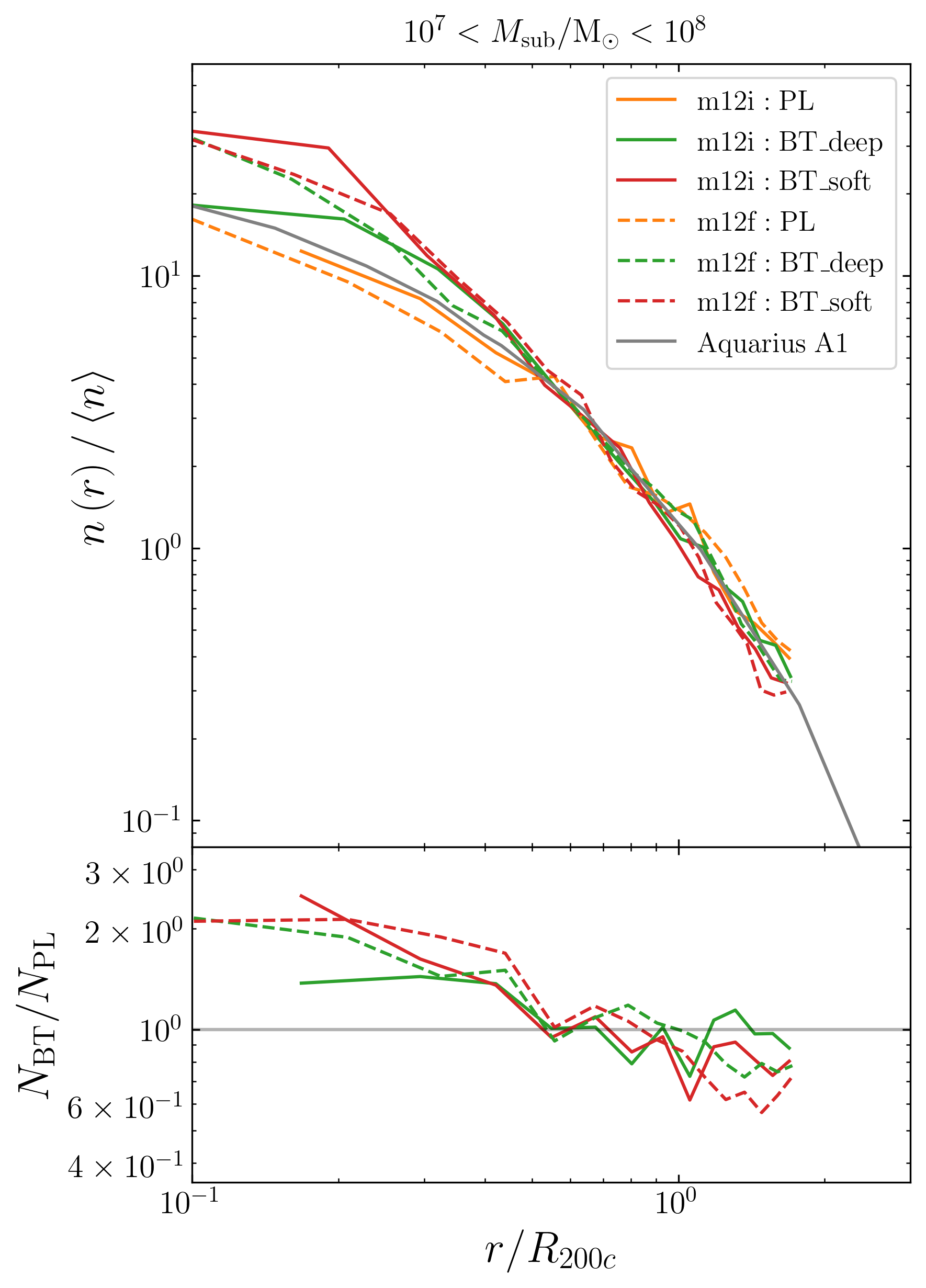}

    \includegraphics[width=0.45\textwidth]{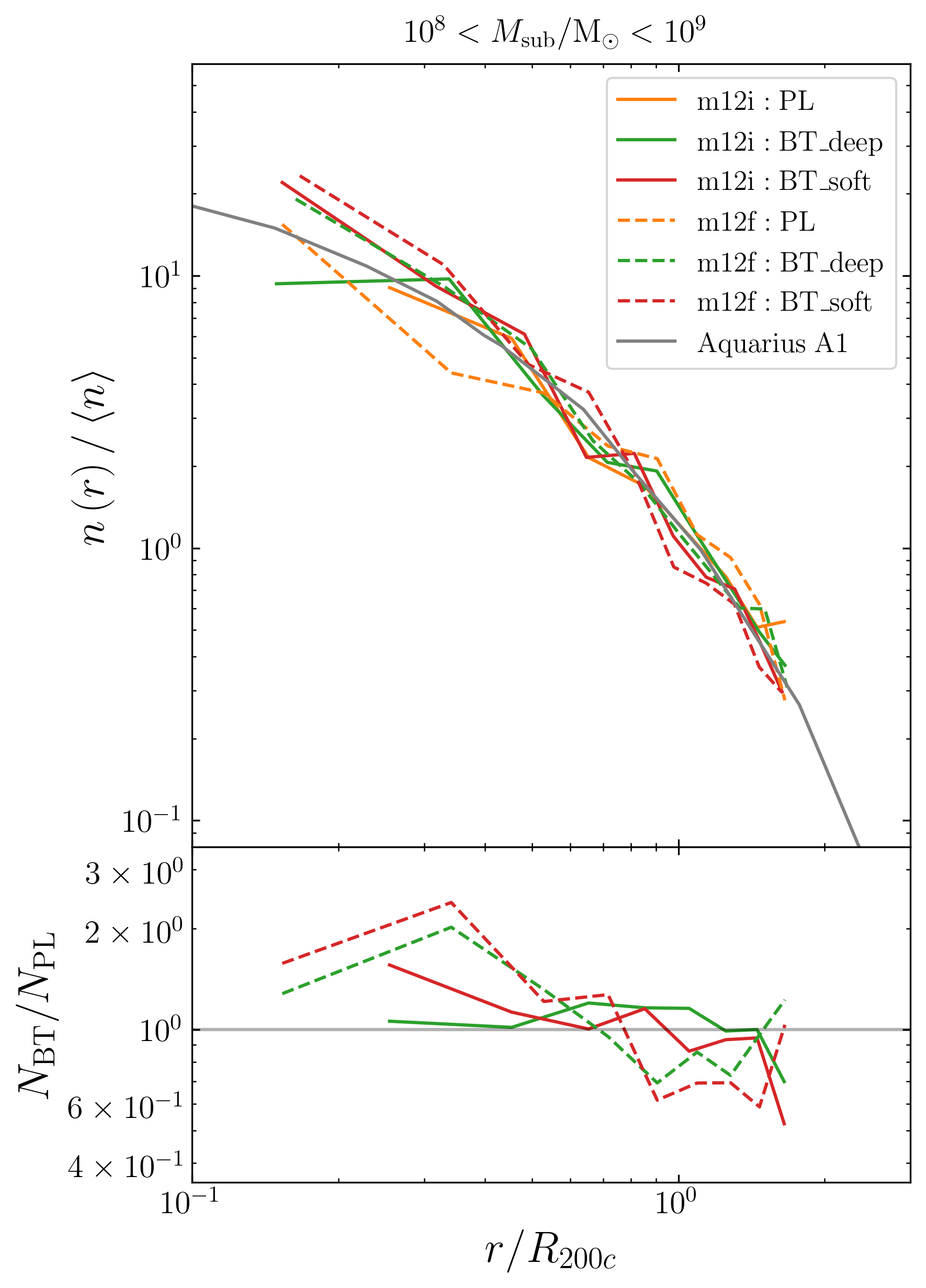}
    \includegraphics[width=0.45\textwidth]{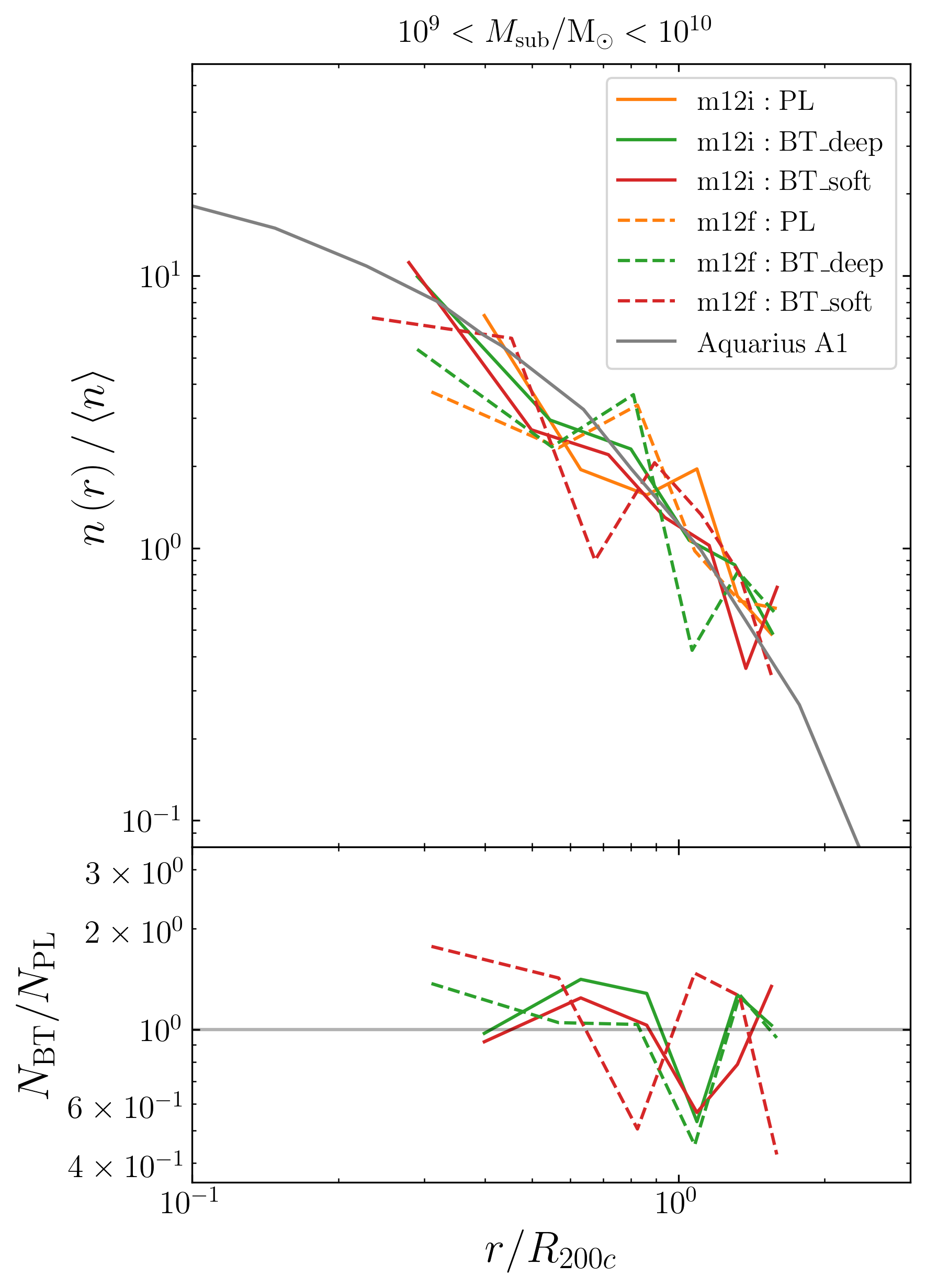}
    \vspace{-2mm}
    \caption{The subhalo radial number density profiles at four different mass bins for m12i (colored solid) and m12f (colored dashed). In each subfigure (also in each mass bin): The \textit{upper panel} shows the radial number density $n(r)$ normalized by $\left< n \right>$ in \autoref{eq:nbar_def}, for the PL (orange), the BT\_deep (green), and the BT\_soft (red) models. The grey solid line shows the equivalent substructure profile from Figure 12 of Ref. \cite{Springel08Aquarius}. The \textit{bottom panel} shows the ratios of the BT normalized number density to the PL normalized number density.}
    \label{fig:HRF_Aquarius}
\end{figure*}

In \autoref{fig:HRF_Aquarius}, we present the subhalo radial number density profile: the normalized radial number density, $n(r)/\left<n\right>$, as a function of ${r}/{R_{200c}}$, which is the subhalo's distance to the main halo center in the units of its main halo's $R_{200c}$. Here, $n(r)$ is the subhalo number density in the specific radial bin; $\left<n\right>$ is defined by:
\begin{align}
    \left<n\right> = \frac{N_{50c}}{\frac{4}{3} \pi R_{50c}^3},
    \label{eq:nbar_def}
\end{align}
where $N_{50c}$ is the number of subhalos within $R_{50c}$ in the corresponding mass bin. The grey ``Aquarius A1'' line is the highest resolution simulation result of the Aquarius project \cite{Springel08Aquarius}, with which the COCO project has shown its fitting. Although the original line uses kpc as a unit, we divide r over Aquarius A1's $R_{200c} \sim 245~{\rm kpc}$ to get the scaled radius. Ref. \cite{Springel08Aquarius} suggested that if splitting subhalos with different mass bins across from $10^5~ \msun$ to $10^{10}~ \msun$, the radial distribution at every bin would all follow this line. Due to the resolution limit of our simulation, we only show the four mass bins from $10^6 ~\msun$ to $10^{10} ~\msun$.

The upper panels of \autoref{fig:HRF_Aquarius} show the PL results between $10^6 ~\msun$ and $10^9 ~\msun$ fit well with the Aquarius lines. However, for the $10^9 ~\msun$ to $10^{10} ~\msun$ mass bin, we have too few subhalos (only $\sim$ 20 within $R_{50c}$), so the plotting has a substantial scatter around the Aquarius A1 line. 

In the bottom panels, for the whole mass range (or at least $10^6 ~\msun \sim 10^{9} ~\msun$ which noise effect is not too much), the BT models boost the normalized radial number density near the center by a factor of two. 

As suggested by \cite{Springel08Aquarius}, the subhalo radial normalized number density in the PL model follows a universal profile across different mass bins, i.e. the grey Aquarius A1 line in the four panels of \autoref{fig:HRF_Aquarius}. However, in more massive bins, the subhalos appear to exist only in the outer region. It is likely due to the low number density of subhalos near the center \cite{Springel08Aquarius} and lack of statistics using only two zoom-in simulations.

\subsubsection{Cumulative Substructure Mass Fraction\label{subsub:CMF}}

\begin{figure}[H]
    \centering
    \includegraphics[width={0.45\textwidth}]{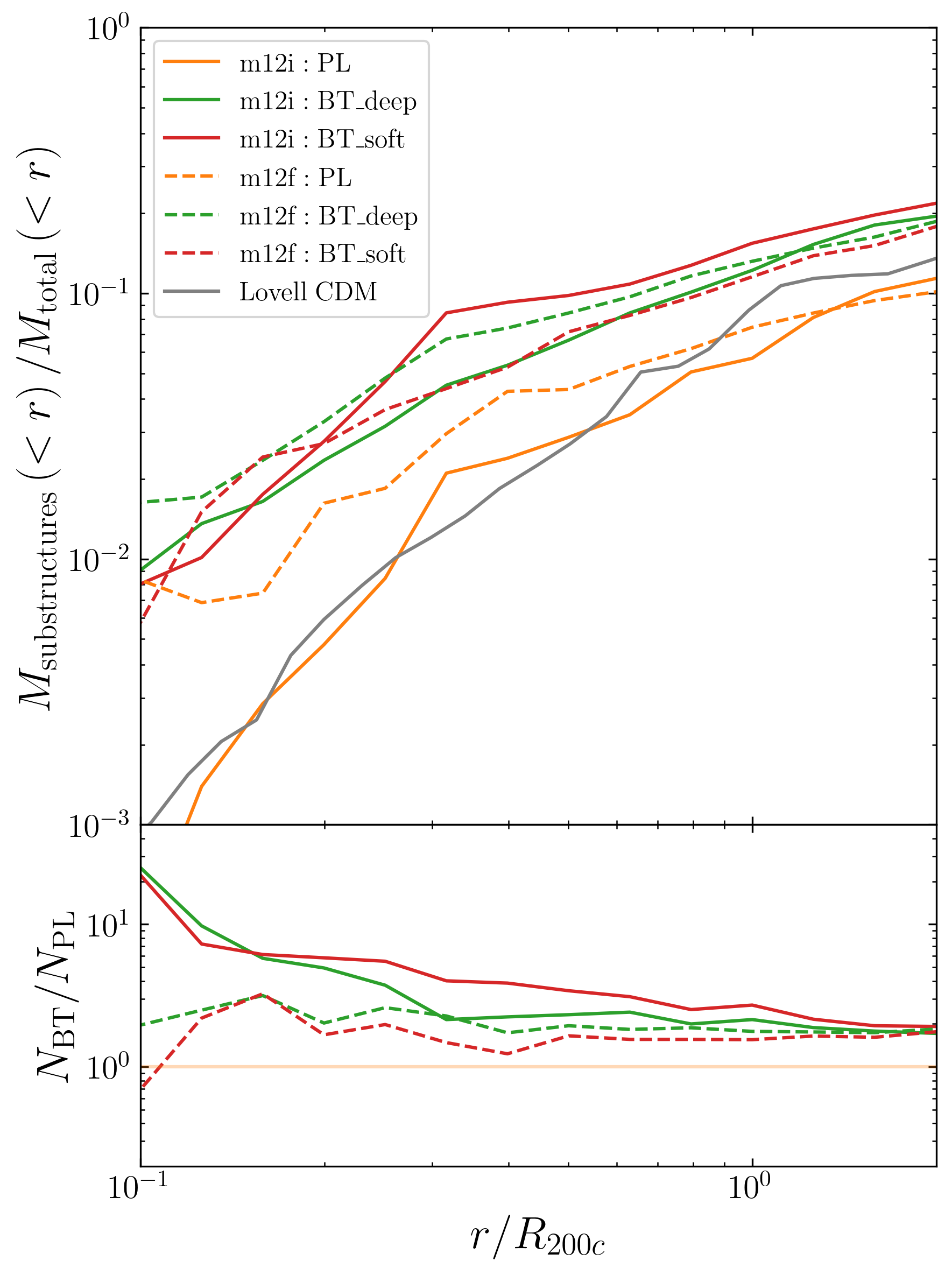}
    \caption{The cumulative mass fraction in substructures as a function of the radius for PL (orange), BT\_deep (green), and BT\_soft (red). The \textit{upper panel} shows the cumulative substructure fraction function, with $r$ (the subhalo distance from the main-halo center) scaled by $R_{200c}$ of the main halo. The grey solid line shows the equivalent substructure profile from Ref. \cite{Love14WDM}. The \textit{bottom panel} shows the ratios of the BT substructure fractions to the PL substructure fractions.}
    \label{fig:CMF}
\end{figure}

In \autoref{fig:CMF}, we present the cumulative substructure mass fraction: the mass of particles belonging to subhalos over the total mass, within a certain radius to the main halo center $r$. This substructure fraction could be useful for future strong lensing studies \cite{Vegetti09StrongLensingToStudyMassFrac} on the BT models. We take the cumulative substructure mass fraction function for the PL LCDM model from Ref. \cite{Love14WDM} as the Lovell CDM result. Additionally, we rescale the radius by $R_{200c}$ of the main halo, to eliminate the influence of the different main halo sizes.

The upper panel of \autoref{fig:CMF} shows a good match between the Lovell CDM result and our PL simulation for m12i. But, unexpectedly, the PL simulation of m12f is higher than the Lovell CDM result. This is because, in m12f, there is a subhalo with $7.7 \times 10^8 ~\msun$ within $0.1 R_{200c}$, which can account for the unusually high substructure fraction for the m12f PL simulation. 

We find that the BT\_deep and BT\_soft lines are higher than the Lovell CDM result and our BT simulations, implying that there are more substructures with the BT than PL PPS. The enhancement is especially strong in the inner region, where strong lensing is likely to probe.

To quantify the difference, in the bottom panel, we show the ratio of the cumulative substructure mass fraction between the BT and PL simulations. For BT\_deep, the fraction is enhanced by a factor of ten at the inner region ($0.1 R_{200c}$). As the radius increases, the enhancement ratio drops and finally becomes stable around a factor of two at the outer region. This is also consistent with our result in \autoref{subsub:HMF}: the BT models can double the total mass residing in resolved subhalos.

\subsubsection{Subhalo \texorpdfstring{$R_{\rm max}-V_{\rm max}$}{Rmax-Vmax} relationship}
\label{subsub:Rmax-Vmax}

\begin{figure*}
    \centering
    \includegraphics[width=0.45\linewidth]{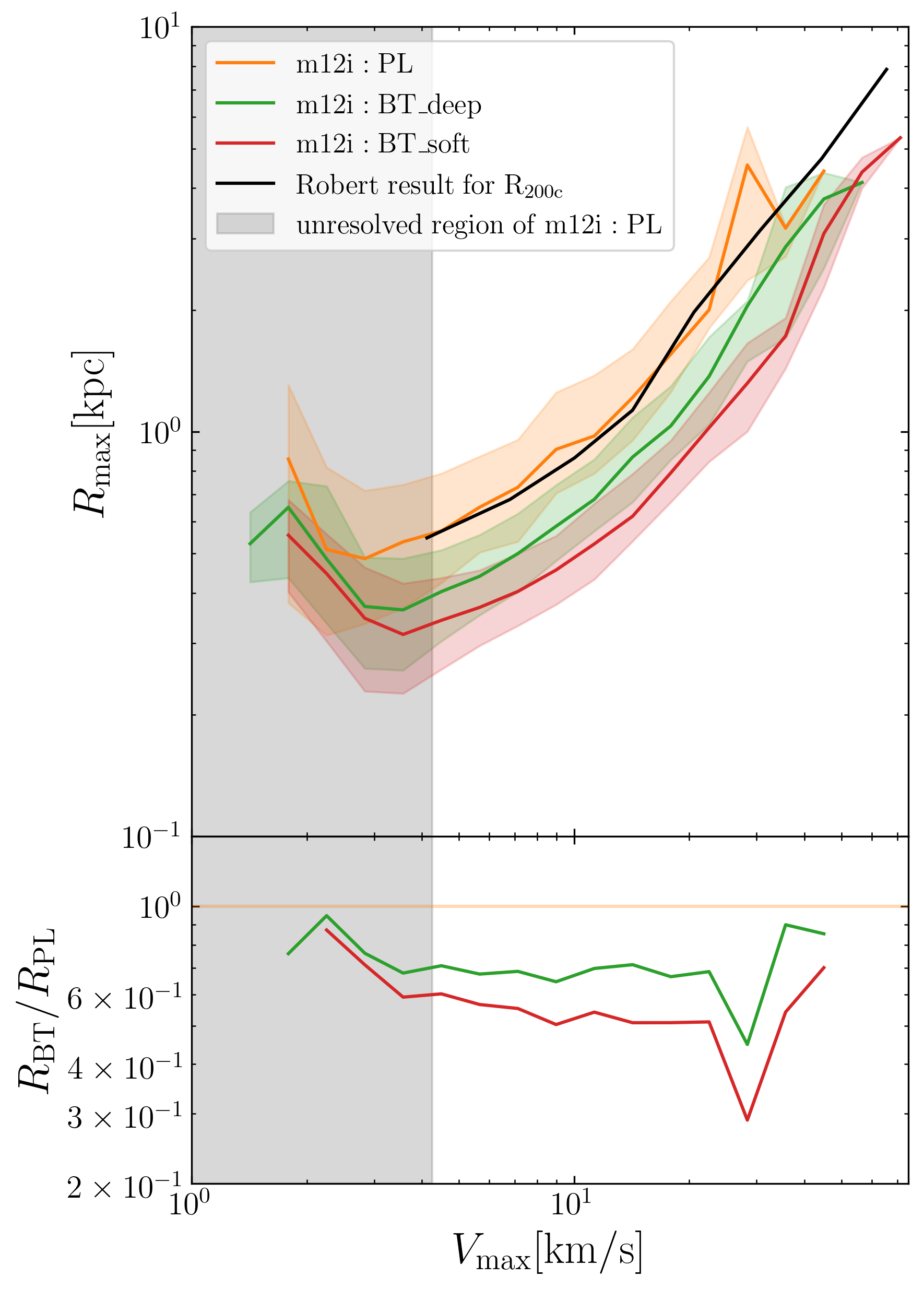}
    \includegraphics[width=0.45\linewidth]{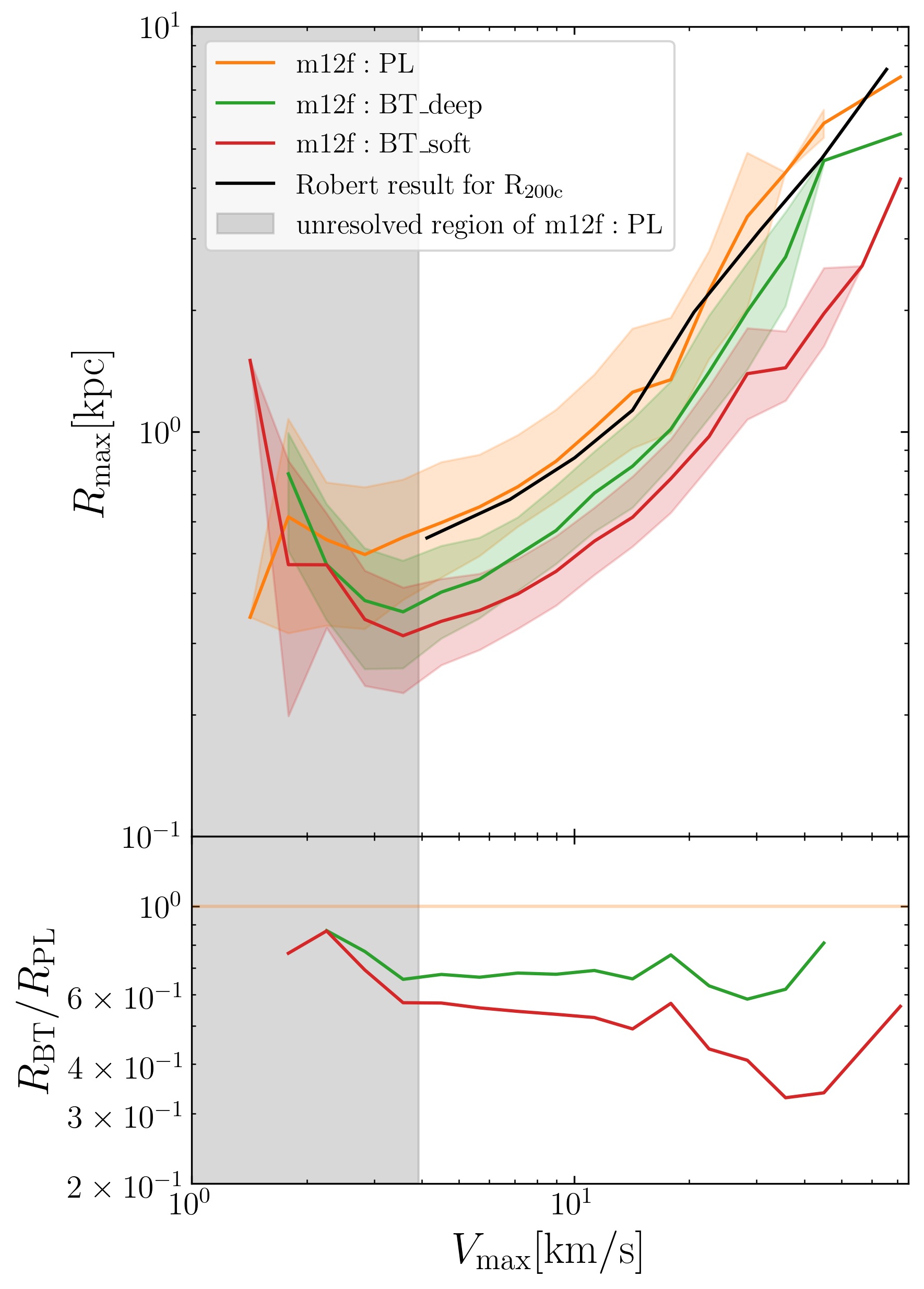}
    \caption{The subhalo $V_{\rm max}$-$R_{\rm max}$ relations for m12i (left figure) and m12f (right figure) respectively. For both figures:\protect\\
    The \textit{upper panel} shows the subhalo \(R_{\rm max}-V_{\rm max}\) relations for all the subhalos within $R_{200c}$ from the main halo center with the PL (orange), BT\_deep (green), and BT\_soft (red) models. The colored solid lines depict the median relation found by binning the subhalos according to their $V_{\rm max}$ values. The colored regions surrounding the colored lines illustrate the 16th-84th percentiles around the medians. The black solid line is the result of Ref. \protect\cite{Robert21RmaxVmax} for the PL LCDM model, taking account of all the subhalos within $R_{200c}$ of the main halo.
    The \textit{bottom panel} shows the ratios of \(R_{\rm max}\) between the BT and PL models. 
    The \textit{shaded area} is the same as that in \autoref{fig:HVF_cautun}.}
    \label{fig:RmaxVmax}
\end{figure*}

We present the subhalo $R_{\rm max}-V_{\rm max}$ relationship in \autoref{fig:RmaxVmax} for all the subhalos within $R_{200c}$ from the main halo center, with m12i and m12f, respectively. It is a measurement of the subhalo central density, thus astrophysically interesting: for a specific $V_{\rm max}$ bin, a larger(smaller) $R_{\rm max}$ usually means a lower(higher) central density. The black line is taken from Ref. \cite{Robert21RmaxVmax} (Robert) for the PL LCDM cosmology, considering all the subhalos within $R_{200c}$ of the main halos.

The upper panel of \autoref{fig:RmaxVmax} shows a good match between the Robert result and our PL simulation, for both m12i and m12f. The two BT models have much smaller $R_{\rm max}$ than the PL model. The BT\_soft model has a lower $R_{\rm max}$ than the BT\_deep model. This implies while subhalos in BT\_deep are more concentrated than in PL, the BT\_soft model is even more concentrated than the BT\_deep model. The unresolved regions for m12i PL and m12f PL are indicated with shaded areas, following the same definition as in \autoref{subsub:HVF}.

In the bottom panels, the BT models suppress $R_{\rm max}$ at nearly the whole resolved $V_{\rm max}$ range. The effects in m12i and m12f are similar: for the BT\_deep model, $R_{\rm max}$ decreases by $\sim 40\%$; for the BT\_soft model, $R_{\rm max}$ decreases nearly by half.

\section{CONCLUSIONS}
\label{sec:conclusion}
The standard model of cosmology -- the Power Law Primordial Power Spectrum + the Lambda Cold Dark Matter -- has been successful over the past several decades, especially on the large scales of the universe \cite{Planck20CMB,Planck20cmbcosmology,Troxel17DesCosmicShear, Blanton17SDSSsurvey,Chabanier19LyalphaPS}. However, there are still possible tensions between the observational results and the theoretical simulations. 

JWST's high redshift observations revealed that galaxies could form earlier than those in the standard model of cosmology \cite{Labbe23highzgalaxiesreport,Boyl23JWSTCDM,Love23JWSTCDM}. The strong lensing observations prefer more substructures \cite{Macc06CDMlensing,Xu09AquariusLensing,Xu15lensingCDM}. Furthermore, the recent observations \cite{Mull24M83,Homm23MWmanydwarf} and theories \cite{Kim18NoMSP} suggested there could be a ``too many satellites'' problem in the nearby universe. These motivate us to consider alternative cosmological models with an enhanced Primordial Power Spectrum at small scales ($>1~{\rm cMpc}^{-1}$), namely the blue-tilted (BT) model.

In this study, we have conducted cosmological zoom-in simulations of MW host mass halos with the blue-tilted primordial power spectrum. We have considered two blue-tilted models from Ref. \cite{Hira24bluetilt}: BT\_soft with the enhancement at $\gtrsim 1~{\rm cMpc}^{-1}$ and BT\_deep with the enhancement at $\gtrsim 4~{\rm cMpc}^{-1}$. They are all within the accepted parameter space according to the JWST observation \cite{Hira24bluetilt} and the dwarf galaxy central density observation \cite{Dekker24dwarfgalaxyBT}.

Our main results are summarized in the following:
\begin{enumerate}

    \item We have found that BT\_soft can enhance the mass, radius, $V_{\rm max}$, concentration (\autoref{tab:result_simulations}), and the radial density profile (\autoref{subsub:RadialDensityProfile}) of the main halo. However, the effects of the BT\_deep model on the main halo are much smaller, since it is blue-tilted at a smaller scale in the primordial power spectrum, which corresponds to a smaller mass scale (\autoref{eq:Mp}).

    \item  The BT models enhance the subhalo mass function by $\sim 2$ and the subhalo $V_{\rm max}$ function by $\gtrsim 3$ for a wide range of halo masses ($\sim 10^6 ~\msun - 10^9 ~\msun$) and maximum circular velocities ($\sim 5~{\rm km/s} - 50 ~{\rm km/s}$) (\autoref{subsub:HMF} and \autoref{subsub:HVF}). The enhancement ratios, defined as the numbers of subhalos in BT over that in PL, follow inverse S shape functions (\autoref{eq:inverse_S_shape_HMF}) in both mass and $V_{\rm max}$ functions. 
    
    In both the BT\_deep and BT\_soft simulations, the enhancements in the subhalo mass function are similar. However, the BT\_soft simulation has a stronger enhancement in the subhalo $V_{\rm max}$ function than the BT\_deep simulation.

    \item The BT primordial power spectrum increases the number density of subhalos in the inner region of the main host halo by more than a factor of two, compared to PL (\autoref{subsub:HRF}). In the BT simulations, the number densities of subhalos in the lower mass bins are more centrally concentrated.

    \item We have found that the BT models can boost the mass fraction of the substructures in the inner region ($<0.1 R_{200c}$) by an order of magnitude (\autoref{subsub:CMF}). Even around $R_{200c}$, the BT models could have a two-fold increase in the substructure mass fraction, compared to PL.

    \item At a fixed $V_{\rm max}$, the BT models reduce the mean $R_{\rm max}$ values of the subhalos by $40 \% -50\%$ (\autoref{subsub:Rmax-Vmax}). This reveals that subhalos are more concentrated in BT than PL.

\end{enumerate}

To constrain the primordial power spectrum with our simulations, we require comparisons with observations, including the statistics of nearby satellite galaxies and the strong lensing observations of Milky-Way size host halos. However, at these small scales, baryonic physics cannot be ignored, e.g. the tidal stripping of subhalos with baryonic disks \citep{Garrison-Kimmel17notsolumpy}, the uncertainties in the stellar-to-halo mass relation \citep{Sales22baryonicsolution}, the baryonic effects on the inner density profiles \citep{Chan15}, and the lensing effects of the stellar structures \citep{Gilm17baryonstronglensing,Hsue16disklensing}. Therefore, it is necessary to incorporate regulator-type models \citep{Finl08regulator,Krav22GRUMPY}, semi-analytic models \citep{Cole00GALFORM,Bens12GALACTICUS}, or even cosmological galaxy simulations \citep{FIRE2,Schaye23Flamingo} to disentangle the effects of the primordial power spectrum from baryons.

% ====== code and data availability ====== %

\section*{Code and Data Availability}

The analysis code used to produce the main figures in this paper is publicly available at \url{https://github.com/rushingfox/btmw}~\citep{wu_2026_20805091}. The repository includes intermediate per-figure caches so that the main figures can be reproduced without access to the raw simulation outputs. The raw simulation data (SOAP halo catalogues, HBT-HERONS subhalo catalogues, VELOCIraptor catalogues, and SWIFT snapshots) are also available separately at \url{https://doi.org/10.5281/zenodo.20805009}~\citep{wu_2026_20805009}.

% ====== acknowledgments ====== %

\section*{Acknowledgments}
We thank Matthieu Schaller, Yiming Zhong, Ariane Dekker, and Kin-Wang Ng for their thoughtful comments. The research in this paper used the {\small SWIFT} open-source simulation code (\url{http://www.swiftsim.com}, \cite{ascl18SWIFT}) version 1.0.0. We also used {\small SWIFTSIMIO} \citep{Borrow2020swiftsimio, Borrow2021swiftsimioprojection} to generate the projection maps. This work also made use of matplotlib \citep{Hunt07matplotlib}, numpy \citep{vand11numpy}, scipy \citep{Jone01scipy,Virt20scipy} and NASA’s Astrophysics Data System. We acknowledge the support of the CUHK Central High-Performance Computing Cluster, on which the computation in this work has been performed; and we thank Edward So, Anny Cheung, and Nicky Leung for their assistance.

TKC is supported by the `Improvement on Competitiveness in Hiring New Faculties' Funding Scheme from the Chinese University of Hong Kong (4937210, 4937211, 4937212) and RGC Early Career Scheme (24301524). TKC and JW are partially supported by the Direct Grant project (4053662) from the Chinese University of Hong Kong. VJFM acknowledges support by NWO through the Dark Universe Science Collaboration (OCENW.XL21.XL21.025).

% ====== bib ====== %
\bibliographystyle{apsrev4-1-JHEPfix}
\bibliography{main}

@STRING{araa = "ARA\&A"}

@STRING{as = "Appl. Spectrosc."}

@STRING{an = "Astron. Nachr."}

@STRING{apj = "ApJ"}

@STRING{apjl = "ApJ"}

@STRING{lo = "Laser Optoelektron."}

@STRING{nat = "Nat"}

@STRING{on = "Opt. News"}

@STRING{prd = "Phys. Rev. D"}

@STRING{pasj = "PASJ"}

@STRING{st = "Sidereal Times"}

@STRING{sky = "Sky Telesc."}

@STRING{space = "Space"}

@STRING{ssr = "Space Sci. Rev."}

@ARTICLE{Victor25HBTHERONS,
       author = {{Forouhar Moreno}, Victor J. and {Helly}, John and {McGibbon}, Robert and {Schaye}, Joop and {Schaller}, Matthieu and {Han}, Jiaxin and {Kugel}, Roi and {Bah{\'e}}, Yannick M.},
        title = "{Assessing subhalo finders in cosmological hydrodynamical simulations}",
      journal = {\mnras},
         year = 2025,
        month = oct,
       volume = {543},
       number = {2},
        pages = {1339-1372},
          doi = {10.1093/mnras/staf1478},
archivePrefix = {arXiv},
       eprint = {2502.06932},
 primaryClass = {astro-ph.CO},
       adsurl = {https://ui.adsabs.harvard.edu/abs/2025MNRAS.543.1339F}
}

@ARTICLE{Diem15massconcentration,
       author = {{Diemer}, Benedikt and {Kravtsov}, Andrey V.},
        title = "{A Universal Model for Halo Concentrations}",
      journal = {\apj},
         year = 2015,
        month = jan,
       volume = {799},
       number = {1},
          eid = {108},
        pages = {108},
          doi = {10.1088/0004-637X/799/1/108},
archivePrefix = {arXiv},
       eprint = {1407.4730},
 primaryClass = {astro-ph.CO},
       adsurl = {https://ui.adsabs.harvard.edu/abs/2015ApJ...799..108D}
}

@ARTICLE{Jeth18minhalo,
       author = {{Jethwa}, P. and {Erkal}, D. and {Belokurov}, V.},
        title = "{The upper bound on the lowest mass halo}",
      journal = {\mnras},
         year = 2018,
        month = jan,
       volume = {473},
       number = {2},
        pages = {2060-2083},
          doi = {10.1093/mnras/stx2330},
archivePrefix = {arXiv},
       eprint = {1612.07834},
 primaryClass = {astro-ph.GA},
       adsurl = {https://ui.adsabs.harvard.edu/abs/2018MNRAS.473.2060J}
}

@ARTICLE{Grau19,
       author = {{Graus}, Andrew S. and {Bullock}, James S. and {Kelley}, Tyler and {Boylan-Kolchin}, Michael and {Garrison-Kimmel}, Shea and {Qi}, Yuewen},
        title = "{How low does it go? Too few Galactic satellites with standard reionization quenching}",
      journal = {\mnras},
         year = 2019,
        month = oct,
       volume = {488},
       number = {4},
        pages = {4585-4595},
          doi = {10.1093/mnras/stz1992},
archivePrefix = {arXiv},
       eprint = {1808.03654},
 primaryClass = {astro-ph.GA},
       adsurl = {https://ui.adsabs.harvard.edu/abs/2019MNRAS.488.4585G}
}

@ARTICLE{Laza20coreddarkmatter,
       author = {{Lazar}, Alexandres and {Bullock}, James S. and {Boylan-Kolchin}, Michael and {Chan}, T.~K. and {Hopkins}, Philip F. and {Graus}, Andrew S. and {Wetzel}, Andrew and {El-Badry}, Kareem and {Wheeler}, Coral and {Straight}, Maria C. and {Kere{\v{s}}}, Du{\v{s}}an and {Faucher-Gigu{\`e}re}, Claude-Andr{\'e} and {Fitts}, Alex and {Garrison-Kimmel}, Shea},
        title = "{A dark matter profile to model diverse feedback-induced core sizes of {\ensuremath{\Lambda}}CDM haloes}",
      journal = {\mnras},
         year = 2020,
        month = sep,
       volume = {497},
       number = {2},
        pages = {2393-2417},
          doi = {10.1093/mnras/staa2101},
archivePrefix = {arXiv},
       eprint = {2004.10817},
 primaryClass = {astro-ph.GA},
       adsurl = {https://ui.adsabs.harvard.edu/abs/2020MNRAS.497.2393L}
}

@ARTICLE{Gilm17baryonstronglensing,
       author = {{Gilman}, Daniel and {Agnello}, Adriano and {Treu}, Tommaso and {Keeton}, Charles R. and {Nierenberg}, Anna M.},
        title = "{Strong lensing signatures of luminous structure and substructure in early-type galaxies}",
      journal = {\mnras},
         year = 2017,
        month = jun,
       volume = {467},
       number = {4},
        pages = {3970-3992},
          doi = {10.1093/mnras/stx158},
archivePrefix = {arXiv},
       eprint = {1610.08525},
 primaryClass = {astro-ph.CO},
       adsurl = {https://ui.adsabs.harvard.edu/abs/2017MNRAS.467.3970G}
}

@ARTICLE{Springel01Gadget1,
       author = {{Springel}, Volker and {Yoshida}, Naoki and {White}, Simon D.~M.},
        title = "{GADGET: a code for collisionless and gasdynamical cosmological simulations}",
      journal = {\na},
         year = 2001,
        month = apr,
       volume = {6},
       number = {2},
        pages = {79-117},
          doi = {10.1016/S1384-1076(01)00042-2},
archivePrefix = {arXiv},
       eprint = {astro-ph/0003162},
 primaryClass = {astro-ph},
       adsurl = {https://ui.adsabs.harvard.edu/abs/2001NewA....6...79S}
}

@ARTICLE{Behroozi13Rockstar,
       author = {{Behroozi}, Peter S. and {Wechsler}, Risa H. and {Wu}, Hao-Yi},
        title = "{The ROCKSTAR Phase-space Temporal Halo Finder and the Velocity Offsets of Cluster Cores}",
      journal = {\apj},
         year = 2013,
        month = jan,
       volume = {762},
       number = {2},
          eid = {109},
        pages = {109},
          doi = {10.1088/0004-637X/762/2/109},
archivePrefix = {arXiv},
       eprint = {1110.4372},
 primaryClass = {astro-ph.CO},
       adsurl = {https://ui.adsabs.harvard.edu/abs/2013ApJ...762..109B}
}

@ARTICLE{Lyth99inflation,
       author = {{Lyth}, D.~H.~D.~H. and {Riotto}, A.~A.},
        title = "{Particle physics models of inflation and the cosmological density perturbation}",
      journal = {\physrep},
         year = 1999,
        month = jun,
       volume = {314},
       number = {1-2},
        pages = {1-146},
          doi = {10.1016/S0370-1573(98)00128-8},
archivePrefix = {arXiv},
       eprint = {hep-ph/9807278},
 primaryClass = {hep-ph},
       adsurl = {https://ui.adsabs.harvard.edu/abs/1999PhR...314....1L}
}

@Misc{Jone01scipy,
  author =    {Eric Jones and Travis Oliphant and Pearu Peterson and others},
  title =     {{SciPy}: Open source scientific tools for {Python}},
  year =      {2001},
  url = "http://www.scipy.org/",
  note = {[Online; accessed <today>]}
}

@ARTICLE{Virt20scipy,
  author  = {Virtanen, Pauli and Gommers, Ralf and Oliphant, Travis E. and
            Haberland, Matt and Reddy, Tyler and Cournapeau, David and
            Burovski, Evgeni and Peterson, Pearu and Weckesser, Warren and
            Bright, Jonathan and {van der Walt}, St{\'e}fan J. and
            Brett, Matthew and Wilson, Joshua and Millman, K. Jarrod and
            Mayorov, Nikolay and Nelson, Andrew R. J. and Jones, Eric and
            Kern, Robert and Larson, Eric and Carey, C J and
            Polat, {\.I}lhan and Feng, Yu and Moore, Eric W. and
            {VanderPlas}, Jake and Laxalde, Denis and Perktold, Josef and
            Cimrman, Robert and Henriksen, Ian and Quintero, E. A. and
            Harris, Charles R. and Archibald, Anne M. and
            Ribeiro, Ant{\^o}nio H. and Pedregosa, Fabian and
            {van Mulbregt}, Paul and {SciPy 1.0 Contributors}},
  title   = {{{SciPy} 1.0: Fundamental Algorithms for Scientific
            Computing in Python}},
  journal = {Nature Methods},
  year    = {2020},
  volume  = {17},
  pages   = {261--272},
  adsurl  = {https://rdcu.be/b08Wh},
  doi     = {10.1038/s41592-019-0686-2},
}

@ARTICLE{vand11numpy, 
author={S. van der Walt and S. C. Colbert and G. Varoquaux}, 
journal={Computing in Science Engineering}, 
title={The NumPy Array: A Structure for Efficient Numerical Computation}, 
year={2011}, 
volume={13}, 
number={2}, 
pages={22-30}, 
doi={10.1109/MCSE.2011.37}, 
ISSN={1521-9615}, 
month={March}}

@ARTICLE{Hunt07matplotlib,
   author = {{Hunter}, J.~D.},
    title = "{Matplotlib: A 2D Graphics Environment}",
  journal = {Computing in Science and Engineering},
     year = 2007,
    month = may,
   volume = 9,
    pages = {90-95},
      doi = {10.1109/MCSE.2007.55},
   adsurl = {http://adsabs.harvard.edu/abs/2007CSE.....9...90H}
}

@ARTICLE{Eise98transfer,
       author = {{Eisenstein}, Daniel J. and {Hu}, Wayne},
        title = "{Baryonic Features in the Matter Transfer Function}",
      journal = {\apj},
         year = 1998,
        month = mar,
       volume = {496},
       number = {2},
        pages = {605-614},
          doi = {10.1086/305424},
archivePrefix = {arXiv},
       eprint = {astro-ph/9709112},
 primaryClass = {astro-ph},
       adsurl = {https://ui.adsabs.harvard.edu/abs/1998ApJ...496..605E}
}

@ARTICLE{Boyl23JWSTCDM,
       author = {{Boylan-Kolchin}, Michael},
        title = "{Stress testing {\ensuremath{\Lambda}}CDM with high-redshift galaxy candidates}",
      journal = {Nature Astronomy},
         year = 2023,
        month = jun,
       volume = {7},
        pages = {731-735},
          doi = {10.1038/s41550-023-01937-7},
archivePrefix = {arXiv},
       eprint = {2208.01611},
 primaryClass = {astro-ph.CO},
       adsurl = {https://ui.adsabs.harvard.edu/abs/2023NatAs...7..731B}
}

@ARTICLE{Love23JWSTCDM,
       author = {{Lovell}, Christopher C. and {Harrison}, Ian and {Harikane}, Yuichi and {Tacchella}, Sandro and {Wilkins}, Stephen M.},
        title = "{Extreme value statistics of the halo and stellar mass distributions at high redshift: are JWST results in tension with {\ensuremath{\Lambda}}CDM?}",
      journal = {\mnras},
         year = 2023,
        month = jan,
       volume = {518},
       number = {2},
        pages = {2511-2520},
          doi = {10.1093/mnras/stac3224},
archivePrefix = {arXiv},
       eprint = {2208.10479},
 primaryClass = {astro-ph.GA},
       adsurl = {https://ui.adsabs.harvard.edu/abs/2023MNRAS.518.2511L}
}

@ARTICLE{Planck20cmbcosmology,
       author = {{Planck Collaboration} and {Aghanim}, N. and {Akrami}, Y. and {Ashdown}, M. and {Aumont}, J. and {Baccigalupi}, C. and {Ballardini}, M. and {Banday}, A.~J. and {Barreiro}, R.~B. and {Bartolo}, N. and {Basak}, S. and {Battye}, R. and {Benabed}, K. and {Bernard}, J. -P. and {Bersanelli}, M. and {Bielewicz}, P. and {Bock}, J.~J. and {Bond}, J.~R. and {Borrill}, J. and {Bouchet}, F.~R. and {Boulanger}, F. and {Bucher}, M. and {Burigana}, C. and {Butler}, R.~C. and {Calabrese}, E. and {Cardoso}, J. -F. and {Carron}, J. and {Challinor}, A. and {Chiang}, H.~C. and {Chluba}, J. and {Colombo}, L.~P.~L. and {Combet}, C. and {Contreras}, D. and {Crill}, B.~P. and {Cuttaia}, F. and {de Bernardis}, P. and {de Zotti}, G. and {Delabrouille}, J. and {Delouis}, J. -M. and {Di Valentino}, E. and {Diego}, J.~M. and {Dor{\'e}}, O. and {Douspis}, M. and {Ducout}, A. and {Dupac}, X. and {Dusini}, S. and {Efstathiou}, G. and {Elsner}, F. and {En{\ss}lin}, T.~A. and {Eriksen}, H.~K. and {Fantaye}, Y. and {Farhang}, M. and {Fergusson}, J. and {Fernandez-Cobos}, R. and {Finelli}, F. and {Forastieri}, F. and {Frailis}, M. and {Fraisse}, A.~A. and {Franceschi}, E. and {Frolov}, A. and {Galeotta}, S. and {Galli}, S. and {Ganga}, K. and {G{\'e}nova-Santos}, R.~T. and {Gerbino}, M. and {Ghosh}, T. and {Gonz{\'a}lez-Nuevo}, J. and {G{\'o}rski}, K.~M. and {Gratton}, S. and {Gruppuso}, A. and {Gudmundsson}, J.~E. and {Hamann}, J. and {Handley}, W. and {Hansen}, F.~K. and {Herranz}, D. and {Hildebrandt}, S.~R. and {Hivon}, E. and {Huang}, Z. and {Jaffe}, A.~H. and {Jones}, W.~C. and {Karakci}, A. and {Keih{\"a}nen}, E. and {Keskitalo}, R. and {Kiiveri}, K. and {Kim}, J. and {Kisner}, T.~S. and {Knox}, L. and {Krachmalnicoff}, N. and {Kunz}, M. and {Kurki-Suonio}, H. and {Lagache}, G. and {Lamarre}, J. -M. and {Lasenby}, A. and {Lattanzi}, M. and {Lawrence}, C.~R. and {Le Jeune}, M. and {Lemos}, P. and {Lesgourgues}, J. and {Levrier}, F. and {Lewis}, A. and {Liguori}, M. and {Lilje}, P.~B. and {Lilley}, M. and {Lindholm}, V. and {L{\'o}pez-Caniego}, M. and {Lubin}, P.~M. and {Ma}, Y. -Z. and {Mac{\'\i}as-P{\'e}rez}, J.~F. and {Maggio}, G. and {Maino}, D. and {Mandolesi}, N. and {Mangilli}, A. and {Marcos-Caballero}, A. and {Maris}, M. and {Martin}, P.~G. and {Martinelli}, M. and {Mart{\'\i}nez-Gonz{\'a}lez}, E. and {Matarrese}, S. and {Mauri}, N. and {McEwen}, J.~D. and {Meinhold}, P.~R. and {Melchiorri}, A. and {Mennella}, A. and {Migliaccio}, M. and {Millea}, M. and {Mitra}, S. and {Miville-Desch{\^e}nes}, M. -A. and {Molinari}, D. and {Montier}, L. and {Morgante}, G. and {Moss}, A. and {Natoli}, P. and {N{\o}rgaard-Nielsen}, H.~U. and {Pagano}, L. and {Paoletti}, D. and {Partridge}, B. and {Patanchon}, G. and {Peiris}, H.~V. and {Perrotta}, F. and {Pettorino}, V. and {Piacentini}, F. and {Polastri}, L. and {Polenta}, G. and {Puget}, J. -L. and {Rachen}, J.~P. and {Reinecke}, M. and {Remazeilles}, M. and {Renzi}, A. and {Rocha}, G. and {Rosset}, C. and {Roudier}, G. and {Rubi{\~n}o-Mart{\'\i}n}, J.~A. and {Ruiz-Granados}, B. and {Salvati}, L. and {Sandri}, M. and {Savelainen}, M. and {Scott}, D. and {Shellard}, E.~P.~S. and {Sirignano}, C. and {Sirri}, G. and {Spencer}, L.~D. and {Sunyaev}, R. and {Suur-Uski}, A. -S. and {Tauber}, J.~A. and {Tavagnacco}, D. and {Tenti}, M. and {Toffolatti}, L. and {Tomasi}, M. and {Trombetti}, T. and {Valenziano}, L. and {Valiviita}, J. and {Van Tent}, B. and {Vibert}, L. and {Vielva}, P. and {Villa}, F. and {Vittorio}, N. and {Wandelt}, B.~D. and {Wehus}, I.~K. and {White}, M. and {White}, S.~D.~M. and {Zacchei}, A. and {Zonca}, A.},
        title = "{Planck 2018 results. VI. Cosmological parameters}",
      journal = {\aap},
         year = 2020,
        month = sep,
       volume = {641},
          eid = {A6},
        pages = {A6},
          doi = {10.1051/0004-6361/201833910},
archivePrefix = {arXiv},
       eprint = {1807.06209},
 primaryClass = {astro-ph.CO},
       adsurl = {https://ui.adsabs.harvard.edu/abs/2020A&A...641A...6P}
}

@article{Borrow2020swiftsimio,
  doi = {10.21105/joss.02430},
  url = {https://doi.org/10.21105/joss.02430},
  year = {2020},
  publisher = {The Open Journal},
  volume = {5},
  number = {52},
  pages = {2430},
  author = {Josh Borrow and Alexei Borrisov},
  title = {swiftsimio: A Python library for reading SWIFT data},
  journal = {Journal of Open Source Software}
}

@article{Borrow2021swiftsimioprojection,
  title={Projecting SPH Particles in Adaptive Environments},
  author={Josh Borrow and Ashley J. Kelly},
  year={2021},
  eprint={2106.05281},
  archivePrefix={arXiv},
  primaryClass={astro-ph.GA}
}

@ARTICLE{Kravtsov04HMF_scaled,
       author = {{Kravtsov}, Andrey V. and {Berlind}, Andreas A. and {Wechsler}, Risa H. and {Klypin}, Anatoly A. and {Gottl{\"o}ber}, Stefan and {Allgood}, Brandon and {Primack}, Joel R.},
        title = "{The Dark Side of the Halo Occupation Distribution}",
      journal = {\apj},
         year = 2004,
        month = jul,
       volume = {609},
       number = {1},
        pages = {35-49},
          doi = {10.1086/420959},
archivePrefix = {arXiv},
       eprint = {astro-ph/0308519},
 primaryClass = {astro-ph},
       adsurl = {https://ui.adsabs.harvard.edu/abs/2004ApJ...609...35K}
}

@article{Moore99HVF_scaled,
doi = {10.1086/312287},
url = {https://dx.doi.org/10.1086/312287},
year = {1999},
month = {sep},
publisher = {},
volume = {524},
number = {1},
pages = {L19},
author = {Ben Moore and Sebastiano Ghigna and Fabio Governato and George Lake and Thomas Quinn and Joachim Stadel and Paolo Tozzi},
title = {Dark Matter Substructure within Galactic Halos},
journal = {The Astrophysical Journal}
}

@ARTICLE{Hsue16disklensing,
       author = {{Hsueh}, J. -W. and {Fassnacht}, C.~D. and {Vegetti}, S. and {McKean}, J.~P. and {Spingola}, C. and {Auger}, M.~W. and {Koopmans}, L.~V.~E. and {Lagattuta}, D.~J.},
        title = "{SHARP - II. Mass structure in strong lenses is not necessarily dark matter substructure: a flux ratio anomaly from an edge-on disc in B1555+375}",
      journal = {\mnras},
         year = 2016,
        month = nov,
       volume = {463},
       number = {1},
        pages = {L51-L55},
          doi = {10.1093/mnrasl/slw146},
archivePrefix = {arXiv},
       eprint = {1601.01671},
 primaryClass = {astro-ph.CO},
       adsurl = {https://ui.adsabs.harvard.edu/abs/2016MNRAS.463L..51H}
}

@ARTICLE{Macc06CDMlensing,
       author = {{Macci{\`o}}, Andrea V. and {Miranda}, Marco},
        title = "{The effect of low-mass substructures on the cusp lensing relation}",
      journal = {\mnras},
         year = 2006,
        month = may,
       volume = {368},
       number = {2},
        pages = {599-608},
          doi = {10.1111/j.1365-2966.2006.10154.x},
archivePrefix = {arXiv},
       eprint = {astro-ph/0509598},
 primaryClass = {astro-ph},
       adsurl = {https://ui.adsabs.harvard.edu/abs/2006MNRAS.368..599M}
}

@ARTICLE{Xu15lensingCDM,
       author = {{Xu}, Dandan and {Sluse}, Dominique and {Gao}, Liang and {Wang}, Jie and {Frenk}, Carlos and {Mao}, Shude and {Schneider}, Peter and {Springel}, Volker},
        title = "{How well can cold dark matter substructures account for the observed radio flux-ratio anomalies}",
      journal = {\mnras},
         year = 2015,
        month = mar,
       volume = {447},
       number = {4},
        pages = {3189-3206},
          doi = {10.1093/mnras/stu2673},
archivePrefix = {arXiv},
       eprint = {1410.3282},
 primaryClass = {astro-ph.CO},
       adsurl = {https://ui.adsabs.harvard.edu/abs/2015MNRAS.447.3189X}
}

@article{Xu09AquariusLensing,
    author = {Xu, D. D. and Mao, Shude and Wang, Jie and Springel, V. and Gao, Liang and White, S. D. M. and Frenk, Carlos S. and Jenkins, Adrian and Li, Guoliang and Navarro, Julio F.},
    title = {Effects of dark matter substructures on gravitational lensing: results from the Aquarius simulations},
    journal = {Monthly Notices of the Royal Astronomical Society},
    volume = {398},
    number = {3},
    pages = {1235-1253},
    year = {2009},
    month = {09},
    issn = {0035-8711},
    doi = {10.1111/j.1365-2966.2009.15230.x},
    url = {https://doi.org/10.1111/j.1365-2966.2009.15230.x},
    eprint = {https://academic.oup.com/mnras/article-pdf/398/3/1235/4089181/mnras0398-1235.pdf},
}

@article{Mao98LensingSubstructure,
    author = {Mao, Shude and Schneider, Peter},
    title = {Evidence for substructure in lens galaxies?},
    journal = {Monthly Notices of the Royal Astronomical Society},
    volume = {295},
    number = {3},
    pages = {587-594},
    year = {1998},
    month = {04},
    issn = {0035-8711},
    doi = {10.1046/j.1365-8711.1998.01319.x},
    url = {https://doi.org/10.1046/j.1365-8711.1998.01319.x},
    eprint = {https://academic.oup.com/mnras/article-pdf/295/3/587/18408388/295-3-587.pdf},
}

@ARTICLE{Mao92SmoothPotentialLensing,
       author = {{Mao}, Shude},
        title = "{Gravitational Microlensing by a Single Star plus External Shear}",
      journal = {\apj},
         year = 1992,
        month = apr,
       volume = {389},
        pages = {63},
          doi = {10.1086/171188},
       adsurl = {https://ui.adsabs.harvard.edu/abs/1992ApJ...389...63M}
}

@ARTICLE{Planck20CMB,
       author = {{Planck Collaboration} and {Akrami}, Y. and {Arroja}, F. and {Ashdown}, M. and {Aumont}, J. and {Baccigalupi}, C. and {Ballardini}, M. and {Banday}, A.~J. and {Barreiro}, R.~B. and {Bartolo}, N. and {Basak}, S. and {Benabed}, K. and {Bernard}, J. -P. and {Bersanelli}, M. and {Bielewicz}, P. and {Bock}, J.~J. and {Bond}, J.~R. and {Borrill}, J. and {Bouchet}, F.~R. and {Boulanger}, F. and {Bucher}, M. and {Burigana}, C. and {Butler}, R.~C. and {Calabrese}, E. and {Cardoso}, J. -F. and {Carron}, J. and {Challinor}, A. and {Chiang}, H.~C. and {Colombo}, L.~P.~L. and {Combet}, C. and {Contreras}, D. and {Crill}, B.~P. and {Cuttaia}, F. and {de Bernardis}, P. and {de Zotti}, G. and {Delabrouille}, J. and {Delouis}, J. -M. and {Di Valentino}, E. and {Diego}, J.~M. and {Donzelli}, S. and {Dor{\'e}}, O. and {Douspis}, M. and {Ducout}, A. and {Dupac}, X. and {Dusini}, S. and {Efstathiou}, G. and {Elsner}, F. and {En{\ss}lin}, T.~A. and {Eriksen}, H.~K. and {Fantaye}, Y. and {Fergusson}, J. and {Fernandez-Cobos}, R. and {Finelli}, F. and {Forastieri}, F. and {Frailis}, M. and {Franceschi}, E. and {Frolov}, A. and {Galeotta}, S. and {Galli}, S. and {Ganga}, K. and {Gauthier}, C. and {G{\'e}nova-Santos}, R.~T. and {Gerbino}, M. and {Ghosh}, T. and {Gonz{\'a}lez-Nuevo}, J. and {G{\'o}rski}, K.~M. and {Gratton}, S. and {Gruppuso}, A. and {Gudmundsson}, J.~E. and {Hamann}, J. and {Handley}, W. and {Hansen}, F.~K. and {Herranz}, D. and {Hivon}, E. and {Hooper}, D.~C. and {Huang}, Z. and {Jaffe}, A.~H. and {Jones}, W.~C. and {Keih{\"a}nen}, E. and {Keskitalo}, R. and {Kiiveri}, K. and {Kim}, J. and {Kisner}, T.~S. and {Krachmalnicoff}, N. and {Kunz}, M. and {Kurki-Suonio}, H. and {Lagache}, G. and {Lamarre}, J. -M. and {Lasenby}, A. and {Lattanzi}, M. and {Lawrence}, C.~R. and {Le Jeune}, M. and {Lesgourgues}, J. and {Levrier}, F. and {Lewis}, A. and {Liguori}, M. and {Lilje}, P.~B. and {Lindholm}, V. and {L{\'o}pez-Caniego}, M. and {Lubin}, P.~M. and {Ma}, Y. -Z. and {Mac{\'\i}as-P{\'e}rez}, J.~F. and {Maggio}, G. and {Maino}, D. and {Mandolesi}, N. and {Mangilli}, A. and {Marcos-Caballero}, A. and {Maris}, M. and {Martin}, P.~G. and {Mart{\'\i}nez-Gonz{\'a}lez}, E. and {Matarrese}, S. and {Mauri}, N. and {McEwen}, J.~D. and {Meerburg}, P.~D. and {Meinhold}, P.~R. and {Melchiorri}, A. and {Mennella}, A. and {Migliaccio}, M. and {Mitra}, S. and {Miville-Desch{\^e}nes}, M. -A. and {Molinari}, D. and {Moneti}, A. and {Montier}, L. and {Morgante}, G. and {Moss}, A. and {M{\"u}nchmeyer}, M. and {Natoli}, P. and {N{\o}rgaard-Nielsen}, H.~U. and {Pagano}, L. and {Paoletti}, D. and {Partridge}, B. and {Patanchon}, G. and {Peiris}, H.~V. and {Perrotta}, F. and {Pettorino}, V. and {Piacentini}, F. and {Polastri}, L. and {Polenta}, G. and {Puget}, J. -L. and {Rachen}, J.~P. and {Reinecke}, M. and {Remazeilles}, M. and {Renzi}, A. and {Rocha}, G. and {Rosset}, C. and {Roudier}, G. and {Rubi{\~n}o-Mart{\'\i}n}, J.~A. and {Ruiz-Granados}, B. and {Salvati}, L. and {Sandri}, M. and {Savelainen}, M. and {Scott}, D. and {Shellard}, E.~P.~S. and {Shiraishi}, M. and {Sirignano}, C. and {Sirri}, G. and {Spencer}, L.~D. and {Sunyaev}, R. and {Suur-Uski}, A. -S. and {Tauber}, J.~A. and {Tavagnacco}, D. and {Tenti}, M. and {Toffolatti}, L. and {Tomasi}, M. and {Trombetti}, T. and {Valiviita}, J. and {Van Tent}, B. and {Vielva}, P. and {Villa}, F. and {Vittorio}, N. and {Wandelt}, B.~D. and {Wehus}, I.~K. and {White}, S.~D.~M. and {Zacchei}, A. and {Zibin}, J.~P. and {Zonca}, A.},
        title = "{Planck 2018 results. X. Constraints on inflation}",
      journal = {\aap},
         year = 2020,
        month = sep,
       volume = {641},
          eid = {A10},
        pages = {A10},
          doi = {10.1051/0004-6361/201833887},
archivePrefix = {arXiv},
       eprint = {1807.06211},
 primaryClass = {astro-ph.CO},
       adsurl = {https://ui.adsabs.harvard.edu/abs/2020A&A...641A..10P}
}

@ARTICLE{Gardner06JWST,
       author = {{Gardner}, Jonathan P. and {Mather}, John C. and {Clampin}, Mark and {Doyon}, Rene and {Greenhouse}, Matthew A. and {Hammel}, Heidi B. and {Hutchings}, John B. and {Jakobsen}, Peter and {Lilly}, Simon J. and {Long}, Knox S. and {Lunine}, Jonathan I. and {McCaughrean}, Mark J. and {Mountain}, Matt and {Nella}, John and {Rieke}, George H. and {Rieke}, Marcia J. and {Rix}, Hans-Walter and {Smith}, Eric P. and {Sonneborn}, George and {Stiavelli}, Massimo and {Stockman}, H.~S. and {Windhorst}, Rogier A. and {Wright}, Gillian S.},
        title = "{The James Webb Space Telescope}",
      journal = {\ssr},
         year = 2006,
        month = apr,
       volume = {123},
       number = {4},
        pages = {485-606},
          doi = {10.1007/s11214-006-8315-7},
archivePrefix = {arXiv},
       eprint = {astro-ph/0606175},
 primaryClass = {astro-ph},
       adsurl = {https://ui.adsabs.harvard.edu/abs/2006SSRv..123..485G}
}

@article{Vegetti09StrongLensingToStudyMassFrac,
    author = {Vegetti, Simona and Koopmans, L. V. E.},
    title = {Statistics of mass substructure from strong gravitational lensing: quantifying the mass fraction and mass function},
    journal = {Monthly Notices of the Royal Astronomical Society},
    volume = {400},
    number = {3},
    pages = {1583-1592},
    year = {2009},
    month = {12},
    issn = {0035-8711},
    doi = {10.1111/j.1365-2966.2009.15559.x},
    url = {https://doi.org/10.1111/j.1365-2966.2009.15559.x},
    eprint = {https://academic.oup.com/mnras/article-pdf/400/3/1583/12396031/mnras0400-1583.pdf},
}

@ARTICLE{Labbe23highzgalaxiesreport,
       author = {{Labb{\'e}}, Ivo and {van Dokkum}, Pieter and {Nelson}, Erica and {Bezanson}, Rachel and {Suess}, Katherine A. and {Leja}, Joel and {Brammer}, Gabriel and {Whitaker}, Katherine and {Mathews}, Elijah and {Stefanon}, Mauro and {Wang}, Bingjie},
        title = "{A population of red candidate massive galaxies  600 Myr after the Big Bang}",
      journal = {\nat},
         year = 2023,
        month = apr,
       volume = {616},
       number = {7956},
        pages = {266-269},
          doi = {10.1038/s41586-023-05786-2},
archivePrefix = {arXiv},
       eprint = {2207.12446},
 primaryClass = {astro-ph.GA},
       adsurl = {https://ui.adsabs.harvard.edu/abs/2023Natur.616..266L}
}

@article{Troxel17DesCosmicShear,
    author = "Troxel, M. A. and others",
    collaboration = "DES",
    title = "{Dark Energy Survey Year 1 results: Cosmological constraints from cosmic shear}",
    eprint = "1708.01538",
    archivePrefix = "arXiv",
    primaryClass = "astro-ph.CO",
    reportNumber = "FERMILAB-PUB-17-279-PPD",
    doi = "10.1103/PhysRevD.98.043528",
    journal = "Phys. Rev. D",
    volume = "98",
    number = "4",
    pages = "043528",
    year = "2018"
}

@article{Chabanier19LyalphaPS,
    author = "Chabanier, Sol\`ene and Millea, Marius and Palanque-Delabrouille, Nathalie",
    title = "{Matter power spectrum: from Ly$\alpha$ forest to CMB scales}",
    eprint = "1905.08103",
    archivePrefix = "arXiv",
    primaryClass = "astro-ph.CO",
    reportNumber = "Volume: 489, pages = 2247-2253",
    doi = "10.1093/mnras/stz2310",
    journal = "Mon. Not. Roy. Astron. Soc.",
    volume = "489",
    number = "2",
    pages = "2247--2253",
    year = "2019"
}

@article{Blanton17SDSSsurvey,
    author = "Blanton, Michael R. and others",
    collaboration = "eBOSS",
    title = "{Sloan Digital Sky Survey IV: Mapping the Milky Way, Nearby Galaxies and the Distant Universe}",
    eprint = "1703.00052",
    archivePrefix = "arXiv",
    primaryClass = "astro-ph.GA",
    doi = "10.3847/1538-3881/aa7567",
    journal = "Astron. J.",
    volume = "154",
    number = "1",
    pages = "28",
    year = "2017"
}

@ARTICLE{Bullock17SmallScaleCrisis,
       author = {{Bullock}, James S. and {Boylan-Kolchin}, Michael},
        title = "{Small-Scale Challenges to the {\ensuremath{\Lambda}}CDM Paradigm}",
      journal = {\araa},
         year = 2017,
        month = aug,
       volume = {55},
       number = {1},
        pages = {343-387},
          doi = {10.1146/annurev-astro-091916-055313},
archivePrefix = {arXiv},
       eprint = {1707.04256},
 primaryClass = {astro-ph.CO},
       adsurl = {https://ui.adsabs.harvard.edu/abs/2017ARA&A..55..343B}
}

@ARTICLE{Dekker24dwarfgalaxyBT,
       author = {{Dekker}, Ariane and {Kravtsov}, Andrey},
        title = "{Constraints on blue and red tilted primordial power spectra using dwarf galaxy properties}",
      journal = {arXiv e-prints},
         year = 2024,
        month = jul,
          eid = {arXiv:2407.04198},
        pages = {arXiv:2407.04198},
          doi = {10.48550/arXiv.2407.04198},
archivePrefix = {arXiv},
       eprint = {2407.04198},
 primaryClass = {astro-ph.CO},
       adsurl = {https://ui.adsabs.harvard.edu/abs/2024arXiv240704198D}
}

@INPROCEEDINGS{Kim18NoMSP,
       author = {{Kim}, Stacy and {Peter}, Annika and {Hargis}, Jonathan},
        title = "{There is No Missing Satellites Problem}",
    booktitle = {APS April Meeting Abstracts},
         year = 2018,
       series = {APS Meeting Abstracts},
       volume = {2018},
        month = jan,
          eid = {K15.005},
        pages = {K15.005},
       adsurl = {https://ui.adsabs.harvard.edu/abs/2018APS..APRK15005K}
}

@ARTICLE{Klypin99MSP,
       author = {{Klypin}, Anatoly and {Kravtsov}, Andrey V. and {Valenzuela}, Octavio and {Prada}, Francisco},
        title = "{Where Are the Missing Galactic Satellites?}",
      journal = {\apj},
         year = 1999,
        month = sep,
       volume = {522},
       number = {1},
        pages = {82-92},
          doi = {10.1086/307643},
archivePrefix = {arXiv},
       eprint = {astro-ph/9901240},
 primaryClass = {astro-ph},
       adsurl = {https://ui.adsabs.harvard.edu/abs/1999ApJ...522...82K}
}

@article{Yang24SIDM,
doi = {10.1088/1475-7516/2024/02/032},
url = {https://dx.doi.org/10.1088/1475-7516/2024/02/032},
year = {2024},
month = {feb},
publisher = {IOP Publishing},
volume = {2024},
number = {02},
pages = {032},
author = {Daneng Yang and Ethan O. Nadler and Hai-Bo Yu and Yi-Ming Zhong},
title = {A parametric model for self-interacting dark matter halos},
journal = {Journal of Cosmology and Astroparticle Physics}
}

@ARTICLE{Toll16NIHAOcore,
       author = {{Tollet}, Edouard and {Macci{\`o}}, Andrea V. and {Dutton}, Aaron A. and {Stinson}, Greg S. and {Wang}, Liang and {Penzo}, Camilla and {Gutcke}, Thales A. and {Buck}, Tobias and {Kang}, Xi and {Brook}, Chris and {Di Cintio}, Arianna and {Keller}, Ben W. and {Wadsley}, James},
        title = "{NIHAO - IV: core creation and destruction in dark matter density profiles across cosmic time}",
      journal = {\mnras},
         year = 2016,
        month = mar,
       volume = {456},
       number = {4},
        pages = {3542-3552},
          doi = {10.1093/mnras/stv2856},
archivePrefix = {arXiv},
       eprint = {1507.03590},
 primaryClass = {astro-ph.GA},
       adsurl = {https://ui.adsabs.harvard.edu/abs/2016MNRAS.456.3542T}
}

@ARTICLE{Roch13SIDM,
       author = {{Rocha}, Miguel and {Peter}, Annika H.~G. and {Bullock}, James S. and {Kaplinghat}, Manoj and {Garrison-Kimmel}, Shea and {O{\~n}orbe}, Jose and {Moustakas}, Leonidas A.},
        title = "{Cosmological simulations with self-interacting dark matter - I. Constant-density cores and substructure}",
      journal = {\mnras},
         year = 2013,
        month = mar,
       volume = {430},
       number = {1},
        pages = {81-104},
          doi = {10.1093/mnras/sts514},
archivePrefix = {arXiv},
       eprint = {1208.3025},
 primaryClass = {astro-ph.CO},
       adsurl = {https://ui.adsabs.harvard.edu/abs/2013MNRAS.430...81R}
}

@ARTICLE{Moor99MSP,
       author = {{Moore}, Ben and {Ghigna}, Sebastiano and {Governato}, Fabio and {Lake}, George and {Quinn}, Thomas and {Stadel}, Joachim and {Tozzi}, Paolo},
        title = "{Dark Matter Substructure within Galactic Halos}",
      journal = {\apjl},
         year = 1999,
        month = oct,
       volume = {524},
       number = {1},
        pages = {L19-L22},
          doi = {10.1086/312287},
archivePrefix = {arXiv},
       eprint = {astro-ph/9907411},
 primaryClass = {astro-ph},
       adsurl = {https://ui.adsabs.harvard.edu/abs/1999ApJ...524L..19M}
}

@ARTICLE{Baylan11TBTF,
       author = {{Boylan-Kolchin}, Michael and {Bullock}, James S. and {Kaplinghat}, Manoj},
        title = "{Too big to fail? The puzzling darkness of massive Milky Way subhaloes}",
      journal = {\mnras},
         year = 2011,
        month = jul,
       volume = {415},
       number = {1},
        pages = {L40-L44},
          doi = {10.1111/j.1745-3933.2011.01074.x},
archivePrefix = {arXiv},
       eprint = {1103.0007},
 primaryClass = {astro-ph.CO},
       adsurl = {https://ui.adsabs.harvard.edu/abs/2011MNRAS.415L..40B}
}

@article{Salucci2000CoreCuspyObserve,
doi = {10.1086/312747},
url = {https://dx.doi.org/10.1086/312747},
year = {2000},
month = {jun},
publisher = {},
volume = {537},
number = {1},
pages = {L9},
author = {P. Salucci and A. Burkert},
title = {Dark Matter Scaling Relations},
journal = {The Astrophysical Journal}
}

@ARTICLE{Flores94CoreCuspySimulation,
       author = {{Flores}, Ricardo A. and {Primack}, Joel R.},
        title = "{Observational and Theoretical Constraints on Singular Dark Matter Halos}",
      journal = {\apjl},
         year = 1994,
        month = may,
       volume = {427},
        pages = {L1},
          doi = {10.1086/187350},
archivePrefix = {arXiv},
       eprint = {astro-ph/9402004},
 primaryClass = {astro-ph},
       adsurl = {https://ui.adsabs.harvard.edu/abs/1994ApJ...427L...1F}
}

@ARTICLE{Sales22baryonicsolution,
       author = {{Sales}, Laura V. and {Wetzel}, Andrew and {Fattahi}, Azadeh},
        title = "{Baryonic solutions and challenges for cosmological models of dwarf galaxies}",
      journal = {Nature Astronomy},
         year = 2022,
        month = jun,
       volume = {6},
        pages = {897-910},
          doi = {10.1038/s41550-022-01689-w},
archivePrefix = {arXiv},
       eprint = {2206.05295},
 primaryClass = {astro-ph.GA},
       adsurl = {https://ui.adsabs.harvard.edu/abs/2022NatAs...6..897S}
}

@ARTICLE{Lidd1992inflation,
       author = {{Liddle}, Andrew R. and {Lyth}, David H.},
        title = "{COBE, gravitational waves, inflation and extended inflation}",
      journal = {Physics Letters B},
         year = 1992,
        month = oct,
       volume = {291},
       number = {4},
        pages = {391-398},
          doi = {10.1016/0370-2693(92)91393-N},
archivePrefix = {arXiv},
       eprint = {astro-ph/9208007},
 primaryClass = {astro-ph},
       adsurl = {https://ui.adsabs.harvard.edu/abs/1992PhLB..291..391L}
}

@ARTICLE{Salopek1990inflation,
       author = {{Salopek}, D.~S. and {Bond}, J.~R.},
        title = "{Nonlinear evolution of long-wavelength metric fluctuations in inflationary models}",
      journal = {\prd},
         year = 1990,
        month = dec,
       volume = {42},
       number = {12},
        pages = {3936-3962},
          doi = {10.1103/PhysRevD.42.3936},
       adsurl = {https://ui.adsabs.harvard.edu/abs/1990PhRvD..42.3936S}
}

@ARTICLE{Stei84inflation,
       author = {{Steinhardt}, Paul J. and {Turner}, Michael S.},
        title = "{Prescription for successful new inflation}",
      journal = {\prd},
         year = 1984,
        month = may,
       volume = {29},
       number = {10},
        pages = {2162-2171},
          doi = {10.1103/PhysRevD.29.2162},
       adsurl = {https://ui.adsabs.harvard.edu/abs/1984PhRvD..29.2162S}
}

@ARTICLE{Lidd93inflation,
       author = {{Liddle}, Andrew R. and {Lyth}, David H.},
        title = "{The cold dark matter density perturbation}",
      journal = {\physrep},
         year = 1993,
        month = aug,
       volume = {231},
       number = {1-2},
        pages = {1-105},
          doi = {10.1016/0370-1573(93)90114-S},
archivePrefix = {arXiv},
       eprint = {astro-ph/9303019},
 primaryClass = {astro-ph},
       adsurl = {https://ui.adsabs.harvard.edu/abs/1993PhR...231....1L}
}

@article{Kopo15UFS,
doi = {10.1088/0004-637X/805/2/130},
url = {https://dx.doi.org/10.1088/0004-637X/805/2/130},
year = {2015},
month = {may},
publisher = {The American Astronomical Society},
volume = {805},
number = {2},
pages = {130},
author = {Sergey E. Koposov and Vasily Belokurov and Gabriel Torrealba and N. Wyn Evans},
title = {BEASTS OF THE SOUTHERN WILD: DISCOVERY OF NINE ULTRA FAINT SATELLITES IN THE VICINITY OF THE MAGELLANIC CLOUDS},
journal = {The Astrophysical Journal}
}

@article{Bech15DES,
doi = {10.1088/0004-637X/807/1/50},
url = {https://dx.doi.org/10.1088/0004-637X/807/1/50},
year = {2015},
month = {jun},
publisher = {The American Astronomical Society},
volume = {807},
number = {1},
pages = {50},
author = {K. Bechtol and A. Drlica-Wagner and E. Balbinot and A. Pieres and J. D. Simon and B. Yanny and B. Santiago and R. H. Wechsler and J. Frieman and A. R. Walker and P. Williams and E. Rozo and E. S. Rykoff and A. Queiroz and E. Luque and A. Benoit-Lévy and D. Tucker and I. Sevilla and R. A. Gruendl and L. N. da Costa and A. Fausti Neto and M. A. G. Maia and T. Abbott and S. Allam and R. Armstrong and A. H. Bauer and G. M. Bernstein and R. A. Bernstein and E. Bertin and D. Brooks and E. Buckley-Geer and D. L. Burke and A. Carnero Rosell and F. J. Castander and R. Covarrubias and C. B. D’Andrea and D. L. DePoy and S. Desai and H. T. Diehl and T. F. Eifler and J. Estrada and A. E. Evrard and E. Fernandez and D. A. Finley and B. Flaugher and E. Gaztanaga and D. Gerdes and L. Girardi and M. Gladders and D. Gruen and G. Gutierrez and J. Hao and K. Honscheid and B. Jain and D. James and S. Kent and R. Kron and K. Kuehn and N. Kuropatkin and O. Lahav and T. S. Li and H. Lin and M. Makler and M. March and J. Marshall and P. Martini and K. W. Merritt and C. Miller and R. Miquel and J. Mohr and E. Neilsen and R. Nichol and B. Nord and R. Ogando and J. Peoples and D. Petravick and A. A. Plazas and A. K. Romer and A. Roodman and M. Sako and E. Sanchez and V. Scarpine and M. Schubnell and R. C. Smith and M. Soares-Santos and F. Sobreira and E. Suchyta and M. E. C. Swanson and G. Tarle and J. Thaler and D. Thomas and W. Wester and J. Zuntz and (The DES Collaboration)},
title = {EIGHT NEW MILKY WAY COMPANIONS DISCOVERED IN FIRST-YEAR DARK ENERGY SURVEY DATA},
journal = {The Astrophysical Journal}
}

@article{Drli15DES,
doi = {10.1088/0004-637X/813/2/109},
url = {https://dx.doi.org/10.1088/0004-637X/813/2/109},
year = {2015},
month = {nov},
publisher = {The American Astronomical Society},
volume = {813},
number = {2},
pages = {109},
author = {A. Drlica-Wagner and K. Bechtol and E. S. Rykoff and E. Luque and A. Queiroz and Y.-Y. Mao and R. H. Wechsler and J. D. Simon and B. Santiago and B. Yanny and E. Balbinot and S. Dodelson and A. Fausti Neto and D. J. James and T. S. Li and M. A. G. Maia and J. L. Marshall and A. Pieres and K. Stringer and A. R. Walker and T. M. C. Abbott and F. B. Abdalla and S. Allam and A. Benoit-Lévy and G. M. Bernstein and E. Bertin and D. Brooks and E. Buckley-Geer and D. L. Burke and A. Carnero Rosell and M. Carrasco Kind and J. Carretero and M. Crocce and L. N. da Costa and S. Desai and H. T. Diehl and J. P. Dietrich and P. Doel and T. F. Eifler and A. E. Evrard and D. A. Finley and B. Flaugher and P. Fosalba and J. Frieman and E. Gaztanaga and D. W. Gerdes and D. Gruen and R. A. Gruendl and G. Gutierrez and K. Honscheid and K. Kuehn and N. Kuropatkin and O. Lahav and P. Martini and R. Miquel and B. Nord and R. Ogando and A. A. Plazas and K. Reil and A. Roodman and M. Sako and E. Sanchez and V. Scarpine and M. Schubnell and I. Sevilla-Noarbe and R. C. Smith and M. Soares-Santos and F. Sobreira and E. Suchyta and M. E. C. Swanson and G. Tarle and D. Tucker and V. Vikram and W. Wester and Y. Zhang and J. Zuntz and (The DES Collaboration)},
title = {EIGHT ULTRA-FAINT GALAXY CANDIDATES DISCOVERED IN YEAR TWO OF THE DARK ENERGY SURVEY},
journal = {The Astrophysical Journal}
}

@BOOK{Hock88PM,
       author = {{Hockney}, R.~W. and {Eastwood}, J.~W.},
        title = "{Computer simulation using particles}",
         year = 1988,
       adsurl = {https://ui.adsabs.harvard.edu/abs/1988csup.book.....H},
      publisher = {CRC Press}
}

@article{Dehn00FMM,
doi = {10.1086/312724},
url = {https://dx.doi.org/10.1086/312724},
year = {2000},
month = {jun},
publisher = {},
volume = {536},
number = {1},
pages = {L39},
author = {Walter Dehnen},
title = {A Very Fast and Momentum-conserving Tree
Code},
journal = {The Astrophysical Journal}
}

@book{Mo10galaxybook,
place={Cambridge}, 
title={Galaxy Formation and Evolution}, 
publisher={Cambridge University Press}, 
author={Mo, Houjun and van den Bosch, Frank and White, Simon}, 
year={2010}}

@BOOK{Peeb93cosmology,
       author = {{Peebles}, P.~J.~E.},
        title = "{Principles of Physical Cosmology}",
         year = 1993,
          doi = {10.1515/9780691206721},
       adsurl = {https://ui.adsabs.harvard.edu/abs/1993ppc..book.....P},
      publisher = {Princeton University Press}
}

@ARTICLE{Fren1985LCDMsim,
       author = {{Frenk}, C.~S. and {White}, S.~D.~M. and {Efstathiou}, G. and {Davis}, M.},
        title = "{Cold dark matter, the structure of galactic haloes and the origin of the Hubble sequence}",
      journal = {\nat},
         year = 1985,
        month = oct,
       volume = {317},
       number = {6038},
        pages = {595-597},
          doi = {10.1038/317595a0},
       adsurl = {https://ui.adsabs.harvard.edu/abs/1985Natur.317..595F}
}

@article{Guth81inflation,
  title   = {Inflationary universe: A possible solution to the horizon and flatness problems},
  author  = {Guth, Alan H.},
  journal = {Phys. Rev. D},
  volume  = {23},
  number  = {2},
  pages   = {347--356},
  year    = {1981},
  month   = jan,
  doi     = {10.1103/PhysRevD.23.347}
}

@ARTICLE{Scha23SWIFT,
       author = {{Schaller}, Matthieu and {Borrow}, Josh and {Draper}, Peter W. and {Ivkovic}, Mladen and {McAlpine}, Stuart and {Vandenbroucke}, Bert and {Bah{\'e}}, Yannick and {Chaikin}, Evgenii and {Chalk}, Aidan B.~G. and {Chan}, Tsang Keung and {Correa}, Camila and {van Daalen}, Marcel and {Elbers}, Willem and {Gonnet}, Pedro and {Hausammann}, Lo{\"\i}c and {Helly}, John and {Hu{\v{s}}ko}, Filip and {Kegerreis}, Jacob A. and {Nobels}, Folkert S.~J. and {Ploeckinger}, Sylvia and {Revaz}, Yves and {Roper}, William J. and {Ruiz-Bonilla}, Sergio and {Sandnes}, Thomas D. and {Uyttenhove}, Yolan and {Willis}, James S. and {Xiang}, Zhen},
        title = "{SWIFT: A modern highly-parallel gravity and smoothed particle hydrodynamics solver for astrophysical and cosmological applications}",
      journal = {\mnras},
         year = 2024,
        month = may,
       volume = {530},
       number = {2},
        pages = {2378-2419},
          doi = {10.1093/mnras/stae922},
archivePrefix = {arXiv},
       eprint = {2305.13380},
 primaryClass = {astro-ph.IM},
       adsurl = {https://ui.adsabs.harvard.edu/abs/2024MNRAS.530.2378S}
}

@ARTICLE{Garr17m12f,
       author = {{Garrison-Kimmel}, Shea and {Wetzel}, Andrew and {Bullock}, James S. and {Hopkins}, Philip F. and {Boylan-Kolchin}, Michael and {Faucher-Gigu{\`e}re}, Claude-Andr{\'e} and {Kere{\v{s}}}, Du{\v{s}}an and {Quataert}, Eliot and {Sanderson}, Robyn E. and {Graus}, Andrew S. and {Kelley}, Tyler},
        title = "{Not so lumpy after all: modelling the depletion of dark matter subhaloes by Milky Way-like galaxies}",
      journal = {\mnras},
         year = 2017,
        month = oct,
       volume = {471},
       number = {2},
        pages = {1709-1727},
          doi = {10.1093/mnras/stx1710},
archivePrefix = {arXiv},
       eprint = {1701.03792},
 primaryClass = {astro-ph.GA},
       adsurl = {https://ui.adsabs.harvard.edu/abs/2017MNRAS.471.1709G}
}

@ARTICLE{Wetz16m12i,
       author = {{Wetzel}, Andrew R. and {Hopkins}, Philip F. and {Kim}, Ji-hoon and {Faucher-Gigu{\`e}re}, Claude-Andr{\'e} and {Kere{\v{s}}}, Du{\v{s}}an and {Quataert}, Eliot},
        title = "{Reconciling Dwarf Galaxies with {\ensuremath{\Lambda}}CDM Cosmology: Simulating a Realistic Population of Satellites around a Milky Way-mass Galaxy}",
      journal = {\apjl},
         year = 2016,
        month = aug,
       volume = {827},
       number = {2},
          eid = {L23},
        pages = {L23},
          doi = {10.3847/2041-8205/827/2/L23},
archivePrefix = {arXiv},
       eprint = {1602.05957},
 primaryClass = {astro-ph.GA},
       adsurl = {https://ui.adsabs.harvard.edu/abs/2016ApJ...827L..23W}
}

@ARTICLE{Wetz23FIREpublic,
       author = {{Wetzel}, Andrew and {Hayward}, Christopher C. and {Sanderson}, Robyn E. and {Ma}, Xiangcheng and {Angl{\'e}s-Alc{\'a}zar}, Daniel and {Feldmann}, Robert and {Chan}, T.~K. and {El-Badry}, Kareem and {Wheeler}, Coral and {Garrison-Kimmel}, Shea and {Nikakhtar}, Farnik and {Panithanpaisal}, Nondh and {Arora}, Arpit and {Gurvich}, Alexander B. and {Samuel}, Jenna and {Sameie}, Omid and {Pandya}, Viraj and {Hafen}, Zachary and {Hummels}, Cameron and {Loebman}, Sarah and {Boylan-Kolchin}, Michael and {Bullock}, James S. and {Faucher-Gigu{\`e}re}, Claude-Andr{\'e} and {Kere{\v{s}}}, Du{\v{s}}an and {Quataert}, Eliot and {Hopkins}, Philip F.},
        title = "{Public Data Release of the FIRE-2 Cosmological Zoom-in Simulations of Galaxy Formation}",
      journal = {\apjs},
         year = 2023,
        month = apr,
       volume = {265},
       number = {2},
          eid = {44},
        pages = {44},
          doi = {10.3847/1538-4365/acb99a},
archivePrefix = {arXiv},
       eprint = {2202.06969},
 primaryClass = {astro-ph.GA},
       adsurl = {https://ui.adsabs.harvard.edu/abs/2023ApJS..265...44W}
}

@article{Schaye23Flamingo,
    author = {Schaye, Joop and Kugel, Roi and Schaller, Matthieu and Helly, John C and Braspenning, Joey and Elbers, Willem and McCarthy, Ian G and van Daalen, Marcel P and Vandenbroucke, Bert and Frenk, Carlos S and Kwan, Juliana and Salcido, Jaime and Bahé, Yannick M and Borrow, Josh and Chaikin, Evgenii and Hahn, Oliver and Huško, Filip and Jenkins, Adrian and Lacey, Cedric G and Nobels, Folkert S J},
    title = "{The FLAMINGO project: cosmological hydrodynamical simulations for large-scale structure and galaxy cluster surveys}",
    journal = {Monthly Notices of the Royal Astronomical Society},
    volume = {526},
    number = {4},
    pages = {4978-5020},
    year = {2023},
    month = {08},
    issn = {0035-8711},
    doi = {10.1093/mnras/stad2419},
    url = {https://doi.org/10.1093/mnras/stad2419},
    eprint = {https://academic.oup.com/mnras/article-pdf/526/4/4978/52313196/stad2419.pdf},
}

@article{HanJiaxin17HBTplus,
    author = {Han, Jiaxin and Cole, Shaun and Frenk, Carlos S. and Benitez-Llambay, Alejandro and Helly, John},
    title = "{hbt+: an improved code for finding subhaloes and building merger trees in cosmological simulations}",
    journal = {Monthly Notices of the Royal Astronomical Society},
    volume = {474},
    number = {1},
    pages = {604-617},
    year = {2017},
    month = {10},
    issn = {0035-8711},
    doi = {10.1093/mnras/stx2792},
    url = {https://doi.org/10.1093/mnras/stx2792},
    eprint = {https://academic.oup.com/mnras/article-pdf/474/1/604/22141268/stx2792.pdf},
}

@ARTICLE{Robert21RmaxVmax,
       author = {{Grand}, Robert J.~J. and {White}, Simon D.~M.},
        title = "{Baryonic effects on the detectability of annihilation radiation from dark matter subhaloes around the Milky Way}",
      journal = {\mnras},
         year = 2021,
        month = mar,
       volume = {501},
       number = {3},
        pages = {3558-3567},
          doi = {10.1093/mnras/staa3993},
archivePrefix = {arXiv},
       eprint = {2012.07846},
 primaryClass = {astro-ph.GA},
       adsurl = {https://ui.adsabs.harvard.edu/abs/2021MNRAS.501.3558G}
}

@ARTICLE{Puebla16BolshoiPMDPL,
       author = {{Rodr{\'\i}guez-Puebla}, Aldo and {Behroozi}, Peter and {Primack}, Joel and {Klypin}, Anatoly and {Lee}, Christoph and {Hellinger}, Doug},
        title = "{Halo and subhalo demographics with Planck cosmological parameters: Bolshoi-Planck and MultiDark-Planck simulations}",
      journal = {\mnras},
         year = 2016,
        month = oct,
       volume = {462},
       number = {1},
        pages = {893-916},
          doi = {10.1093/mnras/stw1705},
archivePrefix = {arXiv},
       eprint = {1602.04813},
 primaryClass = {astro-ph.CO},
       adsurl = {https://ui.adsabs.harvard.edu/abs/2016MNRAS.462..893R}
}

@ARTICLE{Cautun14r100fit,
       author = {{Cautun}, Marius and {Hellwing}, Wojciech A. and {van de Weygaert}, Rien and {Frenk}, Carlos S. and {Jones}, Bernard J.~T. and {Sawala}, Till},
        title = "{Subhalo statistics of galactic haloes: beyond the resolution limit}",
      journal = {\mnras},
         year = 2014,
        month = dec,
       volume = {445},
       number = {2},
        pages = {1820-1835},
          doi = {10.1093/mnras/stu1829},
archivePrefix = {arXiv},
       eprint = {1405.7700},
 primaryClass = {astro-ph.CO},
       adsurl = {https://ui.adsabs.harvard.edu/abs/2014MNRAS.445.1820C}
}

@ARTICLE{Springel08Aquarius,
       author = {{Springel}, V. and {Wang}, J. and {Vogelsberger}, M. and {Ludlow}, A. and {Jenkins}, A. and {Helmi}, A. and {Navarro}, J.~F. and {Frenk}, C.~S. and {White}, S.~D.~M.},
        title = "{The Aquarius Project: the subhaloes of galactic haloes}",
      journal = {\mnras},
         year = 2008,
        month = dec,
       volume = {391},
       number = {4},
        pages = {1685-1711},
          doi = {10.1111/j.1365-2966.2008.14066.x},
archivePrefix = {arXiv},
       eprint = {0809.0898},
 primaryClass = {astro-ph},
       adsurl = {https://ui.adsabs.harvard.edu/abs/2008MNRAS.391.1685S}
}

@ARTICLE{Komatsu12wmap7,
       author = {{Komatsu}, E. and {Smith}, K.~M. and {Dunkley}, J. and {Bennett}, C.~L. and {Gold}, B. and {Hinshaw}, G. and {Jarosik}, N. and {Larson}, D. and {Nolta}, M.~R. and {Page}, L. and {Spergel}, D.~N. and {Halpern}, M. and {Hill}, R.~S. and {Kogut}, A. and {Limon}, M. and {Meyer}, S.~S. and {Odegard}, N. and {Tucker}, G.~S. and {Weiland}, J.~L. and {Wollack}, E. and {Wright}, E.~L.},
        title = "{Seven-year Wilkinson Microwave Anisotropy Probe (WMAP) Observations: Cosmological Interpretation}",
      journal = {\apjs},
         year = 2011,
        month = feb,
       volume = {192},
       number = {2},
          eid = {18},
        pages = {18},
          doi = {10.1088/0067-0049/192/2/18},
archivePrefix = {arXiv},
       eprint = {1001.4538},
 primaryClass = {astro-ph.CO},
       adsurl = {https://ui.adsabs.harvard.edu/abs/2011ApJS..192...18K}
}

@article{Hellwing16cocoproject,
    author = {Hellwing, Wojciech A. and Frenk, Carlos S. and Cautun, Marius and Bose, Sownak and Helly, John and Jenkins, Adrian and Sawala, Till and Cytowski, Maciej},
    title = "{The Copernicus Complexio: a high-resolution view of the small-scale Universe}",
    journal = {Monthly Notices of the Royal Astronomical Society},
    volume = {457},
    number = {4},
    pages = {3492-3509},
    year = {2016},
    month = {02},
    issn = {0035-8711},
    doi = {10.1093/mnras/stw214},
    url = {https://doi.org/10.1093/mnras/stw214},
    eprint = {https://academic.oup.com/mnras/article-pdf/457/4/3492/18512891/stw214.pdf},
}

@article{Garrison-Kimmel17notsolumpy,
    author = {Garrison-Kimmel, Shea and Wetzel, Andrew and Bullock, James S. and Hopkins, Philip F. and Boylan-Kolchin, Michael and Faucher-Giguère, Claude-André and Kereš, Dušan and Quataert, Eliot and Sanderson, Robyn E. and Graus, Andrew S. and Kelley, Tyler},
    title = "{Not so lumpy after all: modelling the depletion of dark matter subhaloes by Milky Way-like galaxies}",
    journal = {Monthly Notices of the Royal Astronomical Society},
    volume = {471},
    number = {2},
    pages = {1709-1727},
    year = {2017},
    month = {07},
    issn = {0035-8711},
    doi = {10.1093/mnras/stx1710},
    url = {https://doi.org/10.1093/mnras/stx1710},
    eprint = {https://academic.oup.com/mnras/article-pdf/471/2/1709/19407370/stx1710.pdf},
}

@article{Elahi19VELOCIraptor,
    author = "Elahi, Pascal J. and Ca\~nas, Rodrigo and Poulton, Rhys J. J. and Tobar, Rodrigo J. and Willis, James S. and Lagos, Claudia del P. and Power, Chris and Robotham, Aaron S. G.",
    title = "{Hunting for galaxies and halos in simulations with VELOCIraptor}",
    eprint = "1902.01010",
    archivePrefix = "arXiv",
    primaryClass = "astro-ph.CO",
    doi = "10.1017/pasa.2019.12",
    journal = "Publ. Astron. Soc. Austral.",
    volume = "36",
    pages = "e021",
    year = "2019"
}

@MISC{ascl18SWIFT,
  author = {{Schaller}, M. and others},
  title = "{SWIFT: SPH With Inter-dependent Fine-grained Tasking}",
  howpublished = {Astrophysics Source Code Library},
  year = 2018,
  month = may,
  eid = {ascl:1805.020},
  pages = {ascl:1805.020},
  archivePrefix = {ascl},
  eprint = {1805.020},
  adsurl = {https://ui.adsabs.harvard.edu/abs/2018ascl.soft05020S}
}

@ARTICLE{Covi99BTPS,
       author = {{Covi}, Laura and {Lyth}, David H.},
        title = "{Running-mass models of inflation and their observational constraints}",
      journal = {\prd},
         year = 1999,
        month = mar,
       volume = {59},
       number = {6},
          eid = {063515},
        pages = {063515},
          doi = {10.1103/PhysRevD.59.063515},
archivePrefix = {arXiv},
       eprint = {hep-ph/9809562},
 primaryClass = {hep-ph},
       adsurl = {https://ui.adsabs.harvard.edu/abs/1999PhRvD..59f3515C}
}

@ARTICLE{Martin01BTPS,
       author = {{Martin}, J{\'e}r{\^o}me and {Brandenberger}, Robert H.},
        title = "{Trans-Planckian problem of inflationary cosmology}",
      journal = {\prd},
         year = 2001,
        month = jun,
       volume = {63},
       number = {12},
          eid = {123501},
        pages = {123501},
          doi = {10.1103/PhysRevD.63.123501},
archivePrefix = {arXiv},
       eprint = {hep-th/0005209},
 primaryClass = {hep-th},
       adsurl = {https://ui.adsabs.harvard.edu/abs/2001PhRvD..63l3501M}
}

@ARTICLE{Gong11BTPS,
       author = {{Gong}, Jinn-Ouk and {Sasaki}, Misao},
        title = "{Waterfall field in hybrid inflation and curvature perturbation}",
      journal = {\jcap},
         year = 2011,
        month = mar,
       volume = {2011},
       number = {3},
          eid = {028},
        pages = {028},
          doi = {10.1088/1475-7516/2011/03/028},
archivePrefix = {arXiv},
       eprint = {1010.3405},
 primaryClass = {astro-ph.CO},
       adsurl = {https://ui.adsabs.harvard.edu/abs/2011JCAP...03..028G}
}

@ARTICLE{Reed07hmf,
       author = {{Reed}, Darren S. and {Bower}, Richard and {Frenk}, Carlos S. and {Jenkins}, Adrian and {Theuns}, Tom},
        title = "{The halo mass function from the dark ages through the present day}",
      journal = {\mnras},
         year = 2007,
        month = jan,
       volume = {374},
       number = {1},
        pages = {2-15},
          doi = {10.1111/j.1365-2966.2006.11204.x},
archivePrefix = {arXiv},
       eprint = {astro-ph/0607150},
 primaryClass = {astro-ph},
       adsurl = {https://ui.adsabs.harvard.edu/abs/2007MNRAS.374....2R}
}

@article{Zhang19OptimalSoftLength,
    author = {Zhang, Tianchi and Liao, Shihong and Li, Ming and Gao, Liang},
    title = "{The optimal gravitational softening length for cosmological N-body simulations}",
    journal = {Monthly Notices of the Royal Astronomical Society},
    volume = {487},
    number = {1},
    pages = {1227-1232},
    year = {2019},
    month = {05},
    issn = {0035-8711},
    doi = {10.1093/mnras/stz1370},
    url = {https://doi.org/10.1093/mnras/stz1370},
    eprint = {https://academic.oup.com/mnras/article-pdf/487/1/1227/28753069/stz1370.pdf},
}

@ARTICLE{Tkac24bumpyps,
       author = {{Tkachev}, M.~V. and {Pilipenko}, S.~V. and {Mikheeva}, E.~V. and {Lukash}, V.~N.},
        title = "{Excess of high-z galaxies as a test for bumpy power spectrum of density perturbations}",
      journal = {\mnras},
         year = 2024,
        month = jan,
       volume = {527},
       number = {1},
        pages = {1381-1388},
          doi = {10.1093/mnras/stad3279},
archivePrefix = {arXiv},
       eprint = {2307.13774},
 primaryClass = {astro-ph.CO},
       adsurl = {https://ui.adsabs.harvard.edu/abs/2024MNRAS.527.1381T}
}

@ARTICLE{Homm23MWmanydwarf,
       author = {{Homma}, Daisuke and {Chiba}, Masashi and {Komiyama}, Yutaka and {Tanaka}, Masayuki and {Okamoto}, Sakurako and {Tanaka}, Mikito and {Ishigaki}, Miho N. and {Hayashi}, Kohei and {Arimoto}, Nobuo and {Lupton}, Robert H. and {Strauss}, Michael A. and {Miyazaki}, Satoshi and {Wang}, Shiang-Yu and {Murayama}, Hitoshi},
        title = "{Final results of the search for new Milky Way satellites in the Hyper Suprime-Cam Subaru Strategic Program survey: Discovery of two more candidates}",
      journal = {\pasj},
         year = 2024,
        month = aug,
       volume = {76},
       number = {4},
        pages = {733-752},
          doi = {10.1093/pasj/psae044},
archivePrefix = {arXiv},
       eprint = {2311.05439},
 primaryClass = {astro-ph.GA},
       adsurl = {https://ui.adsabs.harvard.edu/abs/2024PASJ...76..733H}
}

@ARTICLE{Hira24bluetilt,
       author = {{Hirano}, Shingo and {Yoshida}, Naoki},
        title = "{Early Structure Formation from Primordial Density Fluctuations with a Blue, Tilted Power Spectrum: High-redshift Galaxies}",
      journal = {\apj},
         year = 2024,
        month = mar,
       volume = {963},
       number = {1},
          eid = {2},
        pages = {2},
          doi = {10.3847/1538-4357/ad22e0},
archivePrefix = {arXiv},
       eprint = {2306.11993},
 primaryClass = {astro-ph.GA},
       adsurl = {https://ui.adsabs.harvard.edu/abs/2024ApJ...963....2H}
}

@ARTICLE{Love14WDM,
       author = {{Lovell}, Mark R. and {Frenk}, Carlos S. and {Eke}, Vincent R. and {Jenkins}, Adrian and {Gao}, Liang and {Theuns}, Tom},
        title = "{The properties of warm dark matter haloes}",
      journal = {\mnras},
         year = 2014,
        month = mar,
       volume = {439},
       number = {1},
        pages = {300-317},
          doi = {10.1093/mnras/stt2431},
archivePrefix = {arXiv},
       eprint = {1308.1399},
 primaryClass = {astro-ph.CO},
       adsurl = {https://ui.adsabs.harvard.edu/abs/2014MNRAS.439..300L}
}

@ARTICLE{Hira15bluetilt,
       author = {{Hirano}, Shingo and {Zhu}, Nick and {Yoshida}, Naoki and {Spergel}, David and {Yorke}, Harold W.},
        title = "{Early Structure Formation from Primordial Density Fluctuations with a Blue, Tilted Power Spectrum}",
      journal = {\apj},
         year = 2015,
        month = nov,
       volume = {814},
       number = {1},
          eid = {18},
        pages = {18},
          doi = {10.1088/0004-637X/814/1/18},
archivePrefix = {arXiv},
       eprint = {1504.05186},
 primaryClass = {astro-ph.CO},
       adsurl = {https://ui.adsabs.harvard.edu/abs/2015ApJ...814...18H}
}

@ARTICLE{DiCi14cuspcore,
       author = {{Di Cintio}, Arianna and {Brook}, Chris B. and {Macci{\`o}}, Andrea V. and {Stinson}, Greg S. and {Knebe}, Alexander and {Dutton}, Aaron A. and {Wadsley}, James},
        title = "{The dependence of dark matter profiles on the stellar-to-halo mass ratio: a prediction for cusps versus cores}",
      journal = {\mnras},
         year = 2014,
        month = jan,
       volume = {437},
       number = {1},
        pages = {415-423},
          doi = {10.1093/mnras/stt1891},
archivePrefix = {arXiv},
       eprint = {1306.0898},
 primaryClass = {astro-ph.CO},
       adsurl = {https://ui.adsabs.harvard.edu/abs/2014MNRAS.437..415D}
}

@ARTICLE{Finl08regulator,
       author = {{Finlator}, Kristian and {Dav{\'e}}, Romeel},
        title = "{The origin of the galaxy mass-metallicity relation and implications for galactic outflows}",
      journal = {\mnras},
         year = 2008,
        month = apr,
       volume = {385},
       number = {4},
        pages = {2181-2204},
          doi = {10.1111/j.1365-2966.2008.12991.x},
archivePrefix = {arXiv},
       eprint = {0704.3100},
 primaryClass = {astro-ph},
       adsurl = {https://ui.adsabs.harvard.edu/abs/2008MNRAS.385.2181F}
}

@ARTICLE{Bens12GALACTICUS,
       author = {{Benson}, Andrew J.},
        title = "{GALACTICUS: A semi-analytic model of galaxy formation}",
      journal = {\na},
         year = 2012,
        month = feb,
       volume = {17},
       number = {2},
        pages = {175-197},
          doi = {10.1016/j.newast.2011.07.004},
archivePrefix = {arXiv},
       eprint = {1008.1786},
 primaryClass = {astro-ph.CO},
       adsurl = {https://ui.adsabs.harvard.edu/abs/2012NewA...17..175B}
}

@ARTICLE{Cole00GALFORM,
       author = {{Cole}, Shaun and {Lacey}, Cedric G. and {Baugh}, Carlton M. and {Frenk}, Carlos S.},
        title = "{Hierarchical galaxy formation}",
      journal = {\mnras},
         year = 2000,
        month = nov,
       volume = {319},
       number = {1},
        pages = {168-204},
          doi = {10.1046/j.1365-8711.2000.03879.x},
archivePrefix = {arXiv},
       eprint = {astro-ph/0007281},
 primaryClass = {astro-ph},
       adsurl = {https://ui.adsabs.harvard.edu/abs/2000MNRAS.319..168C}
}

@ARTICLE{Krav22GRUMPY,
       author = {{Kravtsov}, Andrey and {Manwadkar}, Viraj},
        title = "{GRUMPY: a simple framework for realistic forward modelling of dwarf galaxies}",
      journal = {\mnras},
         year = 2022,
        month = aug,
       volume = {514},
       number = {2},
        pages = {2667-2691},
          doi = {10.1093/mnras/stac1439},
archivePrefix = {arXiv},
       eprint = {2106.09724},
 primaryClass = {astro-ph.GA},
       adsurl = {https://ui.adsabs.harvard.edu/abs/2022MNRAS.514.2667K}
}

@ARTICLE{Este23lumpy,
  title = {Milky Way satellite velocities reveal the dark matter power spectrum at small scales},
  author = {Esteban, Ivan and Peter, Annika H. G. and Kim, Stacy Y.},
  journal = {Phys. Rev. D},
  volume = {110},
  issue = {12},
  pages = {123013},
  numpages = {24},
  year = {2024},
  month = {Dec},
  publisher = {American Physical Society},
  doi = {10.1103/PhysRevD.110.123013},
  url = {https://link.aps.org/doi/10.1103/PhysRevD.110.123013}
}

@ARTICLE{Mull24M83,
       author = {{M{\"u}ller}, Oliver and {Pawlowski}, Marcel S. and {Revaz}, Yves and {Venhola}, Aku and {Rejkuba}, Marina and {Hilker}, Michael and {Lutz}, Katharina},
        title = "{A too-many-dwarf-galaxy-satellites problem in the M 83 group}",
      journal = {\aap},
         year = 2024,
        month = apr,
       volume = {684},
          eid = {L6},
        pages = {L6},
          doi = {10.1051/0004-6361/202348969},
archivePrefix = {arXiv},
       eprint = {2403.08717},
 primaryClass = {astro-ph.GA},
       adsurl = {https://ui.adsabs.harvard.edu/abs/2024A&A...684L...6M}
}

@ARTICLE{FIRE2,
      author = {{Hopkins}, Philip F. and {Wetzel}, Andrew and {Kere{\v{s}}}, Du{\v{s}}an and {Faucher-Gigu{\`e}re}, Claude-Andr{\'e} and {Quataert}, Eliot and {Boylan-Kolchin}, Michael and {Murray}, Norman and {Hayward}, Christopher C. and {Garrison-Kimmel}, Shea and {Hummels}, Cameron and {Feldmann}, Robert and {Torrey}, Paul and {Ma}, Xiangcheng and {Angl{\'e}s-Alc{\'a}zar}, Daniel and {Su}, Kung-Yi and {Orr}, Matthew and {Schmitz}, Denise and {Escala}, Ivanna and {Sanderson}, Robyn and {Grudi{\'c}}, Michael Y. and {Hafen}, Zachary and {Kim}, Ji-Hoon and {Fitts}, Alex and {Bullock}, James S. and {Wheeler}, Coral and {Chan}, T.~K. and {Elbert}, Oliver D. and {Narayanan}, Desika},
        title = "{FIRE-2 simulations: physics versus numerics in galaxy formation}",
      journal = {\mnras},
         year = 2018,
        month = oct,
       volume = {480},
       number = {1},
        pages = {800-863},
          doi = {10.1093/mnras/sty1690},
archivePrefix = {arXiv},
       eprint = {1702.06148},
 primaryClass = {astro-ph.GA},
       adsurl = {https://ui.adsabs.harvard.edu/abs/2018MNRAS.480..800H}
}

@ARTICLE{Gilm22stronglensingPPS,
       author = {{Gilman}, Daniel and {Benson}, Andrew and {Bovy}, Jo and {Birrer}, Simon and {Treu}, Tommaso and {Nierenberg}, Anna},
        title = "{The primordial matter power spectrum on sub-galactic scales}",
      journal = {\mnras},
         year = 2022,
        month = may,
       volume = {512},
       number = {3},
        pages = {3163-3188},
          doi = {10.1093/mnras/stac670},
archivePrefix = {arXiv},
       eprint = {2112.03293},
 primaryClass = {astro-ph.CO},
       adsurl = {https://ui.adsabs.harvard.edu/abs/2022MNRAS.512.3163G}
}

@ARTICLE{Chan15,
   author = {{Chan}, T.~K. and {Kere{\v s}}, D. and {O{\~n}orbe}, J. and 
	{Hopkins}, P.~F. and {Muratov}, A.~L. and {Faucher-Gigu{\`e}re}, C.-A. and 
	{Quataert}, E.},
    title = "{The impact of baryonic physics on the structure of dark matter haloes: the view from the FIRE cosmological simulations}",
  journal = {\mnras},
archivePrefix = "arXiv",
   eprint = {1507.02282},
     year = 2015,
    month = dec,
   volume = 454,
    pages = {2981-3001},
      doi = {10.1093/mnras/stv2165},
   adsurl = {http://adsabs.harvard.edu/abs/2015MNRAS.454.2981C}
}

@article{Powe03,
   author = {{Power}, C. and {Navarro}, J.~F. and {Jenkins}, A. and {Frenk}, C.~S. and 
	{White}, S.~D.~M. and {Springel}, V. and {Stadel}, J. and {Quinn}, T.
	},
    title = "{The inner structure of {$\Lambda$}CDM haloes - I. A numerical convergence study}",
  journal = {\mnras},
   eprint = {astro-ph/0201544},
     year = 2003,
    month = jan,
   volume = 338,
    pages = {14-34},
      doi = {10.1046/j.1365-8711.2003.05925.x},
   adsurl = {http://adsabs.harvard.edu/abs/2003MNRAS.338...14P}
}

@article{NFW,
   author = {{Navarro}, J.~F. and {Frenk}, C.~S. and {White}, S.~D.~M.},
    title = "{A Universal Density Profile from Hierarchical Clustering}",
  journal = {\apj},
   eprint = {astro-ph/9611107},
     year = 1997,
    month = dec,
   volume = 490,
    pages = {493-508},
   adsurl = {http://adsabs.harvard.edu/abs/1997ApJ...490..493N}
}

@ARTICLE{Brya98,
   author = {{Bryan}, G.~L. and {Norman}, M.~L.},
    title = "{Statistical Properties of X-Ray Clusters: Analytic and Numerical Comparisons}",
  journal = {\apj},
   eprint = {astro-ph/9710107},
     year = 1998,
    month = mar,
   volume = 495,
    pages = {80-99},
      doi = {10.1086/305262},
   adsurl = {http://adsabs.harvard.edu/abs/1998ApJ...495...80B}
}

@ARTICLE{Hahn11,
   author = {{Hahn}, O. and {Abel}, T.},
    title = "{Multi-scale initial conditions for cosmological simulations}",
  journal = {\mnras},
archivePrefix = "arXiv",
   eprint = {1103.6031},
 primaryClass = "astro-ph.CO",
     year = 2011,
    month = aug,
   volume = 415,
    pages = {2101-2121},
      doi = {10.1111/j.1365-2966.2011.18820.x},
   adsurl = {http://adsabs.harvard.edu/abs/2011MNRAS.415.2101H}
}

@dataset{wu_2026_20805009,
  author       = {Wu, Jianhao and
                  Chan, Tsang Keung and
                  Forouhar Moreno, Victor J.},
  title        = {btmw: Raw data for Blue-Tilt Milky-Way simulations},
  month        = jun,
  year         = 2026,
  publisher    = {Zenodo},
  doi          = {10.5281/zenodo.20805009},
  url          = {https://doi.org/10.5281/zenodo.20805009},
}

@software{wu_2026_20805091,
  author       = {Wu, Jianhao and
                  Chan, Tsang Keung and
                  Forouhar Moreno, Victor J.},
  title        = {btmw: Analysis code for Blue-Tilt Milky-Way simulations},
  month        = jun,
  year         = 2026,
  publisher    = {Zenodo},
  doi          = {10.5281/zenodo.20805091},
  url          = {https://doi.org/10.5281/zenodo.20805091},
}

%\printbibliography

\clearpage

\appendix

\section{Results from another halo finder: VELOCIraptor}\label{sec:VR}

Apart from HBT-HERONS, we also identify the halos and subhalos with VELOCIraptor (VR) \citep{Elahi19VELOCIraptor}, a code based on the friends-of-friends (FOF) algorithm that identifies peaks in 6D phase-space to find subhalos (both the configuration space and phase space). We compare the HBT-HERONS and VR results on the subhalo mass function, subhalo $V_{\rm max}$ function, and subhalo radial distribution in \autoref{sub:comp_HBT_VR}.

Currently, SOAP (see \autoref{sub:halofinder}) only allows the calculation of spherical overdensity properties for field halos. Besides, HBT-HERONS (without SOAP post process) could only calculate limited kinds of SO properties. Hence, we used VR (without SOAP post process) to calculate the virial masses for all the halos (also including subhalos) and show the $M_{\rm vir}-V_{\rm max}$ relationship in \autoref{sub:Mvir-Vmax}. Usually, the spherical overdensity properties will include all of the particles within the SO radius, regardless of whether they are bound to the main halo. However, for subhalos, VR only considers the bound particles\footnote{See \textit{Mass and radius properties} in \url{https://velociraptor-stf.readthedocs.io/en/latest/output.html}}. Otherwise, the subhalos near the center of the main halo will have unphysically large masses.

Our essential parameters for VR are listed in \autoref{tab:VR_config}, based on the sample configuration for 6D dark matter-only simulation \footnote{\url{https://velociraptor-stf.readthedocs.io/en/latest/_downloads/97ba16973bcff9d4afe940c2e13d12d2/sample_dmcosmological_6dfof_subhalo.cfg}}. We set the most bound particle's position as the reference frame center to be consistent with HBT-HERONS for comparison in \autoref{sub:comp_HBT_VR}.

\begin{table}[H]
\centering
\scalebox{1.0}{
\begin{tabular}{ll}
\toprule
\multicolumn{1}{l}{Parameters} & \multicolumn{1}{c}{Value} \\ \midrule 
FoF\_Field\_search\_type & 3 \\
Minimum\_halo\_size & 20 \\  
Bound\_halos & 1 \\
Kinetic\_reference\_frame\_type & 0 \\
Reference\_frame\_for\_properties & 1 \\
\bottomrule
\end{tabular}
}
\caption{Selected parameter settings in the VELOCIraptor configuration file. For detailed explanations of these parameters, refer to the usage section in the documentation \footnote{\url{https://velociraptor-stf.readthedocs.io/en/latest/usage.html}}.}
\label{tab:VR_config}
\end{table}

\subsection{Comparison between halo finders: HBT-HERONS and VELOCIraptor\label{sub:comp_HBT_VR}}

Here, we choose m12i to analyze the difference between the two halo finders. VR stands for VELOCIraptor in both the text and figures.  In this subsection, we also use SOAP to process VR and HBT-HERONS catalogs in a consistent manner.

However, for the previous literature we fit with, COCO \citep{Hellwing16cocoproject} and Aquarius \citep{Springel08Aquarius} use SUBFIND \citep{Springel01Gadget1} as their halo finder, which is a configuration space finder; Cautun fit line (\cite{Cautun14r100fit}) is derived via ROCKSTAR \citep{Behroozi13Rockstar}, a phase space finder. So there would be inherent offset simply due to the choice of halo finder, for the fitting lines. As a result, the deviation from fitting lines could not imply the capability of the halo finder. The relative ratio between the two halo finders would be more meaningful.

\begin{figure}[H]
    \centering
    \includegraphics[width={0.45\textwidth}]{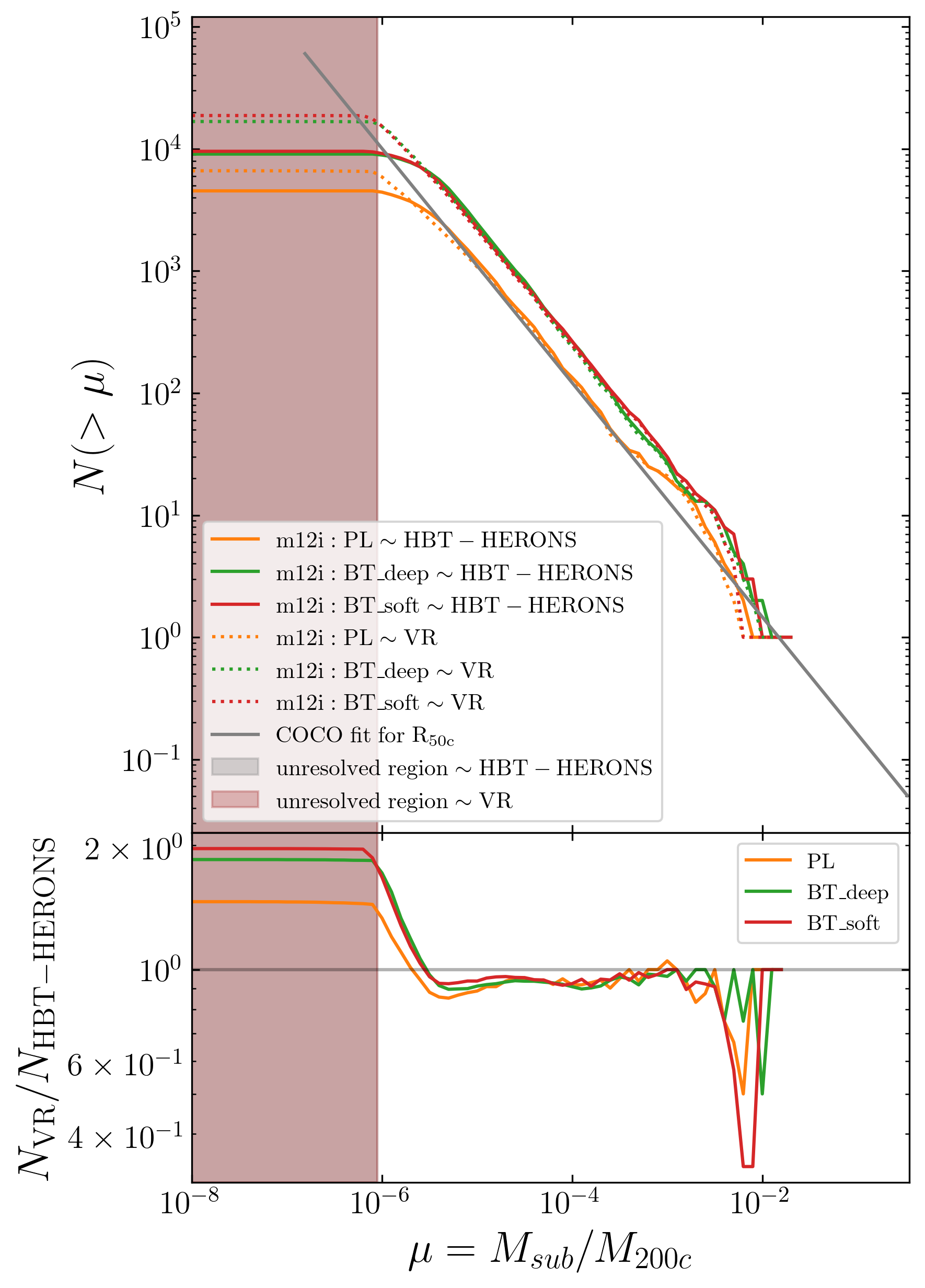}
    \caption{The cumulative scaled subhalo mass functions for m12i, with two halo finders HBT-HERONS (colored solid) and VR (colored dashed), respectively.
    The \textit{upper panel} shows the subhalo mass function for the PL (orange), BT\_deep (green), and BT\_soft (red) models, with the subhalo mass $M_{\rm sub}$ scaled by $M_{200c}$ of the main halo. The \textit{bottom panel} shows the ratios of the VR to HBT-HERONS number. The \textit{shaded areas} are the same as that in \autoref{fig:HMF_coco}. The shaded areas overlap because we set the minimum size of the subhalo to be the same in both halo finders.}
    \label{fig:HMF_VR_comp}
\end{figure}

The \textbf{subhalo mass function} comparison between VR and HBT-HERONS is shown in \autoref{fig:HMF_VR_comp}. These two halo finders agree with each other for $10^{-5} <\mu< 10^{-3}, 10^7~ \msun < M_{\rm sub} < 10^9~  \msun$. However, HBT-HERONS is higher and closer to the COCO fitting line than VR at the high mass end, implying that  HBT-HERONS is better at finding more massive subhalos. But at the low mass end, VR appears to have more subhalos than HBT-HERONS.

\begin{figure}[H]
    \centering
    \includegraphics[width={0.45\textwidth}]{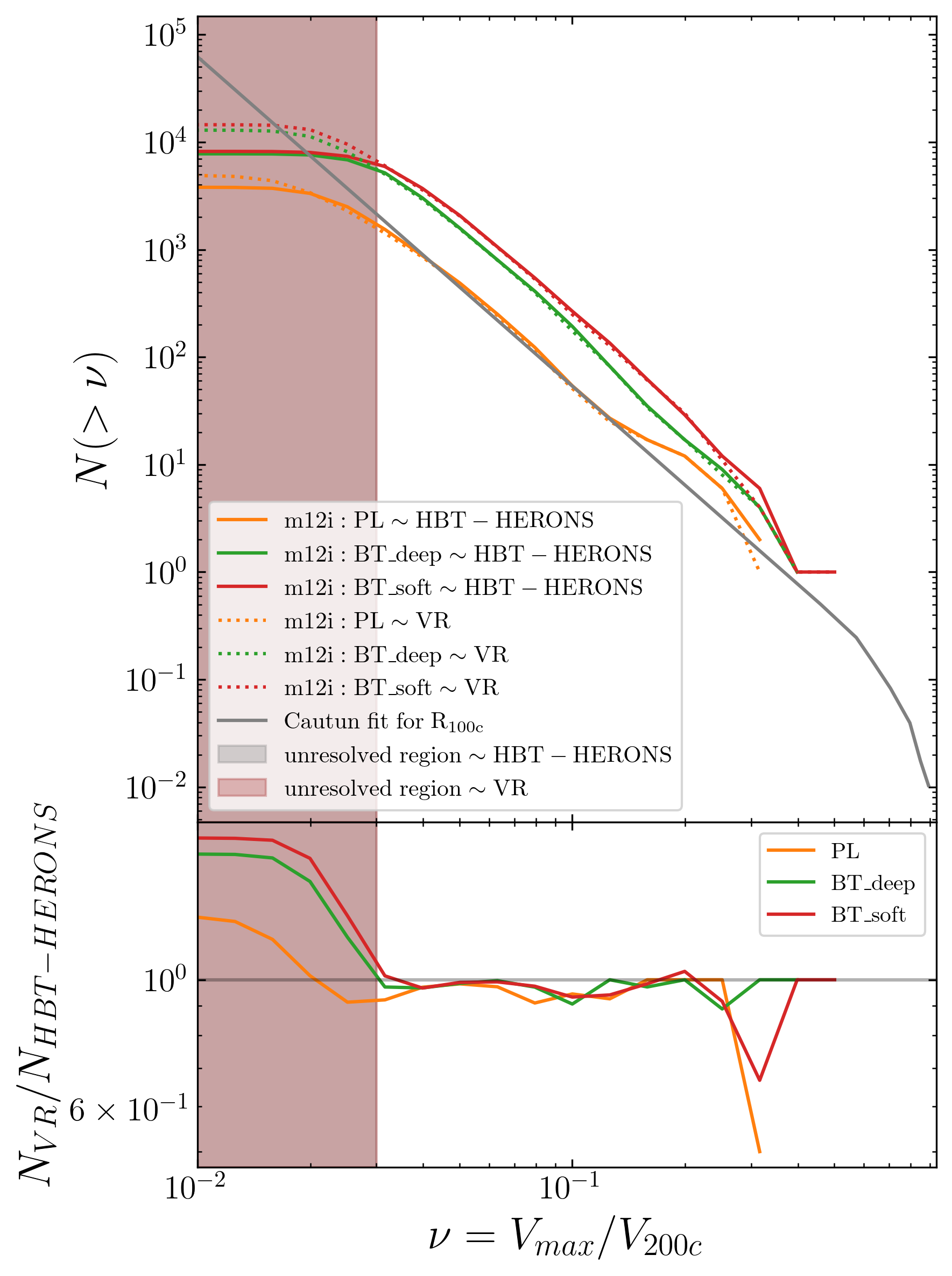}
    \caption{The cumulative scaled subhalo $V_{\rm max}$ functions for m12i, with two halo finders HBT-HERONS (colored solid) and VR (colored dashed), respectively. The \textit{upper panel} shows the subhalo $V_{\rm max}$ function for the PL (orange), BT\_deep (green), and BT\_soft (red) models, with the subhalo $V_{\rm max}$ scaled by $V_{200c}$ of the main halo. The \textit{bottom panel} shows the ratios of the VR to HBT-HERONS number. The \textit{shaded areas} are the same as those in \autoref{fig:HVF_cautun}.   }
    \label{fig:HVF_VR_comp}
\end{figure}

The \textbf{subhalo $V_{\rm max}$ function} comparison between VR and HBT-HERONS is shown in \autoref{fig:HVF_VR_comp}. These two halo finders give similar results except at the low $V_{\rm max}$ end near the resolution limit. When approaching the limit $V_{\rm max}$ value, VR deviates from the power-law behavior later than HBT-HERONS.

\begin{figure*}
    \centering
    \includegraphics[width=0.45\textwidth]{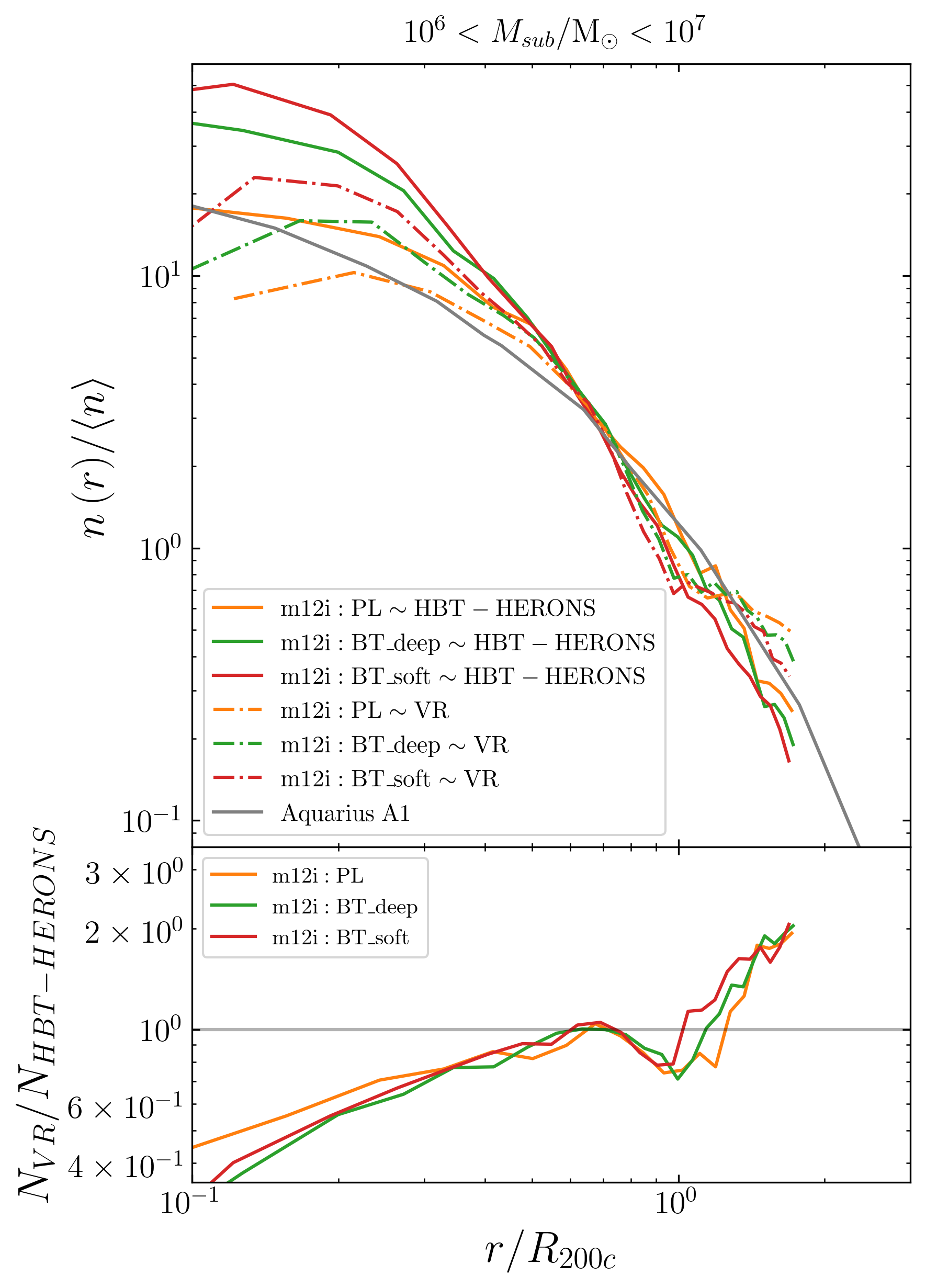}
    \hspace{0.2in}
    \includegraphics[width=0.45\textwidth]{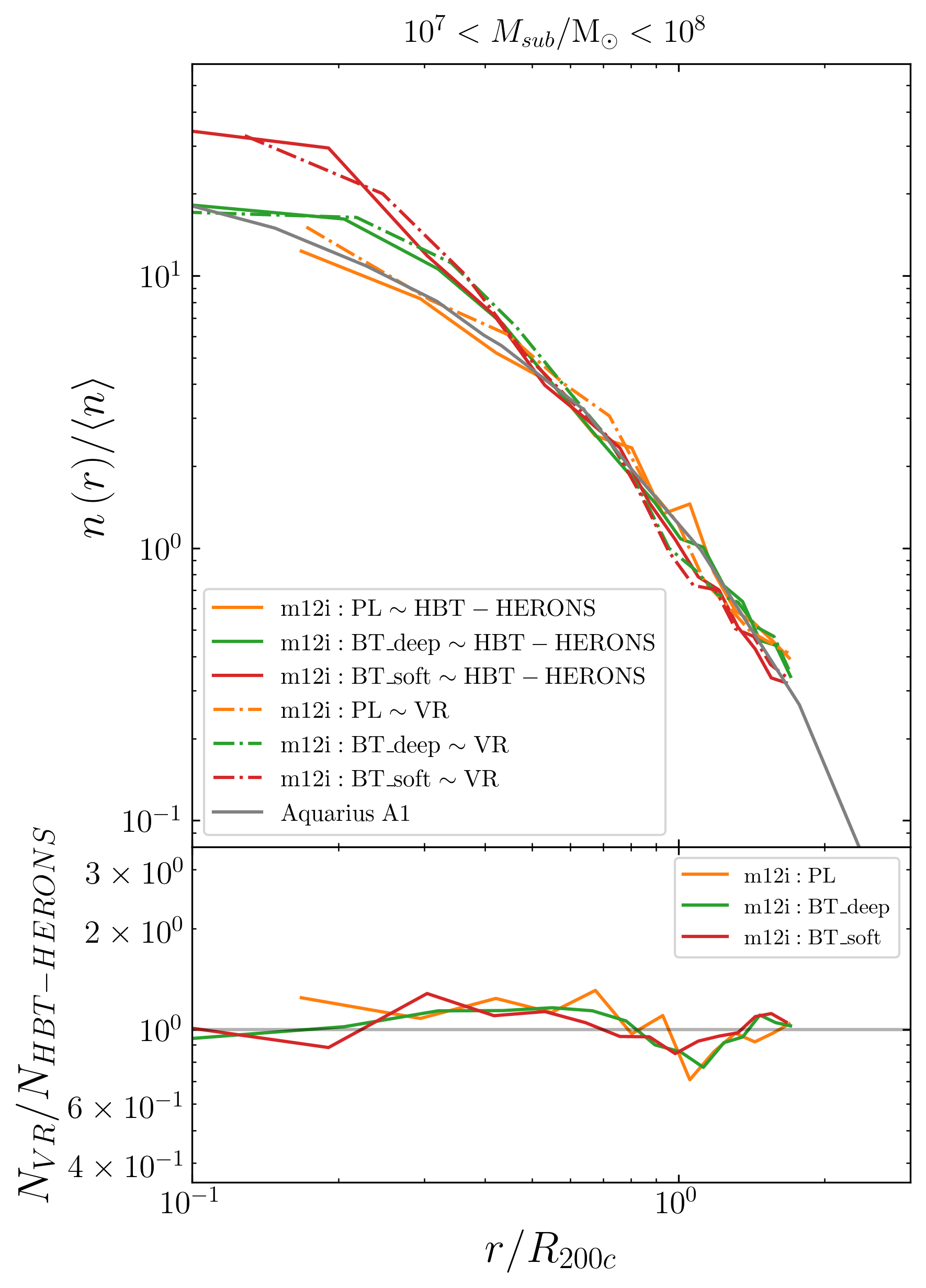}

    \includegraphics[width=0.45\textwidth]{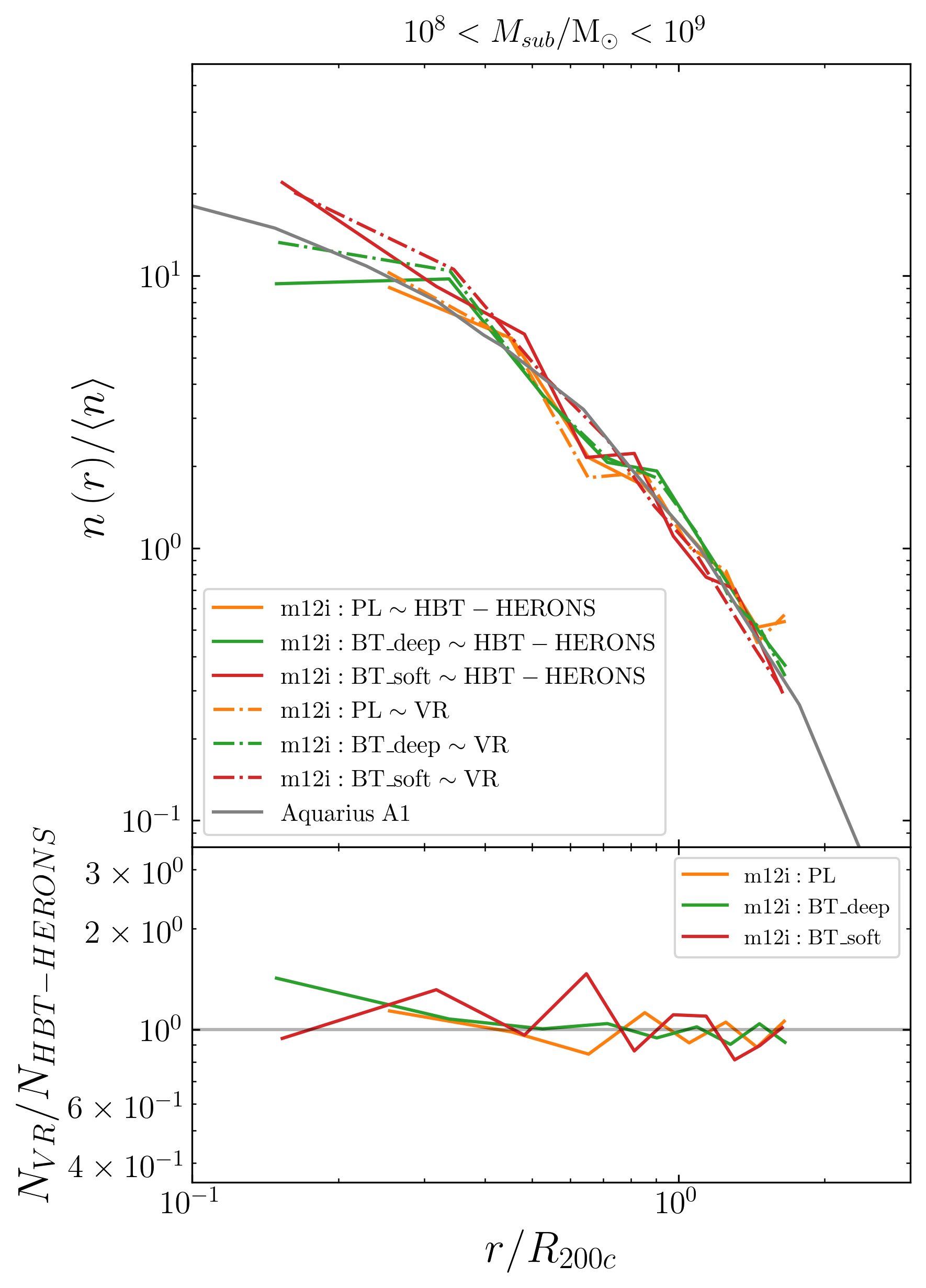}
    \hspace{0.2in}
    \includegraphics[width=0.45\textwidth]{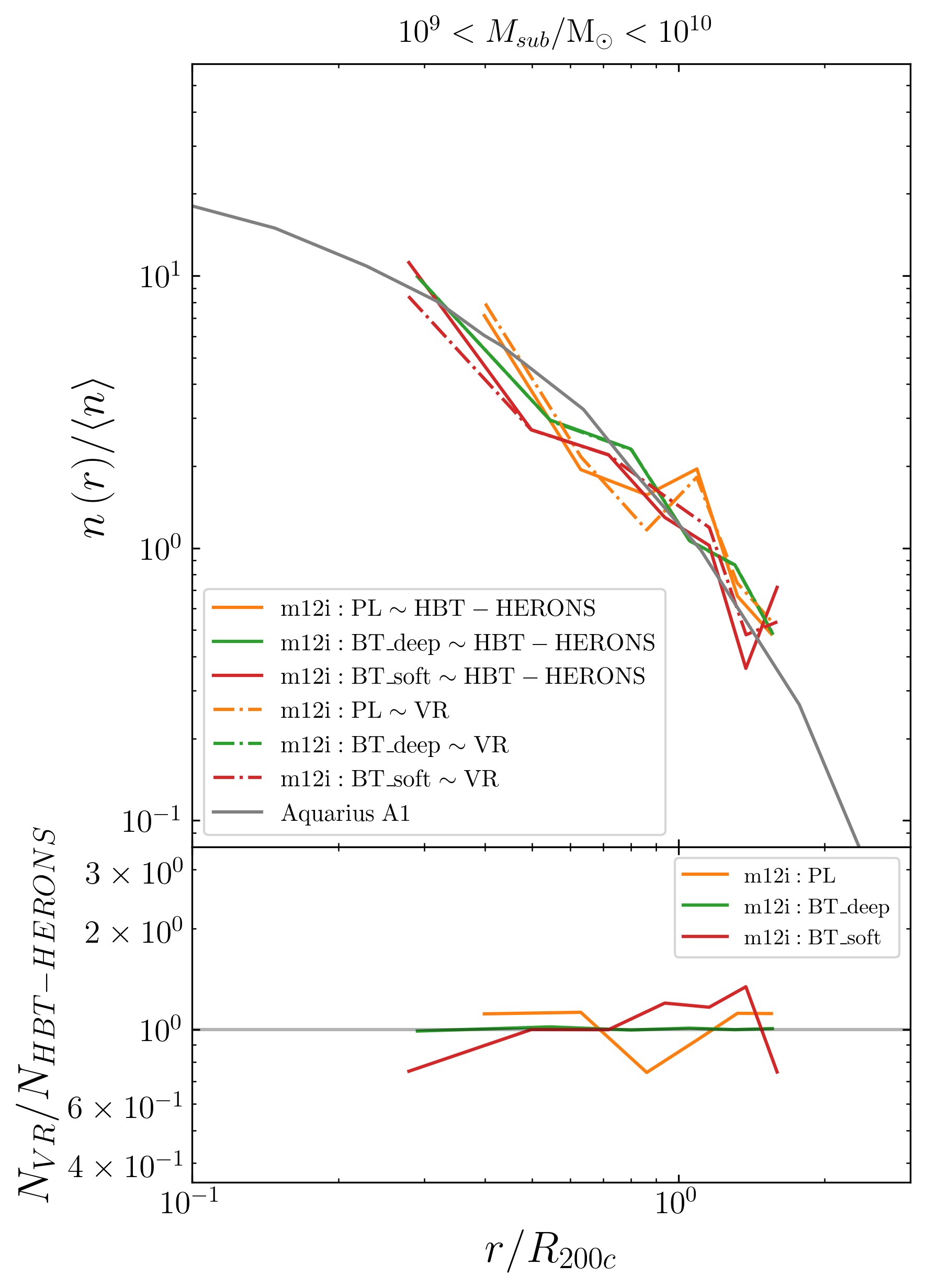}

    \caption{The subhalo radial number density profiles at four different mass bins for m12i, using different halo finders HBT-HERONS (colored solid) and VR (colored dashed). We consider the PL (orange), BT\_deep (green), and BT\_soft (red) models in each subfigure (also in each mass bin): the \textit{upper panel} is the same definition as that in \autoref{fig:HRF_Aquarius}; the \textit{bottom panel} shows the ratios of the VR results to the HBT-HERONS results.
    }
    \label{fig:HRF_Aquarius_comp}
\end{figure*}

The \textbf{subhalo radial distribution} comparison between VR and HBT-HERONS is shown in \autoref{fig:HRF_Aquarius_comp}. In mass bins larger than $10^7~ \msun$, VR and HBT-HERONS have nearly the same results. But, at the lower mass end, VR has much fewer subhalos in the inner region than HBT-HERONS and the Aquarius result. 

To summarize, VR and HBT-HERONS behave similarly for the subhalo mass and  $V_{\rm max}$ functions. However, for the radial properties (the subhalo radial distribution function), VR gaves a lower halo count near the host halo center for the low halo mass bin.

\subsection{Halo \texorpdfstring{$M_{\rm vir}-V_{\rm max}$}{Mvir-Vmax} relationship}
\label{sub:Mvir-Vmax}

\begin{figure*}
    \includegraphics[width=0.45\textwidth]{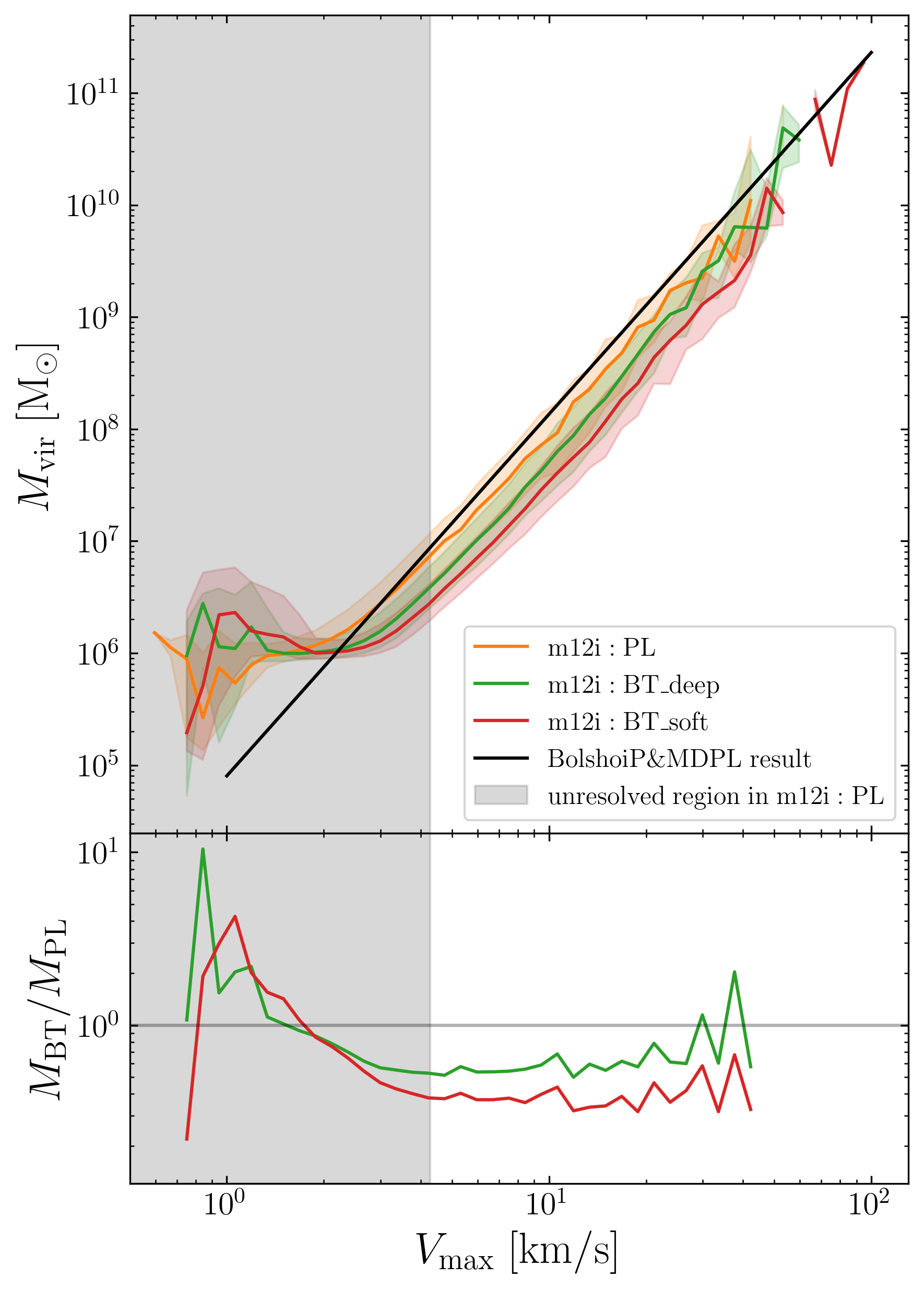}
    \includegraphics[width=0.45\textwidth]{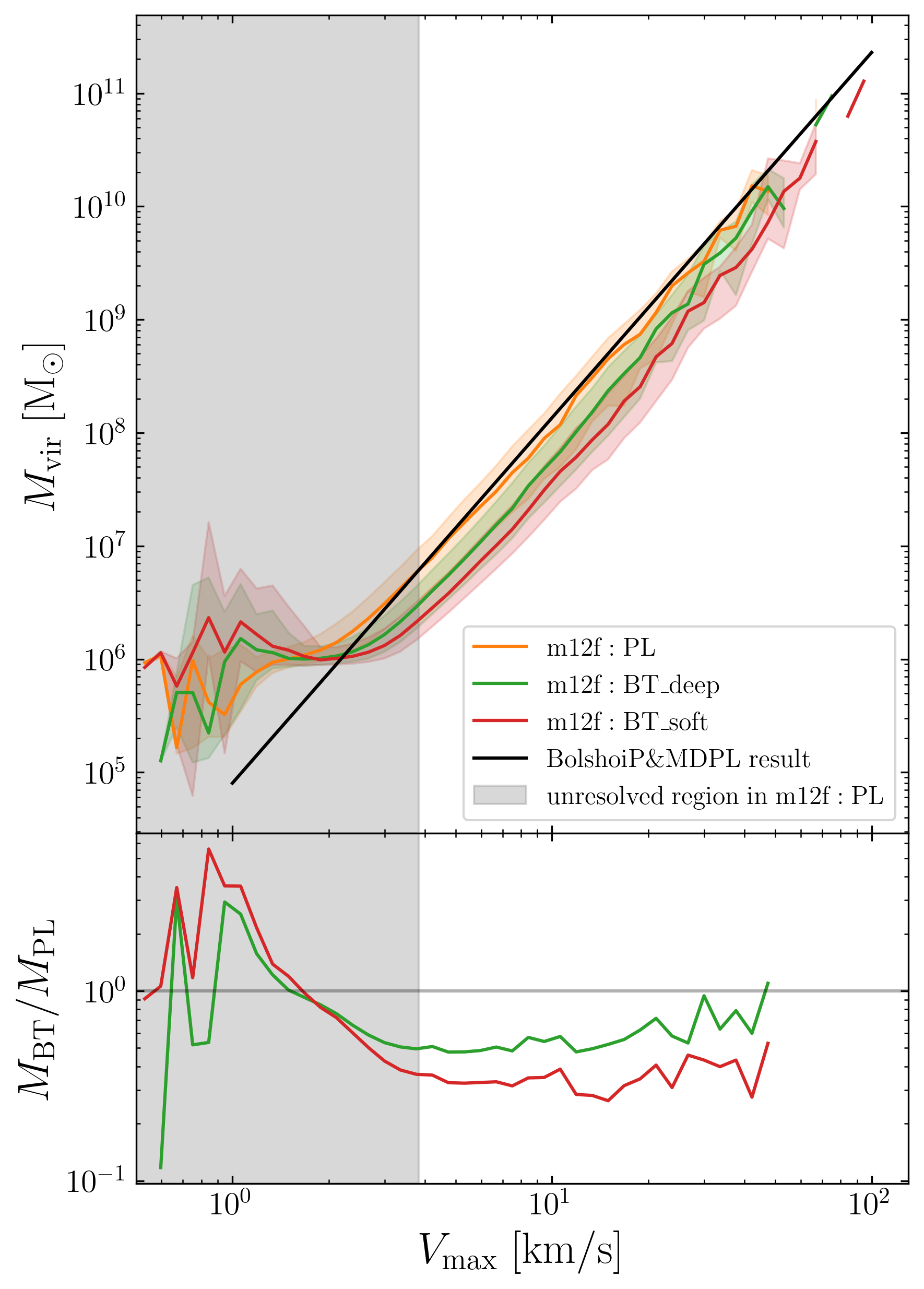}
    \caption{The subhalo $M_{\rm vir}-V_{\rm max}$ relations for m12i (left figure) and m12f (right figure) respectively. For both figures:\protect\\
    The \textit{upper panel} shows the \(M_{\rm vir}-V_{\rm max}\) relationship for all the field halos and subhalos within the simulation domain, with the PL (orange), BT\_deep (green), and BT\_soft (red) models. The colored solid lines depict the median relation by binning the subhalos according to their $V_{\rm max}$ values. The colored regions surrounding the colored lines illustrate the 16th-84th percentiles around the medians. The black solid line results from Ref. \cite{Puebla16BolshoiPMDPL} for the PL LCDM model, including both the field halos and subhalos.
    The \textit{bottom panel} shows the ratios of BT \(M_{\rm vir}\) to PL \(M_{\rm vir}\).
    The \textit{shaded area} is the same as that in \autoref{fig:HVF_cautun}.}
    \label{fig:MvsV}
\end{figure*}

We present the halo $M_{\rm vir}-V_{\rm max}$ relationship in \autoref{fig:MvsV} for all the field halos and subhalos. The black line is the extrapolation of Ref. \citep{Puebla16BolshoiPMDPL} for the PL LCDM cosmology, including both the field halos and subhalos.

The upper panels of \autoref{fig:MvsV} show a good match between the BolshoiP \& MDPL simulation and our PL simulation, for both m12i and m12f. The BT lines are lower than the PL lines, implying the BT halos are more concentrated. We also observe the BT\_soft model has a lower $M_{\rm vir}$ than the BT\_deep model for the same $V_{\rm max}$.

In the bottom panels, we show that the BT models suppress $M_{\rm vir}$ at the resolved $V_{\rm max}$ range. The suppressions in m12i and m12f are similar: for the BT\_deep model, $M_{\rm vir}$ decreases by around 30\%; for the BT\_soft model, $M_{\rm vir}$ decreases by around 50\%. At high velocities, the BT models are closer to the PL model's $M_{\rm vir}$ value. This implies the BT effects are stronger in the lower mass halos.

\clearpage

\section{Resolution limit study\label{sec:res}}

To investigate whether the conclusion still holds at a higher resolution, we take m12i as an example to study the consistency between different resolutions. In the high-resolution run, we reduce the particle mass to one-eighth of the original (and boost the number of particles by a factor of 8 times). The main halo properties for the high-resolution runs are also shown in \autoref{tab:result_simulations}.

\begin{figure}[H]
    \begin{center}
    \includegraphics[width={0.45\textwidth}]{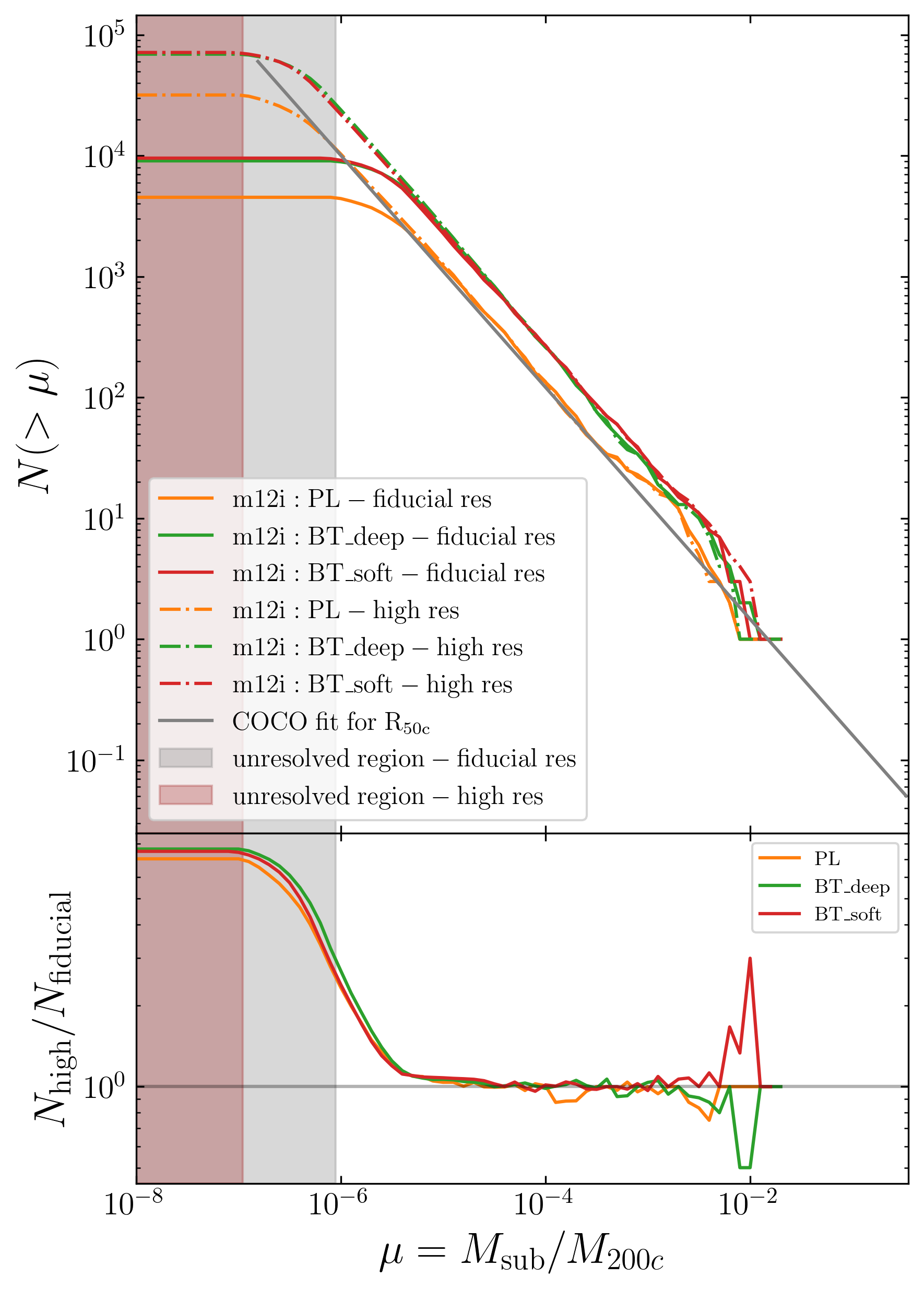}
    \end{center}
    \caption{The cumulative scaled subhalo mass functions for m12i, under fiducial resolution (colored solid) and high resolution (colored dashed), respectively.\protect\\
    The \textit{upper panel} shows the subhalo mass function for the PL (orange), BT\_deep (green), and BT\_soft (red) models, with the subhalo mass $M_{\rm sub}$ scaled by $M_{200c}$ of the main halo. The \textit{bottom panel} shows the ratios of high-resolution to fiducial-resolution numbers. The \textit{shaded areas} are the same definition as that in \autoref{fig:HMF_coco}. However, the high-resolution run has a smaller particle mass, so the lower bound is also lower.}
    \label{fig:HMF_reso}
    \vspace{-2mm}
\end{figure}

The \textbf{subhalo mass function} comparison between the high and fiducial resolution is shown in \autoref{fig:HMF_reso}. From the upper panel, we find that the high-resolution run can extend the power-law behavior to a lower mass range, matching the COCO fitting line. We find the overlapping of the deep and soft models in \autoref{fig:HMF_coco} still holds at a higher resolution. From the bottom panel, by comparing the high and fiducial runs, we find that the fiducial run deviates from the high-resolution run at around $\mu\sim 10^{-5.5}$. This implies that we can trust the results from halos with more than 50 particles. The main conclusions in \autoref{fig:HMF_coco} still hold.

\begin{figure}[H]
    \centering
    \includegraphics[width={0.45\textwidth}]{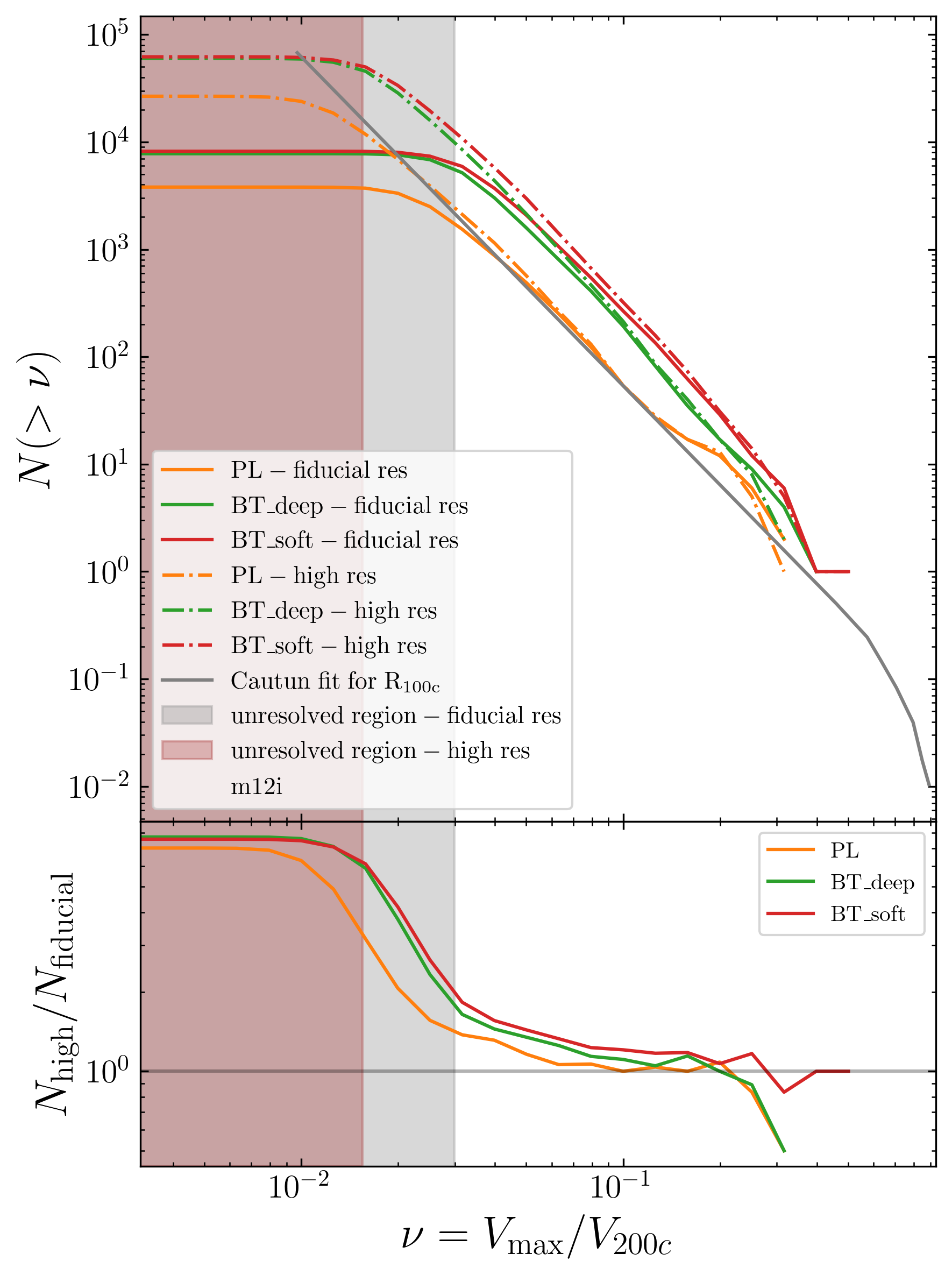}
    \caption{The cumulative scaled subhalo $V_{\rm max}$ functions for m12i, under fiducial resolution (colored solid) and high resolution (colored dashed), respectively.\protect\\
    The \textit{upper panel} shows the subhalo $V_{\rm max}$ function for the PL (orange), BT\_deep (green), and BT\_soft (red) models, with $V_{\rm max}$ of subhalo scaled by $V_{200c}$ of the main halo. The \textit{bottom panel} shows the ratios of the high-resolution to fiducial-resolution numbers. The \textit{shaded areas} are the same definition as that in \autoref{fig:HVF_cautun}. Similar to \autoref{fig:HMF_reso}, the high-resolution run can resolve the subhalos with a lower $V_{\rm max}$. }
    \label{fig:HVF_reso}
    \vspace{-2mm}
\end{figure}

The \textbf{subhalo $V_{\rm max}$ function} comparison between the high and fiducial resolution is shown in \autoref{fig:HVF_reso}. Similar to \autoref{fig:HMF_reso}, we find the power-law behavior to a lower $V_{\rm max}$ range. The BT\_deep line is higher than the BT\_soft line, which is higher than the PL line. Thus, as in \autoref{subsub:HVF}, we find that the subhalo $V_{\rm max}$ function is a better distinguisher for the BT models than the subhalo mass function. From the bottom panel, we find that the $V_{\rm max}$ of subhalo with 100 particles is reliable (with 30\%).

\begin{figure}[H]
    \includegraphics[width={0.45\textwidth}]{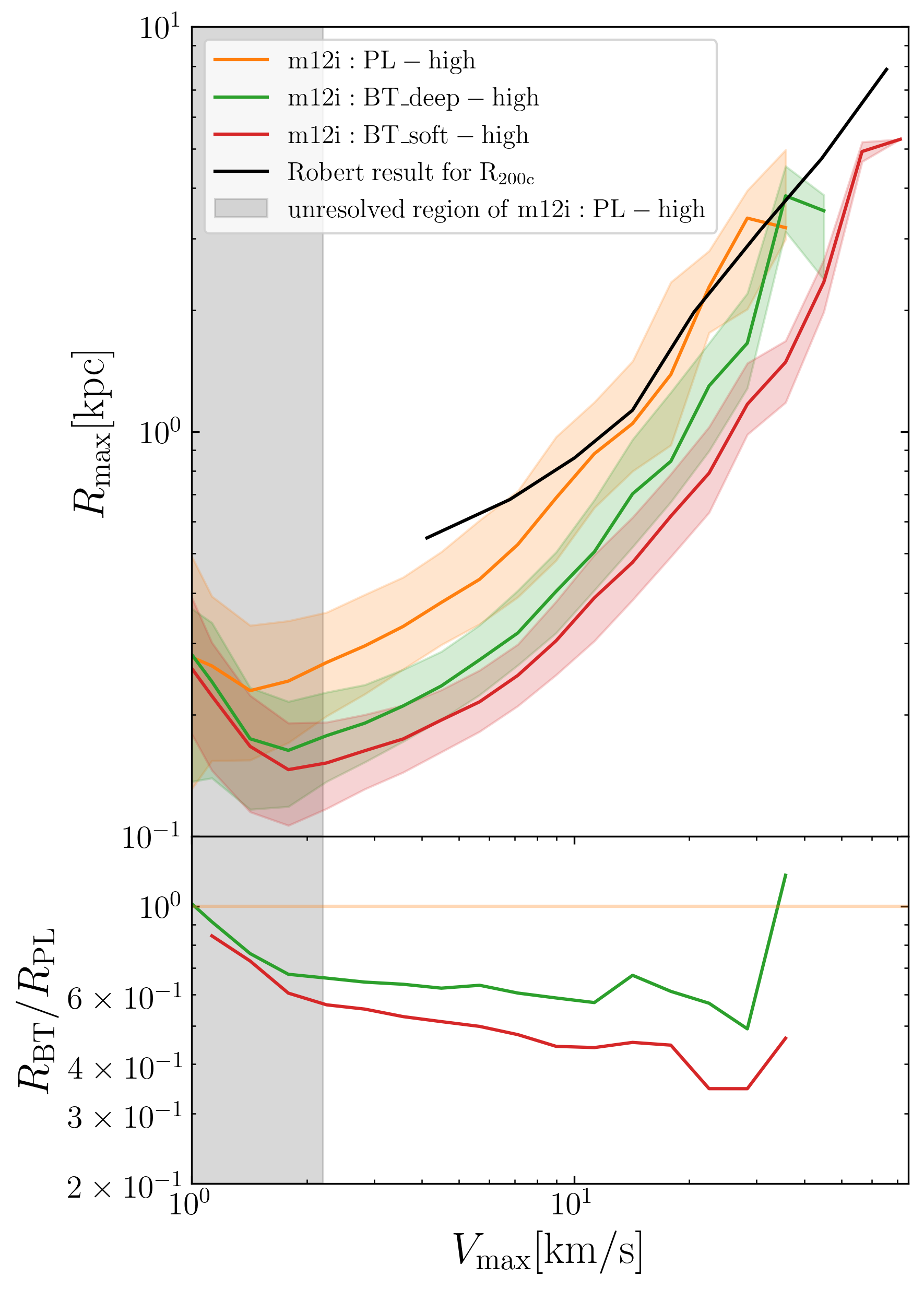}
    \caption{
    This figure is the same as the upper panel of \autoref{fig:RmaxVmax}, but with a higher resolution.
    The \textit{upper panel} shows the subhalo \(R_{\rm max}-V_{\rm max}\) relationship with the PL (orange), BT\_deep (green), and BT\_soft (red) models, in the \textbf{high-resolution run of m12i}; the colored solid lines depict the median relation found by binning the subhalos according to their $V_{\rm max}$ values. The colored regions surrounding the colored lines illustrate the 16th-84th percentiles around the medians. The black solid line is the result of  Ref. \protect\cite{Robert21RmaxVmax} for the PL LCDM model, taking account of all the subhalos within $R_{200c}$ of the main halo.
    The \textit{bottom panel} shows the ratios of \(R_{\rm max}\) between the BT models and the PL model. 
    The \textit{shaded area} is the same definition as in \autoref{fig:HVF_reso}.
    }
    \label{fig:RmaxVmax_reso}
\end{figure}

The \textbf{subhalo \(R_{\rm max} - V_{\rm max}\) relationship} at a higher resolution for m12i is shown in \autoref{fig:RmaxVmax_reso}. From the upper panel, we find that the PL result starts to diverge with the reference line \cite{Robert21RmaxVmax} from $V_{\rm max}\lesssim 10 ~{\rm km/s}$. It suggests that the $R_{\rm max}$ of subhalos at the low- $V_{\rm max}$ end is overestimated in the fiducial-resolution run (see the upper panels of \autoref{fig:RmaxVmax}). However, for the subhalos whose $V_{\rm max}$ is beyond $10 ~{\rm km/s}$, the results on different resolutions are consistent; besides, the relative differences among different PPS models still hold: from the bottom panel, there is a $40\%$ ($50\%$) decrease in BT\_deep (BT\_soft) in $R_{\rm max}$ at the same $V_{\rm max}$ bin.

\begin{figure*}
    \includegraphics[width=0.45\textwidth]{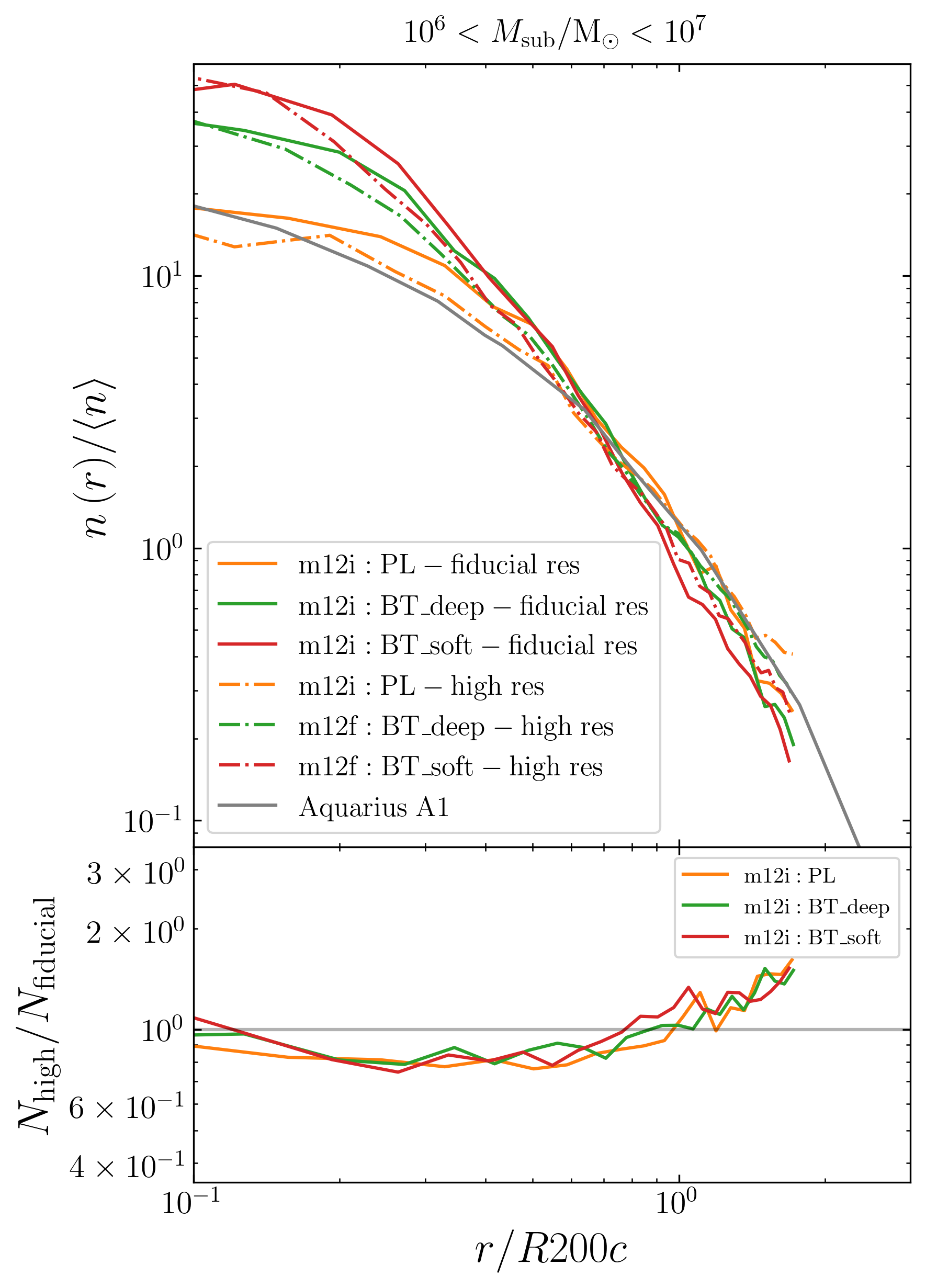}
    \hspace{0.2in}
    \includegraphics[width=0.45\textwidth]{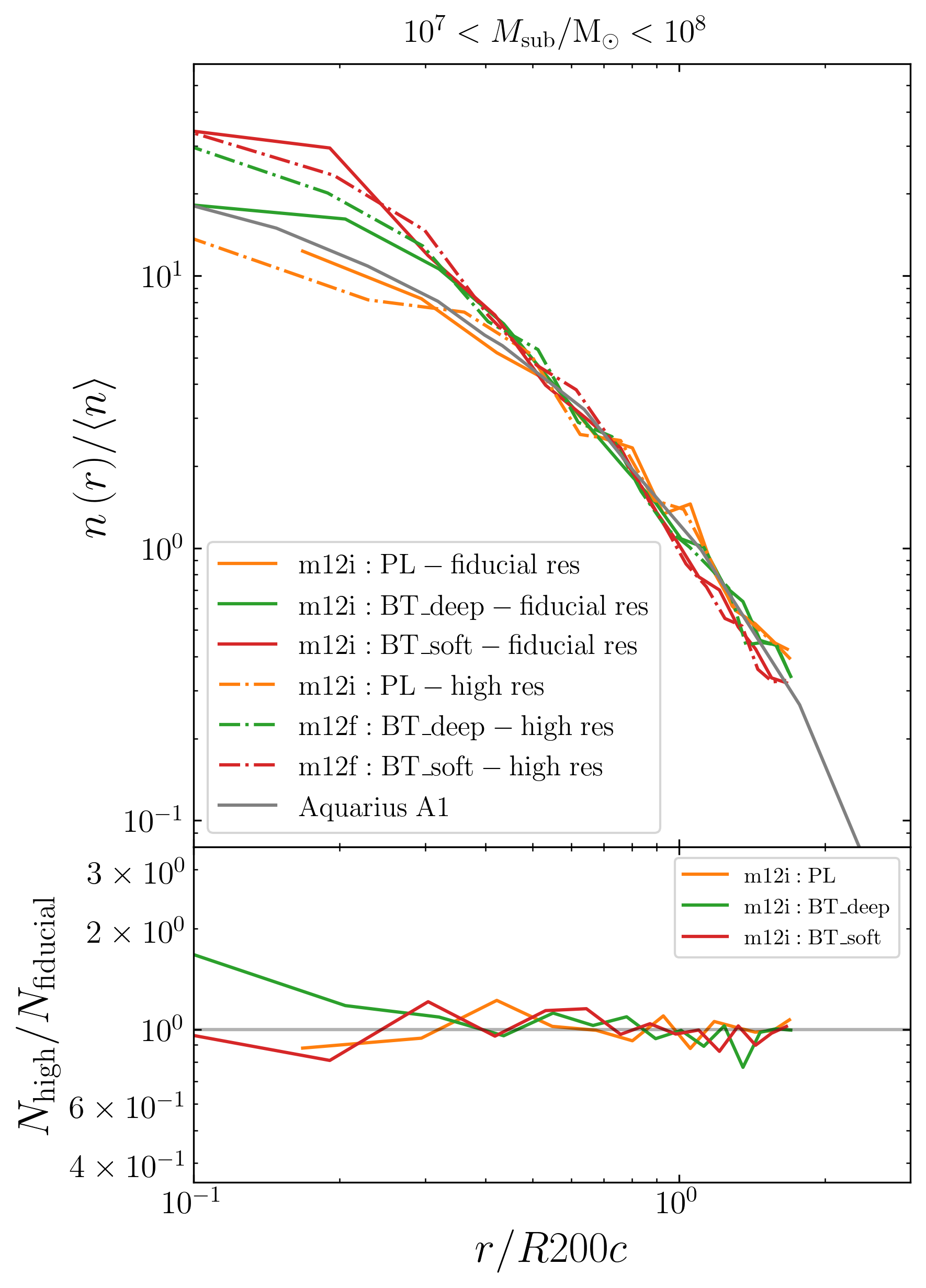}

    \includegraphics[width=0.45\textwidth]{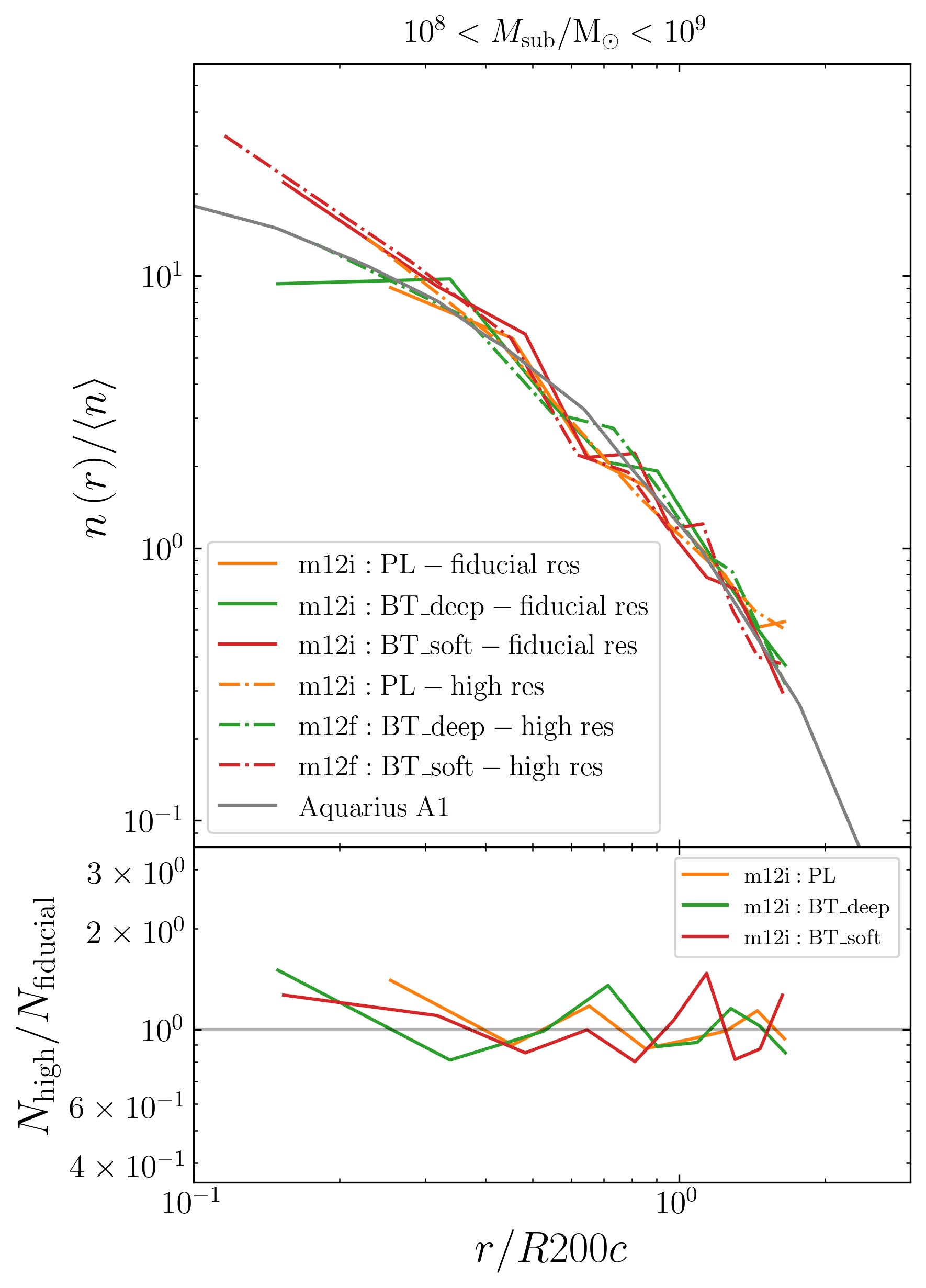}
    \hspace{0.2in}
    \includegraphics[width=0.45\textwidth]{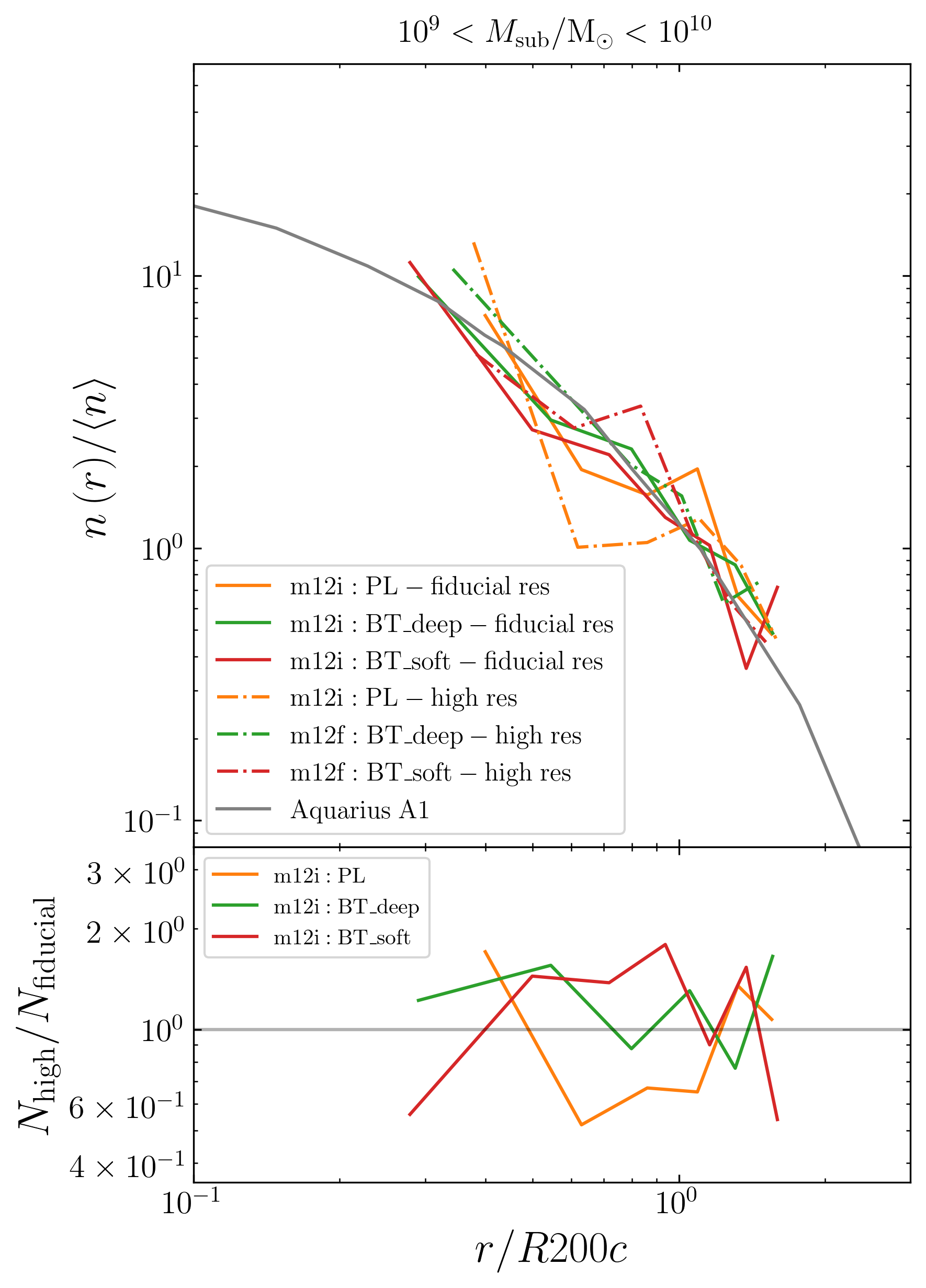}
    \caption{The subhalo radial number density profiles at four different mass bins for m12i, with the fiducial (colored solid) and high resolution (colored dashed), respectively. We consider the PL (orange), BT\_deep (green), and BT\_soft (red) models in each subfigure (also in each mass bin):
    The \textit{upper panel} is the same definition as that in \autoref{fig:HRF_Aquarius}.
    The \textit{bottom panel} shows the ratios of the result under the high resolution to the result with the fiducial resolution.}
    \label{fig:HRF_reso}
\end{figure*}

The \textbf{subhalo radial distribution} comparison between high resolution and fiducial resolution is shown in \autoref{fig:HRF_reso}. The upper panels show that the relative differences among different PPS models still hold: at the inner region $BT\_soft > BT\_deep > PL$. The bottom panels show that for the mass bins larger than $10^7~ \msun$, the result is quite close at these two resolutions within scatters. However, consistent with what we see in the mass function comparison, the mass bin $10^6 ~\msun < M_{\rm sub} < 10^7 ~\msun$, part of which is close to the minimum halo size of the fiducial-resolution run, shows the largest difference at two resolutions: the fiducial case has over-predicted subhalo number by a quarter in the inner region, while underestimated that by around 30\% in the outer region.

\textbf{Summary}: We have proven the fiducial-resolution simulation's conclusions still hold at a higher resolution. We can trust results from halos with $\gtrsim 50$ particles for the subhalo mass function and $V_{\rm max} \gtrsim 10 km/s$ for the subhalo $R_{\rm max}-V_{\rm max}$ relation and the subhalo $V_{\rm max}$ function.

\end{document}